\documentclass[12pt]{article}
\usepackage{epsfig}
\usepackage{color}
\usepackage{amssymb,amsmath}
\usepackage{graphicx}
\usepackage{here}
\newcommand{\ba}{\begin{align}}
\newcommand{\ea}{\end{align}}

 \makeatletter
\def\alt{\mathrel{\mathpalette\gl@align<}}
\def\agt{\mathrel{\mathpalette\gl@align>}}
\def\gl@align#1#2{\lower.6ex\vbox{\baselineskip\z@skip\lineskip\z@
\ialign{$\m@th#1\hfil##\hfil$\crcr#2\crcr\sim\crcr}}} \makeatother

\begin{document}
\begin{flushright}
\end{flushright}
\vspace*{1.0cm}

\begin{center}
\baselineskip 20pt 
{\Large\bf 
Heavy neutrino mixing in the T2HK, the T2HKK and \\
an extension of the T2HK with a detector at Oki Islands
}
\vspace{1cm}

{\large 
Yugo Abe$^{a,b}$, \ Yusuke Asano$^a$, \ Naoyuki Haba$^a$ \ and \ Toshifumi Yamada$^a$
} \vspace{.5cm}

{\baselineskip 20pt \it
$^a$ Graduate School of Science and Engineering, Shimane University, Matsue 690-8504, Japan
\\
$^b$ National Institute of Technology, Miyakonojo College, Miyakonojo-shi Miyazaki 885-8567, Japan
}

\vspace{.5cm}

\vspace{1.5cm} {\bf Abstract} \end{center}

We study the discovery potential for the mixing of heavy isospin-singlet neutrinos in extensions of the Tokai-to-Kamioka (T2K) experiment,
 the Tokai-to-Hyper-Kamiokande (T2HK), the Tokai-to-Hyper-Kamiokande-to-Korea (T2HKK) with a Korea detector with $\simeq1000$~km baseline length and 1$^\circ$ off-axis angle, and a plan of adding a small detector at Oki Islands to the T2HK.
We further pursue the possibility of measuring the neutrino mass hierarchy and the standard $CP$-violating phase $\delta_{CP}$ in the presence of heavy neutrino mixing by fitting data with the standard oscillation parameters only.
We show that the sensitivity to heavy neutrino mixing is highly dependent on $\delta_{CP}$ and new $CP$-violating phases in the heavy neutrino mixing matrix,
 and deteriorates considerably when these phases conspire to suppress interference between the standard oscillation amplitude and an amplitude arising from heavy neutrino mixing, at the first oscillation peak.
Although this suppression is avoided by the use of a beam with smaller off-axis angle, 
 the T2HKK and the T2HK+small Oki detector do not show improvement over the T2HK.
As for the mass hierarchy measurement, the wrong mass hierarchy is possibly favored in the T2HK because heavy neutrino mixing can mimic matter effects.
In contrast, the T2HKK and the T2HK+small Oki detector are capable of correctly measuring the mass hierarchy despite heavy neutrino mixing,
 as measurements with different baselines resolve degeneracy between heavy neutrino mixing and matter effects.
Notably, adding a small detector at Oki to the T2HK drastically ameliorates the sensitivity, which is the central appeal of this paper.
As for the $\delta_{CP}$ measurement, there can be a sizable discrepancy between the true $\delta_{CP}$ and the value obtained by fitting data with the standard oscillation parameters only, as large as $1\sigma$ resolution of the $\delta_{CP}$ measurement.
Hence, if a hint of heavy neutrino mixing is discovered, it is necessary to incorporate the effects of heavy neutrino mixing to measure $\delta_{CP}$ accurately.

\thispagestyle{empty}

\newpage

\baselineskip 18pt
%

\section{Introduction}

It is a viable possibility that isospin-singlet neutrinos mix with active flavors $\nu_e,\nu_\mu,\nu_\tau$ after electroweak symmetry breaking.
Notably, such a mixing is inherent in the Type-I seesaw model~\cite{seesaw},
 and although the mixing angle is undetectably small in most of the parameter space,
 it can be sizable when there is a cancellation in the seesaw mass.
To see this, suppose, for example, there are three isospin-singlet neutrinos with a Majorana mass $M_N$, having a Dirac mass term $m_D$ with active neutrinos.
Since the tiny neutrino mass is derived as $-m_D M_N^{-1} m_D^T$, one can express $m_D$ as~\cite{ci}
\begin{align}
m_D &= U_{PMNS} \, i\left(
\begin{array}{ccc}
\sqrt{m_1} & 0 & 0 \\
0 & \sqrt{m_2} & 0 \\
0 & 0 & \sqrt{m_3}
\end{array}
\right) \, R_3(\alpha,\beta,\gamma) \, \sqrt{M_N},
\end{align}
 where $U_{PMNS}$ denotes the Pontecorvo-Maki-Nakagawa-Sakata (PMNS) mixing matrix~\cite{mns,p},
 $m_1,m_2,m_3$ are the tiny neutrino mass, and
 $R_3$ is a 3-dimensional rotation matrix with complex-valued rotation angles $\alpha,\beta,\gamma$.
Unless $m_1,m_2,m_3$ are degenerate, one can arbitrarily enhance $m_D$ and the mixing angles between isospin-singlet and active neutrinos
 by taking large imaginary values for $\alpha,\beta,\gamma$.
In this case, a cancellation is taking place in the components of $-m_D^T M_N^{-1} m_D$.
Large mixing between isospin-singlet and active neutrinos is also realized in extensions of the Type-I seesaw model, such as
 the inverse seesaw model~\cite{inverse}.

In the presence of isospin-singlet neutrino mixing,
 transitions among active flavors require for their consistent description extra parameters beyond the standard oscillation parameters,
 the three mixing angles $\theta_{12},\theta_{23},\theta_{13}$, one $CP$-violating phase $\delta_{CP}$ and two mass differences $\Delta m_{12}^2,\Delta m_{13}^2$.
Tension in fitting neutrino oscillation data with the standard parameters would thus be evidence for isospin-singlet neutrino mixing.
If isospin-singlet neutrinos are "heavy", by which we signify that their mass is much above the neutrino beam energy,
 they do not contribute to oscillations and simply alter
 the mixing matrix $U_{PMNS}$ into a non-unitary matrix in the following fashion:
\begin{align}
U_{PMNS} &\to \left(
\begin{array}{ccc}
\alpha_{11} & 0 & 0 \\
\alpha_{21} & \alpha_{22} & 0 \\
\alpha_{31} & \alpha_{32} & \alpha_{33}
\end{array}
\right) U_{PMNS}
\label{nu}
\end{align}
 where $\alpha_{ii}$'s are real and $\alpha_{ij}$'s $(i\neq j)$ are complex numbers.

One purpose of this paper is to estimate the largest possible significance of heavy isospin-singlet neutrino mixing
 in the Tokai-to-Hyper-Kamiokande (T2HK)~\cite{t2hk} and the Tokai-to-Hyper-Kamiokande-to-Korea (T2HKK)~\cite{t2hkk} (for early proposals, see Refs.~\cite{t2kk,t2kok}) experiments, 
 as well as an experiment in which a new detector at Oki Islands is added to the T2HK (for early proposals on a detector at Oki, see Refs.~\cite{t2ko,t2kok}).
Emphasis is on how the significance varies with $CP$-violating phases in the PMNS and the non-unitary mixing matrices
 and to what extent the sensitivity deteriorates with these phases.
The secondary goal of this paper is to pursue the possibility that the neutrino mass hierarchy and the standard $CP$-violating phase in the PMNS matrix $\delta_{CP}$ are measured in the presence of heavy neutrino mixing by fitting neutrino data with the standard oscillation parameters only, namely, without assuming heavy neutrino mixing.
This is achievable if data allow us to discern contributions from the standard oscillation and those from the non-unitary mixing matrix.
Such measurement is of practical importance when a hint of heavy neutrino mixing were discovered but were not conclusive.
The mass hierarchy and $\delta_{CP}$ measurement in the presence of heavy neutrino mixing has been investigated in a general context~\cite{general} and for the T2K, T2HK, NO$\nu$A and DUNE experiments~\cite{ge,dutta1,dutta2,dune}.

The T2HK and the T2HKK are proposed extensions of the on-going T2K experiment~\cite{t2k},
 utilizing an off-axis $\nu_\mu$+$\bar{\nu}_{\mu}$ beam delivered from J-PARC.
In the T2HK, there are two water Cerenkov detectors, each with 187~kton fiducial volume, located solely at Kamioka, where the baseline length is $L=295$~km and the off-axis-angle is 2.5$^\circ$.
In one plan of the T2HKK, one water Cerenkov detector with 187~kton fiducial volume is placed at Kamioka
 and the same detector is also at a site in Korea where the baseline length is $L\simeq1000$~km and the off-axis-angle is 1$^\circ$.
We also consider a plan of placing, in addition to the T2HK, a small 10~kton water Cerenkov detector at Oki Islands, where the baseline length is $L=653$~km and the off-axis-angle is 1.0$^\circ$.

Merits of the above long baseline experiments for the search for heavy neutrino mixing are as follows:
(i) In disappearance experiments, the effects of heavy neutrino mixing as parametrized by Eq.~(\ref{nu}) cancel between measurements at the near and far detectors, while the T2K and its extensions are primarily appearance experiments.
(ii) $\nu_\mu \to \nu_e$ ($\bar{\nu}_\mu \to \bar{\nu}_e$) transition amplitude arising from the non-unitary mixing matrix Eq.~(\ref{nu})
 interferes with the standard oscillation amplitude, enhancing the impact of non-unitary mixing on the transition probability.
The merit (ii), however, bears two subtleties:
\begin{itemize}
\item
$CP$-violating phases in the PMNS and the non-unitary mixing matrices possibly lead to suppression of the interference.
\item
Non-unitary mixing $\alpha_{21}$ in Eq.~(\ref{nu}) can mimic matter effects and the effect of the standard phase $\delta_{CP}$,
 which distorts the measurement of the mass hierarchy and $\delta_{CP}$.
\end{itemize}
These issues become transparent by expressing the transition amplitude in the leading order of $\sin\theta_{13}$, $\Delta m_{21}^2$, $\alpha_{21}$ and matter effects as
\begin{align}
i\langle \nu_e \vert e^{-i\int_0^L {\rm d}x \, H(x)} \vert \nu_\mu \rangle 
&= 2 \, e^{-i\Delta m_{31}^2 L/(4E)} \sin\left( \frac{\Delta m_{31}^2 L}{4E} \right)e^{-i\delta_{CP}}\sin\theta_{13}\sin\theta_{23} 
+ \frac{\Delta m_{21}^2 L}{4E}\sin(2\theta_{12})\cos\theta_{23} 
\nonumber \\
&+ i\alpha_{21}^*
\nonumber \\
&-iV_{cc} \, L \, \left\{ \, \frac{\Delta m_{31}^2 L}{4E}e^{-i\delta_{CP}}\sin\theta_{13}\sin\theta_{23} + \frac{1}{2}\frac{\Delta m_{21}^2 L}{4E}\sin(2\theta_{12})\cos\theta_{23} \, \right\},
\label{amp1}
\\
i\langle \bar{\nu}_e \vert e^{-i\int_0^L {\rm d}x \, H(x)} \vert \bar{\nu}_\mu \rangle
&= 2 \, e^{-i\Delta m_{31}^2 L/(4E)} \sin\left( \frac{\Delta m_{31}^2 L}{4E} \right)e^{i\delta_{CP}}\sin\theta_{13}\sin\theta_{23} 
+ \frac{\Delta m_{21}^2 L}{4E}\sin(2\theta_{12})\cos\theta_{23}
\nonumber \\
&+ i\alpha_{21}
\nonumber \\
&+iV_{cc} \, L \, \left\{ \, \frac{\Delta m_{31}^2 L}{4E}e^{i\delta_{CP}}\sin\theta_{13}\sin\theta_{23} + \frac{1}{2}\frac{\Delta m_{21}^2 L}{4E}\sin(2\theta_{12})\cos\theta_{23} \, \right\},
\label{amp2}
\end{align}
 where $E$ denotes the neutrino energy and $L$ the baseline length, and $V_{cc}$ is the potential due to charged current interaction inside matter.

We notice that the following interference terms,
\begin{align}
&\vert\langle \nu_e \vert \, e^{-i\int_0^L {\rm d}x \, H(x)} \, \vert \nu_\mu \rangle\vert^2
\nonumber \\
&\supset -4\sin\left( \frac{\Delta m_{31}^2 L}{4E} \right)\sin\theta_{13}\sin\theta_{23} \, {\rm Im}\left(\alpha_{21} e^{-i\Delta m_{31}^2 L/(4E)-i\delta_{CP}}\right) + 2\frac{\Delta m_{21}^2 L}{4E}\sin(2\theta_{12})\cos\theta_{23} \, {\rm Im}(\alpha_{21}),
\label{interference1}
\\
&\vert\langle \bar{\nu}_e \vert \, e^{-i\int_0^L {\rm d}x \, H(x)} \, \vert \bar{\nu}_\mu \rangle\vert^2
\nonumber \\
&\supset 4\sin\left( \frac{\Delta m_{31}^2 L}{4E} \right)\sin\theta_{13}\sin\theta_{23} \, {\rm Im}\left(\alpha_{21} e^{i\Delta m_{31}^2 L/(4E)-i\delta_{CP}}\right) - 2\frac{\Delta m_{21}^2 L}{4E}\sin(2\theta_{12})\cos\theta_{23} \, {\rm Im}(\alpha_{21}),
\label{interference2}
\end{align}
 are simultaneously suppressed at the first oscillation peak, $\Delta m_{31}^2 L/(4E)=\pi/2$,
 if Arg($\alpha_{21})\simeq\delta_{CP}\pm\pi/2$ holds accidentally
(note the relation $\Delta m_{21}^2 L/(4E) \simeq 0.1\cdot2\sin\theta_{13}$ at the first oscillation peak).
This suppression is mitigated if the energy distribution is not concentrated at the first oscillation peak.
We thus infer that when Arg($\alpha_{21})\simeq\delta_{CP}\pm\pi/2$ holds, 
 the T2HKK with a Korea detector placed at $1^\circ$ off-axis angle exhibits better sensitivity to $\alpha_{21}$ than the T2HK,
 and that the extension of the T2HK with an Oki detector gives qualitative improvement.
Of course, these anticipations depend on statistics, namely, competition with the loss of statistics in the T2HKK, or
 the size of an Oki detector.

We observe in Eqs.~(\ref{amp1}),~(\ref{amp2}) that when $\delta_{CP}\simeq 0,\pi$, the real part of non-unitary mixing Re($\alpha_{21}$) plays exactly the same role as matter effects, 
 which afflicts the measurement of the mass hierarchy because it is probed through interference with matter effects.
However, when measurements with different baselines and hence different $V_{cc}L$'s are combined, one resolves degeneracy between the non-unitary mixing and matter effects, thereby resurrecting sensitivity to the mass hierarchy.
We thus expect that the mass hierarchy measurement in the T2HK is subject to strong influence from heavy neutrino mixing,
 whereas the T2HKK maintains sensitivity to the mass hierarchy even with heavy neutrino mixing.
Also, the addition of a small detector at Oki to the T2HK leads to qualitative improvement in the mass hierarchy measurement.

Eqs.~(\ref{amp1}),~(\ref{amp2}) tell us that when $\delta_{CP}\simeq$Arg$(\alpha_{21})$,
 non-unitary mixing $\alpha_{21}$ and the first term in the first line of Eqs.~(\ref{amp1}),(\ref{amp2}), which is proportional to $e^{i\delta_{CP}}$, interfere maximally.
In such a case, the presence of $\alpha_{21}$ affects the measurement of $\delta_{CP}$ and causes
 a discrepancy between the true $\delta_{CP}$ and the value measured by fitting data with the standard oscillation parameters only.
This effect is mitigated if the baseline is longer, because more information about the energy dependence of the standard oscillation amplitude
 helps distinguish it from non-unitary mixing $\alpha_{21}$.
We thus expect that the discrepancy is smaller in the T2HKK than in the T2HK.

In the main body of this paper, we demonstrate, through a simulation, the following:
\\

\noindent
(a) The sensitivity to heavy neutrino mixing is suppressed for Arg($\alpha_{21})\simeq\delta_{CP}\pm\pi/2$.
Contrary to our anticipations, the T2HK shows better sensitivity than the T2HKK for any values of $CP$-violating phases, simply because of higher statistics.
Also, adding a small detector at Oki to the T2HK does not make qualitative difference.
\\

\noindent
(b) In the presence of heavy neutrino mixing, the wrong mass hierarchy can be favored in the T2HK.
The T2HKK maintains sensitivity to the mass hierarchy and measures it correctly despite heavy neutrino mixing.
The extension of the T2HK with a small detector at Oki exhibits revived sensitivity to the mass hierarchy, even if statistics at Oki is quite subdominant compared to that at Kamioka.
\\

\noindent
(c) In the presence of heavy neutrino mixing, the true $\delta_{CP}$ and the value obtained by fitting data with the standard oscillation parameters can have a sizable discrepancy that is of the same order as $1\sigma$ resolution of the $\delta_{CP}$ measurement.
The discrepancy tends to be larger in the T2HK than in the T2HKK. 
Adding a small Oki detector to the T2HK does not make any difference due to its limited statistics.
\\

\noindent
Our most prominent finding is related to (b), which is that adding a small 10~kton detector at Oki to the T2HK leads to revived sensitivity to the mass hierarchy.

This paper is organized as follows:
In Section~2, we present the formalism for incorporating the effects of heavy isospin-singlet neutrino mixing with a non-unitary mixing matrix.
In Section~3, we estimate the largest possible significance of heavy neutrino mixing in the T2HK, the T2HKK and the extension of the T2HK with a small detector at OKi,
 and further study the mass hierarchy and $\delta_{CP}$ measurement in the presence of heavy neutrino mixing.
Here, we describe our benchmark model, procedures for simulating the long baseline experiments, and analysis with a $\chi^2$ fitting.
Section~4 summarizes the paper.
\\

\section{Formalism for incorporating heavy isospin-singlet neutrino mixing}

Suppose we have $N-3$ isospin-singlet neutrinos $\nu_{s_f} \, (f=4,5,...,N)$ in addition to the three active flavors $\nu_e,\nu_\mu,\nu_\tau$.
Accordingly, there are $N$ mass eigenstates $\nu_i \, (i=1,2,...,N)$ belonging to mass eigenvalues $m_i^2$ and the neutrino mixing is expressed as
\begin{align}
\vert \nu_\alpha \rangle &= \sum_{i=1}^N \, U_{\alpha i} \vert \nu_i \rangle, \ \ \ \ \ \alpha=e,\mu,\tau,s_f,
\label{mixing}
\end{align}
 where $U_{\alpha i}$ is a $N\times N$ unitary matrix.
In this paper, the $N-3$ isospin-singlet neutrinos are assumed to be heavier than about 10~GeV.

In long baseline neutrino oscillation experiments, only the light mass eigenstates $\nu_1,\nu_2,\nu_3$ are produced and propagate.
Therefore, the probability for a neutrino of active flavor $\alpha$ with energy $E$ to transition to active flavor $\beta$ after traveling a distance $L$ inside matter is
\begin{align}
P(\nu_\alpha \to \nu_\beta) &= \left\vert \langle \nu_\beta \vert U_{3\times3} \, 
\exp\left( -i\int_0^L {\rm d}x \, H(x) \right) \,
(U^\dagger)_{3\times3} \vert \nu_\alpha \rangle \right\vert^2 \ \ \ \ \ (\alpha,\beta=e,\mu,\tau),
\label{prob} \\
H(x) &= {\rm diag}\left( \ -\sqrt{E^2-m_1^2}, \, -\sqrt{E^2-m_2^2}, \, -\sqrt{E^2-m_3^2} \ \right)
\label{mass} \\
&+ \left[ \, U^\dagger \, {\rm diag}\left( \ V_{nc}(x)+V_{cc}(x), \, V_{nc}(x), \, V_{nc}(x), \, 0,..., \, 0 \ \right) \, U \, \right]_{3\times3},
\label{matter}
\end{align}
 where $U_{3\times3}$ denotes the $\alpha=e,\mu,\tau$ and $i=1,2,3$ part of the mixing matrix $U$ in Eq.~(\ref{mixing}),
 and $(U^\dagger)_{3\times3}$ denotes the same part of the matrix $U^\dagger$.
Note that $U_{3\times3}\neq (U^\dagger)_{3\times3}$ in general.
The factor $-\sqrt{E^2-m_i^2}$ $(i=1,2,3)$ is because a phase shift due to time evolution, $E{\rm d}t$, is equivalent to $-\sqrt{E^2-m_i^2}{\rm d}x$ for the mass eigenstate of mass $m_i$.
$V_{nc}(x)$ is the potential of the neutral current interaction inside matter divided by the neutrino velocity,
 and $V_{cc}(x)$ is that of the charged current interaction,
 which are evaluated to be
\begin{align}
V_{nc}(x) \, L &= -\frac{1}{\sqrt{2}}G_F \, n_n \, L = -Y_n \, 0.193 \left(\frac{\rho(x)}{{\rm g/cm}^3}\right) \left(\frac{L}{1000~{\rm km}}\right), 
\nonumber \\
V_{cc}(x) \, L &= \sqrt{2}G_F \, n_e \, L = Y_e \, 0.385 \left(\frac{\rho(x)}{{\rm g/cm}^3}\right) \left(\frac{L}{1000~{\rm km}}\right),
\end{align}
 where $n_n$ and $n_e$ respectively denote the neutron and electron number density, $Y_n$ and $Y_e$ are respectively the number of neutrons and electrons per one nucleon, which are typically $1/2$,
 and $\rho(x)$ is the matter density.
$[...]_{3\times3}$ in Eq.~(\ref{matter}) signifies that one extracts the $i,j=1,2,3$ part of the $N\times N$ matrix.

It is convenient to parametrize $U$ in terms of mixing angles, $\theta_{ij}$, and $CP$-violating phases, $\phi_{ij}$
\footnote{
Only $(N-1)(N-2)/2$ linear combinations out of $N(N-1)/2$ phases $\phi_{ij}$'s are physical.
}
, $(i<j, \, i,j=1,2,...,N)$ as~\cite{schechter,forero} (for the case with $N=6$, see also Ref.~\cite{xing})
\begin{align}
U &= \Omega_{N-1,N} \, \Omega_{N-2,N} \, ... \, \Omega_{1,N} \, \Omega_{N-2,N-1} \, \Omega_{N-3,N-1} \, ... \, \Omega_{1,N-1} \, \Omega_{N-3,N-2} \, ... \, \Omega_{1,2},
\nonumber \\
(\Omega_{n,m})_{kk} &=1 \ \ {\rm for} \ k\neq n,m, \ \ \ (\Omega_{n,m})_{nn}=(\Omega_{n,m})_{mm}=\cos\theta_{nm}, \ \ \ 
\nonumber \\
(\Omega_{n,m})_{nm} &=\sin\theta_{nm} \, e^{-i\phi_{nm}}, \ \ \ (\Omega_{n,m})_{mn}=-\sin\theta_{nm} \, e^{i\phi_{nm}},
\ \ \ (\Omega_{n,m})_{kl}=0 \ \ {\rm otherwise}.
\label{nnmixing}
\end{align}
Note that the PMNS mixing matrix $U_{PMNS}$ corresponds to $(\Omega_{2,3}\Omega_{1,3}\Omega_{1,2})$ with phases $\phi_{12}$, $\phi_{23}$ set to 0 by the phase redefinition of $\vert \nu_1 \rangle, \ \vert \nu_2 \rangle$.
We can thus express $U$ in terms of the PMNS mixing matrix as
\begin{align}
U &= \left(
\begin{array}{cc}
N_{NU} U_{PMNS} & S_{3\times N-3} \\
W_{N-3\times3} & T_{N-3\times N-3}
\end{array}
\right), \ \ \ \ \
N_{NU} = \left(
\begin{array}{ccc}
\alpha_{11} & 0 & 0 \\
\alpha_{21} & \alpha_{22} & 0 \\
\alpha_{31} & \alpha_{32} & \alpha_{33}
\end{array}
\right),
\label{upara}
\end{align}
 where $S_{3\times N-3}$, $W_{N-3\times3}$ and $T_{N-3\times N-3}$ denote $3\times(N-3)$, $(N-3)\times3$ and $(N-3)\times(N-3)$ matrices, respectively,
 $\alpha_{11}$, $\alpha_{22}$ and $\alpha_{33}$ are real numbers and $\alpha_{21}$, $\alpha_{31}$ and $\alpha_{32}$ are complex ones.

Since the mixing of isospin-singlet and active neutrinos is tiny, we make the approximation that
 we ignore terms in the second order of the mixing angles of active and isospin-singlet neutrinos.
This gives $(U^\dagger)_{3\times3}=(U_{3\times3})^\dagger$,
 and also 
\begin{align*} 
&\left[ \, U^\dagger \, {\rm diag}\left( \ V_{nc}(x)+V_{cc}(x), \, V_{nc}(x), \, V_{nc}(x), \, 0,..., \, 0 \ \right) \, U \, \right]_{3\times3}
\\
&= (U_{3\times3})^\dagger \, {\rm diag} \left(\ V_{nc}(x)+V_{cc}(x), \, V_{nc}(x), \, V_{nc}(x) \ \right) \, U_{3\times3}.
\end{align*}
We thus arrive at the following formula for the transition probabilities.
\begin{align}
P(\nu_\alpha \to \nu_\beta) &= \left\vert \langle \nu_\beta \vert \, N_{NU}U_{PMNS} \, \exp\left( -i\int_0^L {\rm d}x \, H(x) \right) \, U_{PMNS}^\dagger N_{NU}^\dagger \, \vert \nu_\alpha \rangle \right\vert^2 \ \ \ \ \ (\alpha,\beta=e,\mu,\tau),
\nonumber \\
H(x) &= \frac{1}{2E}{\rm diag}\left(0, \, \Delta m_{21}^2, \, \Delta m_{31}^2 \right)
\nonumber \\
&+ U_{PMNS}^\dagger N_{NU}^\dagger \left(
\begin{array}{ccc}
V_{nc}(x)+V_{cc}(x) & 0& 0 \\
0 & V_{nc}(x)& 0 \\
0 & 0& V_{nc}(x)
\end{array}
\right) N_{NU}U_{PMNS},
\label{prob3}
\end{align}
 where we have made a further approximation with $E^2 \gg m_1^2,m_2^2,m_3^2$
 and discarded terms proportional to the unit matrix in $H(x)$.
The transition probabilities for antineutrinos, $P(\bar{\nu}_\alpha \to \bar{\nu}_\beta)$, are obtained by changing
 $\delta_{CP} \to -\delta_{CP}$ in $U_{PMNS}$, $\alpha_{ij} \to \alpha_{ij}^*$ in $N_{NU}$, and 
 $V_{cc}(x) \to -V_{cc}(x), \ V_{nc}(x) \to -V_{nc}(x)$ in Eq.~(\ref{prob3}).
\\

\section{Significance of heavy neutrino mixing and its effects on the mass hierarchy and $\delta_{CP}$ measurement}

We estimate the largest possible significance of heavy neutrino mixing and its effects on the mass hierarchy and $\delta_{CP}$ measurement, through the following simulation:
First we introduce a benchmark model where $\alpha_{11},\alpha_{22},\vert\alpha_{21}\vert$ in Eq.~(\ref{nu}) are set
 at the current experimental bounds, in order to evaluate maximal impact of heavy neutrino mixing.
Neutrino detection in the T2HK, the T2HKK, and the extension of the T2HK with a small detector at Oki is simulated based on the above benchmark,
 with Particle Data Group values employed for the standard parameters $\theta_{12},\theta_{23},\theta_{13},\Delta m^2_{21},\vert\Delta m^2_{31}\vert$.
Since no conclusive data on the mass hierarchy and $\delta_{CP}$ exist, and the phase of $\alpha_{21}$ is undetermined in the benchmark,
 we repeat the simulation for both mass hierarchies and for various values of $\delta_{CP}$ and Arg$(\alpha_{21})$.
To assess merits of the extension of the T2HK with an Oki detector, we additionally consider, for comparative study, an experiment where the detector at Oki is instead placed at Kamioka.
Finally, we perform a $\chi^2$ fit of the simulation results under the assumption that no heavy neutrino mixing is present, namley, $N_{NU}=I_3$ in Eq.~(\ref{upara}).
The minimum of $\chi^2$ is the square of the significance of heavy neutrino mixing.
The mass hierarchy and the value of $\delta_{CP}$ with which $\chi^2$ is minimized correspond to
 their measured values obtained without assuming heavy neutrino mixing. 
\\

\subsection{Benchmark model}

Our analysis is based on the following benchmark model:
We employ the parametrization of Eq.~(\ref{upara}).
Since $\nu_\mu \to \nu_e$ process is most affected by $\alpha_{11}, \, \alpha_{22}$ and $\alpha_{21}$, we further set
\begin{align}
N_{NU} &= \left(
\begin{array}{ccc}
\alpha_{11} & 0 & 0 \\
\alpha_{21} & \alpha_{22} & 0 \\
0 & 0 & 1
\end{array}
\right).
\label{benchmark}
\end{align}

At present, the most severe constraint on $\alpha_{11}, \, \alpha_{22}$ and $\alpha_{21}$ derives from experimental tests of lepton flavor universality and a mathematical inequality among $\alpha_{11}, \, \alpha_{22}$ and $\alpha_{21}$.
These are enumerated below:
\begin{itemize}
 \item 
One test of lepton flavor universality has been done in $\pi^\pm$ decays.
When the non-unitary matrix Eq.~(\ref{benchmark}) enters into the neutrino mixing Eq.~(\ref{mixing}), the ratio of the branching fractions becomes
\begin{align}
\frac{\Gamma(\pi^\pm \to e^\pm \nu)}{\Gamma(\pi^\pm \to \mu^\pm \nu)} &= \frac{m_e^2}{m_\mu^2}\frac{(1-m_e^2/m_{\pi^\pm}^2)^2}{(1-m_\mu^2/m_{\pi^\pm}^2)^2}(1+\epsilon_{{\rm SM}}) \, \frac{\alpha_{11}^2}{\alpha_{22}^2+\vert\alpha_{21}\vert^2},
\label{universality}
\end{align}
 where $\epsilon_{{\rm SM}}$ indicates a Standard Model correction taking into account final state radiation.
The experimental value and the Standard Model prediction are respectively given by~\cite{czapek,nardi,abada,antusch,pavon}
\begin{align} 
\frac{\Gamma(\pi^\pm \to e^\pm \nu)}{\Gamma(\pi^\pm \to \mu^\pm \nu)}\vert_{{\rm experimental}} &= (1.2354\pm0.0002)\times10^{-4},
\nonumber \\
\frac{m_e^2}{m_\mu^2}\frac{(1-m_e^2/m_{\pi^\pm}^2)^2}{(1-m_\mu^2/m_{\pi^\pm}^2)^2}(1+\epsilon_{{\rm SM}})
&= (1.230\pm0.004)\times10^{-4},
\end{align}
 from which we obtain the following bound:
\begin{align} 
\frac{\alpha_{11}^2}{\alpha_{22}^2+\vert\alpha_{21}\vert^2} &=0.9956\pm0.0032.
\label{bound1}
\end{align}
 
 \item 
Another test of lepton flavor universality comes from unitarity test of the Cabibbo-Kobayashi-Maskawa (CKM) matrix.
In the presence of the non-unitary matrix Eq.~(\ref{benchmark}),
 the Fermi constant measured in the muon decay becomes $G_\mu = G_F \, \alpha_{11}\sqrt{\alpha_{22}^2+\vert\alpha_{21}\vert^2}$,
 whereas the Fermi constant measured in decays of hadrons involving $e\nu$ becomes $G_\beta = G_F \, \alpha_{11}$.
Deviation of $G_\beta/G_\mu$ from 1 mimics non-unitarity in the first row of the CKM matrix if $\vert V_{ud}\vert$ is measured in $\beta$-decays of nuclei and $\vert V_{us}\vert$ is measured in $K \to \pi e \nu$ decays,
 since $G_\mu$ is regarded as the true Fermi constant in these measurements while it is $G_\beta$ that is involved in the processes, and $\vert V_{ub}\vert^2$ is known to be smaller than experimental error of $\vert V_{us}\vert^2$.
We therefore have
\begin{align}
\frac{1}{\alpha_{22}^2+\vert\alpha_{21}\vert^2} &= \left(\frac{G_\beta}{G_\mu}\right)^2 \simeq \vert V_{ud}\vert^2 + \vert V_{us}\vert^2 \ ({\rm measured \ in \ }K \to \pi e \nu{\rm \ decays}) + \vert V_{ub}\vert^2.
\label{ckm}
\end{align}
Measurements of superallowed $\beta$-decays of nuclei yield the following experimental value~\cite{pdg}:
\begin{align} 
\vert V_{ud} \vert &= 0.97417\pm0.00021.
\label{vud}
\end{align}
Combining measurements of the branching ratios of only three processes $K^\pm \to \pi^0 e \nu$, $K_L \to \pi^\pm e \nu$ and $K_S \to \pi^\pm e \nu$~\cite{pdg}, 
 we obtain the following experimental value for $\vert V_{us}\vert$ times the kaon form factor, $f_+(0)$:
\begin{align} 
\vert V_{us} \vert \ f_+(0) &= 0.21633\pm0.00055.
\label{vusff}
\end{align}
The average of lattice calculations of the kaon form factor $f_+(0)$ in Refs.~\cite{lattice1,lattice2} is given by
\begin{align} 
f_+(0) &= 0.9677\pm0.0037.
\label{ff}
\end{align}
From Eqs.~(\ref{vud}),~(\ref{vusff}),~(\ref{ff}), we obtain the following bound:
\begin{align} 
\frac{1}{\alpha_{22}^2+\vert\alpha_{21}\vert^2} &=0.99898\pm0.00061.
\label{bound2}
\end{align}

 \item 
The mathematical inequality among $\alpha_{11}, \, \alpha_{22}$ and $\alpha_{21}$ stems from the fact that
 the non-unitary mixing matrix $N_{NU}$ is derived from the product of mixing matrices as in Eq.~(\ref{nnmixing}).
The inequality reads~\cite{antusch,pavon,blennow}
\begin{align} 
4(1-\alpha_{11})(1-\alpha_{22}) &> \vert\alpha_{21}\vert^2.
\label{bound3}
\end{align}
\end{itemize}

We combine the 3$\sigma$ experimental bounds in Eqs.~(\ref{bound1}),~(\ref{bound2}) and the mathematical inequality~Eq.~(\ref{bound3}),
 and obtain the following 3$\sigma$ bound
\footnote{
A different bound on $\alpha_{11}, \, \alpha_{22}$ and $\vert\alpha_{21}\vert$ has been derived in Ref.~\cite{forero} and used to determine a benchmark in Refs.~\cite{ge,dutta1,dutta2,dune}, 
 which takes into account experimental bounds on lepton flavor universality similar to Eqs.(\ref{bound1}),~(\ref{bound2}) and the NOMAD bound Eq.~(\ref{nomad}), but does not accommodate the mathematical inequality Eq.~(\ref{bound3}).
}:
\begin{align} 
\alpha_{11} &>0.992573, \ \ \ \alpha_{22} > 0.999589, \ \ \ \vert\alpha_{21}\vert < 0.003494 \ \ \ \ \ {\rm at} \ 3\sigma.
\end{align}
To estimate the maximum significance of heavy neutrino mixing, we set
 the values of $\alpha_{11}, \, \alpha_{22}$ and $\vert\alpha_{21}\vert$ at the above 3$\sigma$ bounds
 and therefore employ a benchmark model with the following non-unitary mixing matrix $N_{NU}$:
\begin{align}
N_{NU} &= \left(
\begin{array}{ccc}
\alpha_{11} & 0 & 0 \\
\alpha_{21} & \alpha_{22} & 0 \\
0 & 0 & 1
\end{array}
\right),
\nonumber \\
\alpha_{11} &= 0.992573, \ \ \ \alpha_{22}=0.999589, \ \ \ \vert\alpha_{21}\vert=0.003494.
\label{ournu}
\end{align}
Since no experimental bound is reported on the phase of $\alpha_{21}$, we leave it as a free parameter.

We comment in passing on constraints from past neutrino oscillation experiments.
Among those experiments, the NOMAD experiment~\cite{nomad}, a short baseline experiment searching for $\nu_\mu \to \nu_e$ and $\bar{\nu}_\mu \to \bar{\nu}_e$ transitions, gives the most severe bound, which is translated into the following:
\begin{align}
2\vert\alpha_{21}\vert^2 \alpha_{11}^2 &\leq 1.4\times 10^{-3} \ \ \ \ \ {\rm at \ 90\% \ confidence \ level}.
\label{nomad}
\end{align}
Our benchmark Eq.~(\ref{ournu}) satisfies the above bound.

Finally, we mention constraints from lepton flavor violating processes.
These constraints can always be evaded by tuning heavy neutrino mass $m_4,m_5,m_6,...,m_N$ around the $W$ boson mass,
 because the amplitude of a lepton flavor violating process involving charged leptons $\ell_\alpha$ and $\ell_\beta$,
 $A_{\alpha\beta}$, depends on $N_{NU}$ through the following combination:
\begin{align}
A_{\alpha\beta} &\propto \sum_{i=1}^3 \, (N_{NU}U_{PMNS})_{\beta i} \, F\left(\frac{m_i}{M_W}\right) \, (U_{PMNS}^\dagger N_{NU}^\dagger)_{i\alpha}
+ \sum_{j=4}^N \, (S_{3\times N-3})_{\beta j} \, F\left(\frac{m_j}{M_W}\right) \, (S_{3\times N-3}^\dagger)_{j\alpha}
\nonumber \\
&\simeq (N_{NU}N_{NU}^\dagger)_{\beta\alpha} \, F(0)
+ \sum_{j=4}^N \, (S_{3\times N-3})_{\beta j} \, F\left(\frac{m_j}{M_W}\right) \, (S_{3\times N-3}^\dagger)_{j\alpha},
\end{align}
 where $m_i,m_j$ are neutrino mass eigenvalues, $M_W$ is the $W$ boson mass, and $F(x)$ is a function depending on the process,
 and in the second line, we make the approximation with $m_1,m_2,m_3 \ll M_W$ as $m_1,m_2,m_3$ correspond to the tiny neutrino mass.
It is evident that one can negate the term $(N_{NU}N_{NU}^\dagger)_{\beta\alpha}F(0)$ by taking appropriate $O(1)$ values for $m_j/M_W$ $(j=4,5,6,...,)$.
We therefore neglect any constraints from lepton flavor violating processes.
\\

In Appendix~C, we consider another benchmark that satisfies the experimental bounds of Eqs.~(\ref{bound1}),~(\ref{bound2}) at 2$\sigma$ level
 as well as the mathematical inequality~Eq.~(\ref{bound3}).
\\

\subsection{Simulation}

We simulate the neutrino propagation and detection based on the benchmark model Eq.~(\ref{ournu}).
The oscillation probability is calculated numerically with the formula Eq.~(\ref{prob3}) without making any approximation.
For the standard oscillation parameters, we employ the central values of $\theta_{12}, \, \theta_{13}, \, \theta_{23}, \, \Delta m^{2}_{21}, \, \vert \Delta m^{2}_{32} \vert$ found in the Particle Data Group~\cite{pdg}.
As the $CP$-violating phase $\delta_{CP}$ and the sign of $\Delta m_{32}^2$ have not been measured conclusively, and the phase of $\alpha_{21}$, Arg$(\alpha_{21})$, is undetermined in the benchmark,
 we repeat the simulation for various values of $\delta_{CP}$ and Arg$(\alpha_{21})$ and for both cases with $\Delta m_{32}^2>0$ and $\Delta m_{32}^2<0$.
The values of parameters used in the simulation are shown in Table~\ref{physpara}.
\begin{table}[H]
\begin{center}
  \caption{Parameters used in our simulation.}
  \begin{tabular}{|c||c|} \hline
    physical parameter & value in our simulation \\ \hline
    $\sin^2\theta_{12}$ & 0.304 \\
    $\sin^2\theta_{13}$ & 0.0219 \\
    $\sin^2\theta_{23}$ (normal hierarchy)   & 0.51 \\
    $\sin^2\theta_{23}$ (inverted hierarchy) & 0.50 \\
    $\Delta m^{2}_{21}$ & $7.53\times10^{-5}$~eV$^2$ \\
    $\vert \Delta m^{2}_{32} \vert$ (normal hierarchy)   & $2.44\times10^{-3}$ \\
    $\vert \Delta m^{2}_{32} \vert$ (inverted hierarchy) & $2.51\times10^{-3}$ \\
    $\delta_{CP}$ & free \\ \hline
    $\alpha_{11}$ & 0.992573 \\
    $\vert\alpha_{21}\vert$ & 0.003494 \\
    $\alpha_{22}$ & 0.999589 \\
    $\alpha_{31}$ & 0 \\
    $\alpha_{32}$ & 0 \\
    $\alpha_{33}$ & 1 \\
    $\varphi_{21}={\rm arg}(\alpha_{21})$ & free \\ \hline
  \end{tabular}
  \label{physpara}
  \end{center}
\end{table}

We assume the baseline length $L$, fiducial volume, matter density along the baseline and neutrino beam off-axis angle~\cite{t2kok,senda} for the water Cerenkov detectors used in the T2HK, the T2HKK, the extension of the T2HK with a detector at Oki (T2HK+Oki), and the extension with the same detector at Kamioka (T2HK+Kami) as Table~\ref{exppara}.
The matter density is approximated to be uniform~\cite{koike,senda}.
\begin{table}[H]
\begin{center}
  \caption{Parameters assumed for the detectors used in the experiments.}
  \begin{tabular}{|c|c||c|c|c|c|c|} \hline
    experiment & site & baseline length $L$ & fiducial volume & matter density $\rho$ & off-axis angle \\ \hline
    T2HK & Kamioka & 295~km & 374~kton & 2.60~g/cm$^3$  & 2.5$^{\circ}$ \\ \hline
    T2HKK & Kamioka & 295~km & 187~kton & 2.60~g/cm$^3$  & 2.5$^{\circ}$ \\
          & Korea & 1000~km & 187~kton & 2.90~g/cm$^3$  & 1.0$^{\circ}$ \\ \hline
    T2HK+Oki & Kamioka & 295~km & 374~kton & 2.60~g/cm$^3$  & 2.5$^{\circ}$ \\
          & Oki & 653~km & 10~kton & 2.75~g/cm$^3$  & 1.0$^{\circ}$ \\ \hline
    T2HK+Kami & Kamoika & 295~km & 384~kton & 2.60~g/cm$^3$  & 2.5$^{\circ}$ \\ \hline
  \end{tabular}
  \label{exppara}
  \end{center}
\end{table}

We assume that J-PARC operates with 1.3~MW beam power for 10~years in the neutrino mode and for another 10~years in the antineutrino mode, 
 delivering $27\times10^{21}$~proton-on-target (POT) flux of neutrino-focusing beam and $27\times10^{21}$~POT flux of antineutrino-focusing beam, as Table~\ref{flux}.
\begin{table}[H]
\begin{center}
  \caption{Flux of neutrino-focusing beam and antineutrino-focusing beam from J-PARC assumed.}
  \begin{tabular}{|c||c|c|} \hline
            &  Neutrino-focusing & Antineutrino-focusing \\ \hline
    Flux & $27\times10^{21}$~POT & $27\times10^{21}$~POT \\ \hline
  \end{tabular}
  \label{flux}
  \end{center}
\end{table}
\noindent
The energy distributions of $\nu_\mu$ and $\bar{\nu}_\mu$ in neutrino-focusing and antineutrino-focusing beams at each site
 are calculated based on Ref.~\cite{beam}.
$\nu_e$ and $\bar{\nu}_e$ components in the beam are ignored in this analysis.
In Appendix~A, we present the energy distributions.

The cross sections for charged current quasi-elastic scattering between a neutrino and a proton, 
 $\nu_\ell \, n \to \ell^- \, p$, and that between an antineutrino and a neutron, $\bar{\nu}_\ell \, p \to \ell^+ \, n$,
 ($p$ and $n$ denote proton and neutron, respectively, and $\ell$ denotes $e$ or $\mu$)
 are quoted from Ref.~\cite{crosssection}.
In Appendix~B, we present the cross sections.

We assume the ideal measurement of $\nu_e$ and $\bar{\nu}_e$ and do not take into account acceptance and efficiency factors and finite energy resolution.
Signals coming from processes other than charged current quasi-elastic scattering, as well as
 backgrounds due to neutral current quasi-elastic scatterings and other processes are not considered in the analysis.

The total numbers of neutrino events at the detectors at Kamioka, Oki and in Korea in the T2HK, T2HKK, T2HK+Oki, and T2HK+Kami
 with the detector size of Table~\ref{exppara} and the beam flux of Table~\ref{flux} are tabulated in Table~\ref{numbers}.
\begin{table}[H]
\begin{center}
  \caption{The total number of neutrino events at each detector, with the detector size of Table~\ref{exppara} and the beam flux of Table~\ref{flux}.
  The numbers for the neutrino-focusing beam and for the antineutrino-focusing beam are shown separately.
  }
  \begin{tabular}{|c|c||c|c|} \hline
      experiment & detector &  number for  & number for \\ 
                        &                  &  neutrino-focusing beam & antineutrino-focusing beam \\ \hline
     T2HK          & Kamioka  &  21$\times10^4$ & 7.5$\times10^4$ \\ \hline
     T2HKK        & Kamioka  &  11$\times10^4$ & 3.7$\times10^4$ \\
                         & Korea      &  3.1$\times10^4$ & 1.7$\times10^4$ \\ \hline
     T2HK+Oki 10~kton        & Kamioka  & 21$\times10^4$ & 7.5$\times10^4$ \\
                                              & Oki      &  0.39$\times10^4$& 0.21$\times10^4$ \\ \hline
     T2HK+Kami 10~kton    & Kamioka  &  22$\times10^4$ & 7.7$\times10^4$   \\ \hline
  \end{tabular}
  \label{numbers}
  \end{center}
\end{table}

We calculate the numbers of $\nu_e$ and $\bar{\nu}_e$ events in 0.05~GeV bins of the neutrino energy $E$ in the range 0.4~GeV$\leq E \leq$3~GeV
 for various values of $\delta_{CP}$ and Arg$(\alpha_{21})$ and for both cases with $\Delta m_{31}^2>0$ and $\Delta m_{31}^2<0$.
We remind that in real experiments, the neutrino flux is determined by the measurement of $\nu_\mu$ and $\bar{\nu}_\mu$ at the near detector, 
 for which the standard oscillation is negligible but the non-unitary mixing reduces the $\nu_\mu$ and $\bar{\nu}_\mu$ flux by the following amount:
\begin{align}
P(\nu_\mu \to \nu_\mu)_{{\rm near \ detector}} &= P(\bar{\nu}_\mu \to \bar{\nu}_\mu)_{{\rm near \ detector}} = 
(N_{NU}N_{NU}^\dagger)_{\mu\mu} = \vert\alpha_{21}\vert^2+\alpha_{22}^2.
\label{nd}
\end{align}
In the analysis, we consider a virtual neutrino flux that is larger than the one found in Ref.~\cite{beam} by $1/(\vert\alpha_{21}\vert^2+\alpha_{22}^2)$, to compensate the reduction of Eq.~(\ref{nd}).
We therefore compute the number of events with a neutrino-focusing beam, $N_{e,i}$, and that with an antineutrino-focusing beam, $\tilde{N}_{e,i}$, as
\begin{align}
{\rm for} \ i&=1,2,3,...,52,
\nonumber \\
N_{e,i} &= \left( \, {\rm Number \ of \ } \nu_e {\rm \ and \ } \bar{\nu}_e \ {\rm with} \ (0.4+i \cdot 0.05){\rm GeV} \geq E \geq (0.4+(i-1) \cdot 0.05){\rm GeV} \, \right)
\nonumber \\
&= \int_{0.4+(i-1) \cdot 0.05~{\rm GeV}}^{0.4+i \cdot 0.05~{\rm GeV}} {\rm d}E \, 
\left\{ \, \frac{\Phi_{\nu_\mu}(E)}{\vert\alpha_{21}\vert^2+\alpha_{22}^2} \ P(\nu_\mu \to \nu_e; \, E) \ N_n \ \sigma(\nu_e n \to e^- p; \, E) \right.
\nonumber \\
&\left. + \frac{\Phi_{\bar{\nu}_\mu}(E)}{\vert\alpha_{21}\vert^2+\alpha_{22}^2} \ P(\bar{\nu}_\mu \to \bar{\nu}_e; \, E) \ N_p \ \sigma(\bar{\nu}_e p \to e^+ n; \, E) \, \right\},
\nonumber \\
\tilde{N}_{e,i} &= ( \, \Phi_{\nu_\mu}\to\tilde{\Phi}_{\nu_\mu} \ {\rm and} \ \Phi_{\bar{\nu}_\mu}\to\tilde{\Phi}_{\bar{\nu}_\mu} \ {\rm in \ above} \, ),
\end{align}
 where $\Phi_{\nu_\mu}(E)$ and $\Phi_{\bar{\nu}_\mu}(E)$ respectively denote the neutrino and antineutrino flux per energy in a neutrino-focusing beam,
 and $\tilde{\Phi}_{\nu_\mu}(E)$ and $\tilde{\Phi}_{\bar{\nu}_\mu}(E)$ are those in an antineutrino-focusing beam.
$P$ denotes the transition probability Eq.~(\ref{prob3}), with $E$ dependence made explicit.
$N_n$ and $N_p$ are respectively the number of neutrons and protons in a water Cerenkov detector.
$\sigma$ denotes the cross sections for $\nu_\ell n \to \ell^- p$ and $\bar{\nu}_\ell p \to \ell^+ n$ processes.
\\

\subsection{$\chi^2$ analysis}

We fit the numbers of events in the bins with neutrino-focusing and antineutrino-focusing beams, detected at Kamioka/Korea/Oki, $N_{e,i,{\rm site}}$ and $\tilde{N}_{e,i,{\rm site}}$ $(i=1,2,...,52)$ (site=Kamioka, Korea, Oki), under the assumption that no heavy neutrino mixing is present, that is, $N_{NU}=I_3$.
This is performed by minimizing $\chi^2(\Pi)$,
\begin{align}
\chi^2(\Pi) &= \chi^2_{{\rm statistical}}(\Pi) + \chi^2_{{\rm parametrical}}(\Pi) + \chi^2_{{\rm systematic}},
\nonumber \\
\chi^2_{{\rm statistical}}(\Pi) &= \sum_{{\rm site}} \, \sum_{i=1}^{52} \, \left\{ \ \frac{( N_{e,i,{\rm site}} - N_{e,i,{\rm site}}^{{\rm unitary}}(\Pi) )^2}{N_{e,i,{\rm site}}} + \frac{( \tilde{N}_{e,i,{\rm site}} - \tilde{N}_{e,i,{\rm site}}^{{\rm unitary}}(\Pi) )^2}{\tilde{N}_{e,i,{\rm site}}} \ \right\},
\label{chi2}
\end{align}
 with respect to the standard oscillation parameters $\Pi=(\theta_{12},\theta_{13},\theta_{23},\Delta m^2_{21},\Delta m^2_{32},\delta_{CP})$.
Here, $\chi^2_{{\rm statistical}}$ represents statistical uncertainty,
 $N_{e,i,{\rm site}}^{{\rm unitary}}(\Pi)$ is the number of $\nu_e$ and $\bar{\nu}_e$ events in a bin with a neutrino-focusing beam when $N_{NU}=I_3$, 
 calculated as a function of the standard oscillation parameters $\Pi$,
 and $\tilde{N}_{e,i,{\rm site}}^{{\rm unitary}}(\Pi)$ is the corresponding number with an antineutrino-focusing beam.

The following simplification is made for $\chi^2_{{\rm parametrical}}(\Pi)$:
We note that $\theta_{12},\theta_{13},\theta_{23},\Delta m^2_{21},\vert\Delta m^2_{32}\vert$ can be accurately measured in disappearance experiments with $\bar{\nu}_e \to \bar{\nu}_e$, $\nu_\mu \to \nu_\mu$ and $\bar{\nu}_\mu \to \bar{\nu}_\mu$,
 which are not influenced by heavy neutrino mixing as parametrized by $N_{NU}$, because $N_{NU}$ equally affects the neutrino flux of the same flavor at the near and far detectors.
($\cos\delta_{CP}$ and the sign of $\Delta m^2_{32}$ are also possibly measured in disappearance experiments, but the sensitivity is limited.)
We therefore assume that the true values of $\theta_{12},\theta_{13},\theta_{23},\Delta m^2_{21},\vert\Delta m^2_{32}\vert$ are known prior to the experiments considered in this paper,
 and accordingly, we fix $\theta_{12},\theta_{13},\theta_{23},\Delta m^2_{21},\vert\Delta m^2_{32}\vert$ in $\chi^2(\Pi)$
 at the values used in the simulation and set $\chi^2_{{\rm parametrical}}(\Pi)=0$.

Finally, we have $\chi^2_{{\rm systematic}}=0$ in the current analysis, because no acceptance, efficiency, energy resolution and background events are considered.

To summarize, $\chi^2(\Pi)$ is approximated as
\begin{align}
\chi^2(\Pi) &= \chi^2_{{\rm statistical}}( \, \delta_{CP}, \, {\rm sgn}(\Delta m^2_{32}) \, ) \ \ \ {\rm with \ } \theta_{12},\theta_{13},\theta_{23},\Delta m^2_{21},\vert\Delta m^2_{32}\vert {\rm \ given \ in \ Table~\ref{physpara}}.
\label{chi22}
\end{align}
\\

\begin{figure}[H]
  \begin{center}
    \includegraphics[width=80mm]{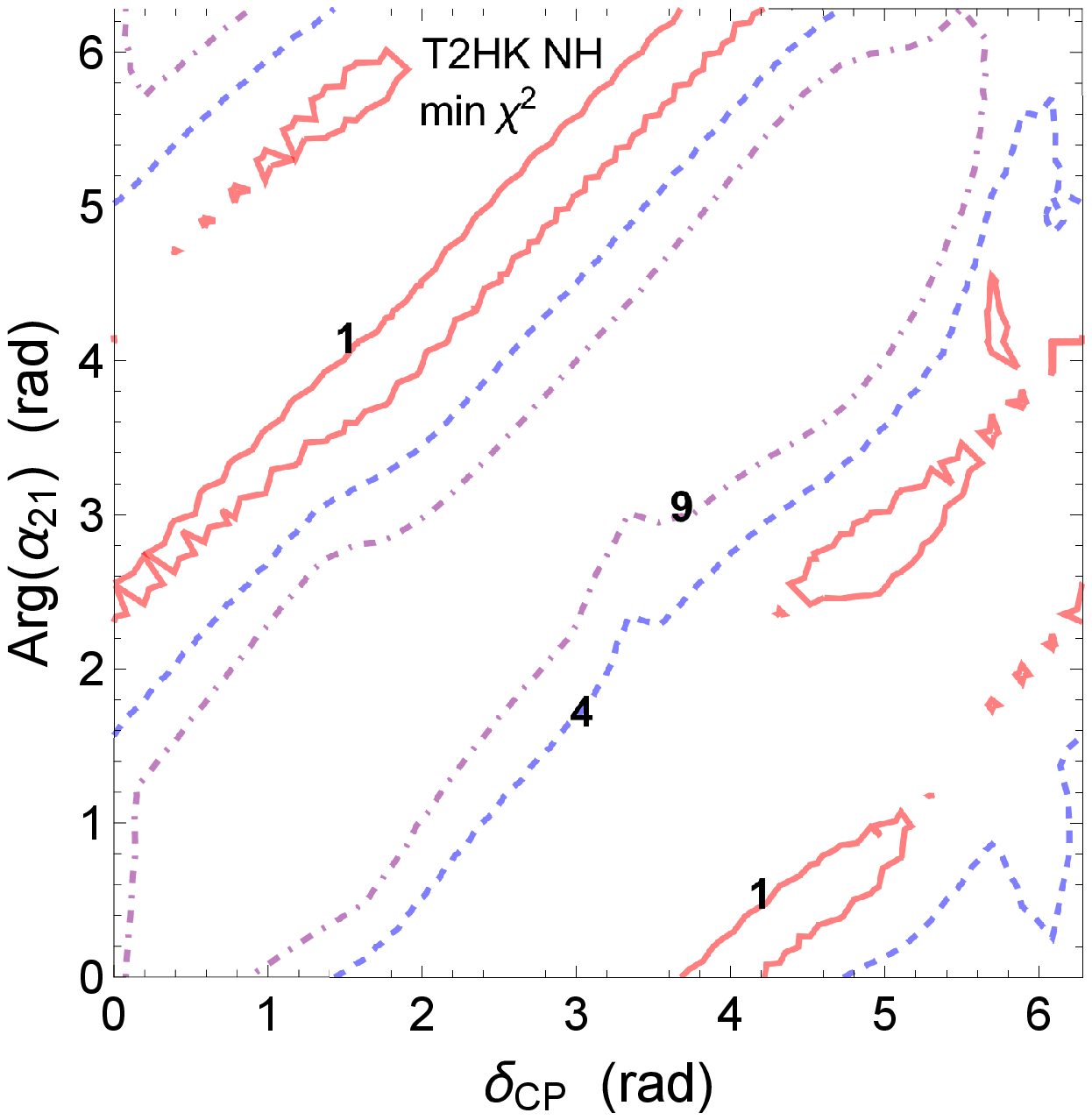} 
    \includegraphics[width=80mm]{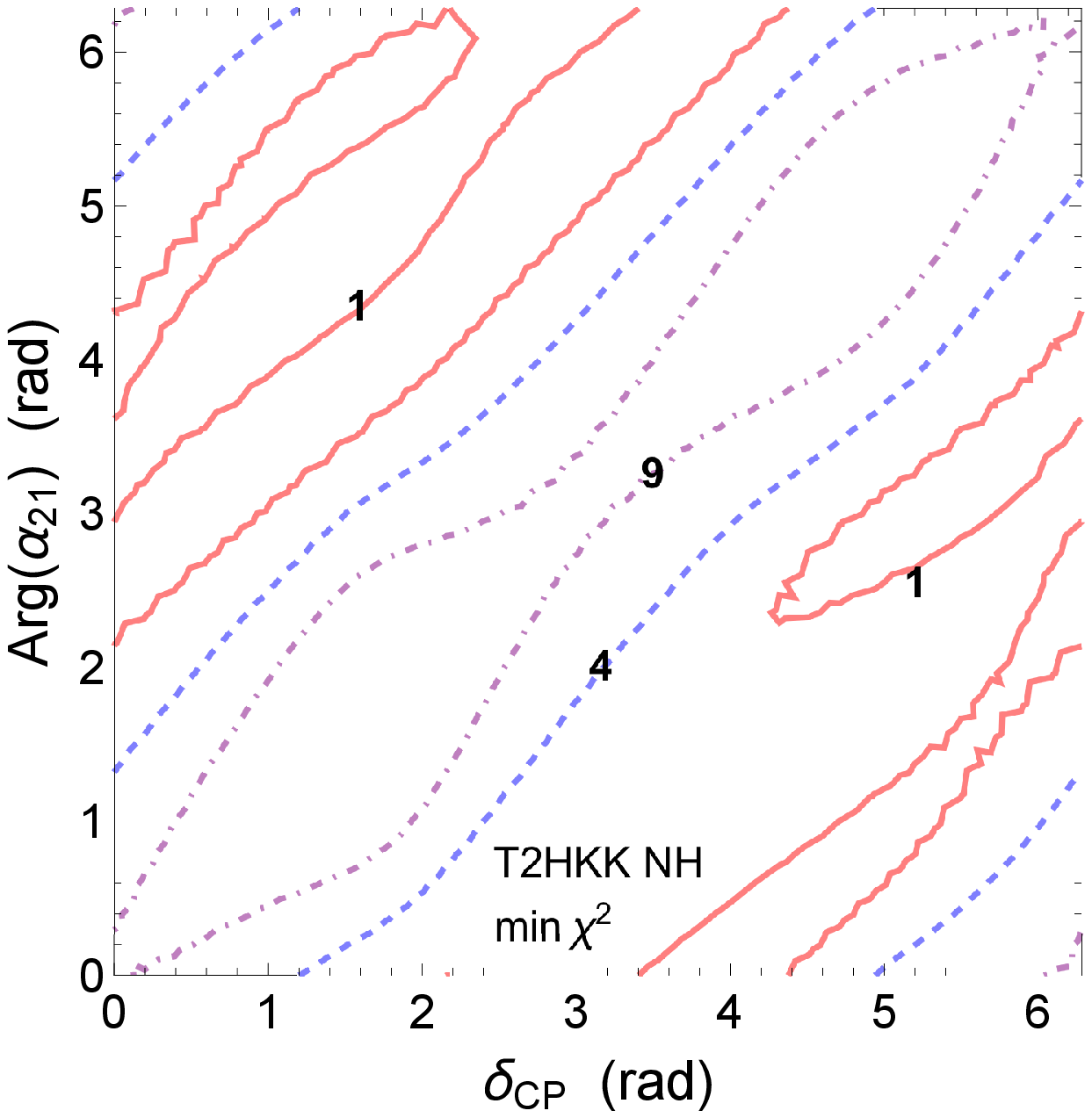}
    \\
    \includegraphics[width=80mm]{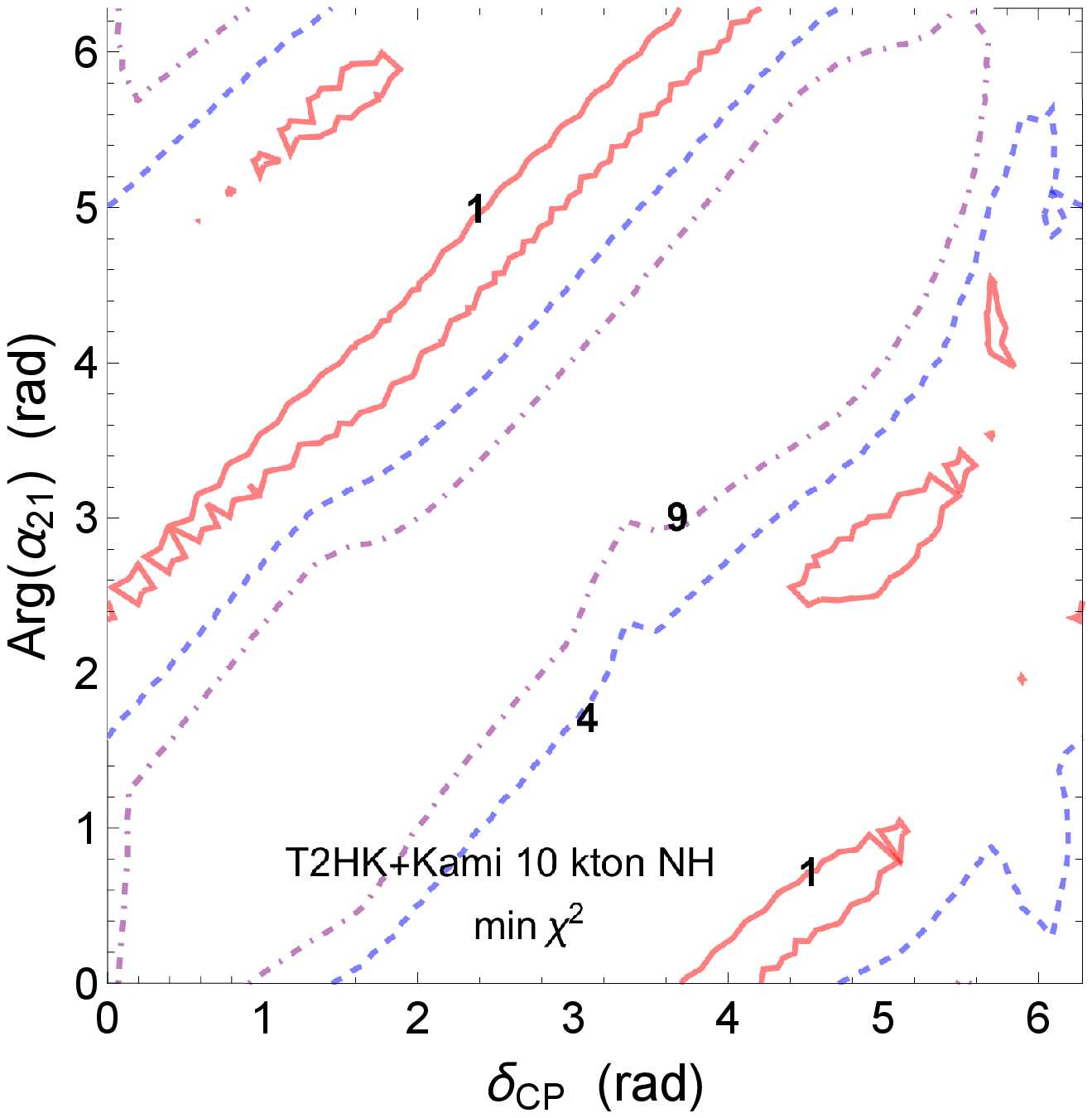}
    \includegraphics[width=80mm]{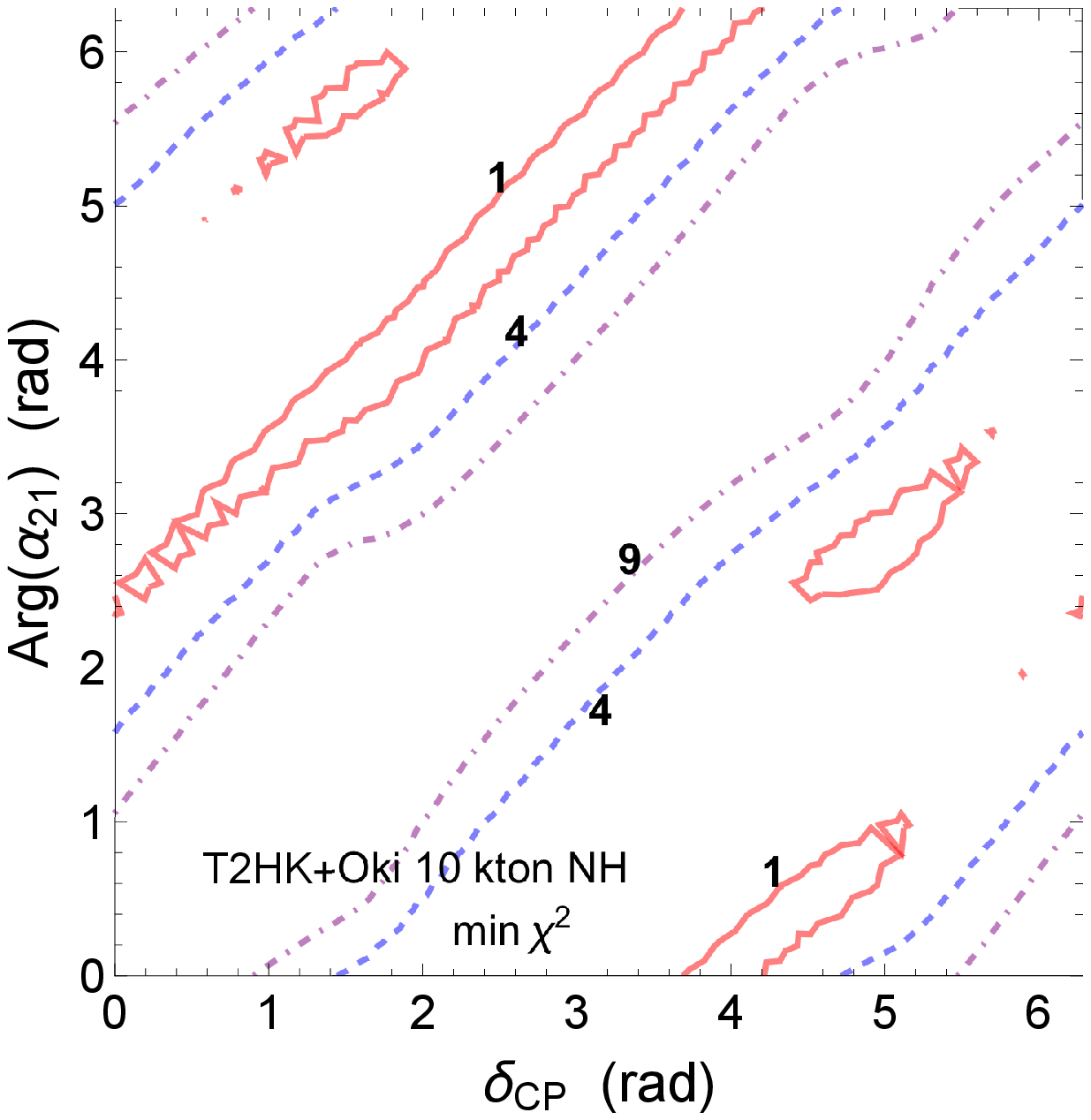}
    \caption{
    The minimum of $\chi^2(\Pi)$ Eq.~(\ref{chi22}), $\min \chi^2$,
     on the plane of the standard $CP$-violating phase $\delta_{CP}$ and the new phase in the non-unitary mixing matrix Arg($\alpha_{21}$).
    The benchmark with Eq.~(\ref{ournu}) is assumed, and the true mass hierarchy is normal.
    The upper-left, upper-right and lower-right subplots correspond to the T2HK, the T2HKK, 
     and the plan of the T2HK plus a 10~kton water Cerenkov detector at Oki, respectively.
    For comparative study, we show in the lower-left a subplot for a plan of the T2HK plus a 10~kton water Cerenkov detector at Kamioka.
    $\min \chi^2=1,\,4,\,9$ on the red solid, blue dashed, and purple dot-dashed contours, respectively.
    }
    \label{nhch}
  \end{center}
\end{figure}

We numerically evaluate the minimum of $\chi^2(\Pi)$ Eq.~(\ref{chi22}), which corresponds to the square of the significance for heavy neutrino mixing.
In Figure~\ref{nhch}, we show contour plots of the minimum of $\chi^2(\Pi)$,
\begin{align}
\min\chi^2 &\equiv \min\left\{ \ \min_{\delta_{CP}}\chi^2(\delta_{CP}, \ \Delta m^2_{32}>0), \ \min_{\delta_{CP}}\chi^2(\delta_{CP}, \ \Delta m^2_{32}<0) \ \right\},
\end{align}
 on the plane spanned by the $CP$-violating phase in the PMNS matrix $\delta_{CP}$ and the phase in the non-unitary mixing matrix Arg($\alpha_{21}$), when the true mass hierarchy is normal.
The four subplots correspond to the T2HK, the T2HKK, the extension of the T2HK with a detector at OKi, and that with the same detector at Kamioka.

\begin{figure}[H]
  \begin{center}
    \includegraphics[width=80mm]{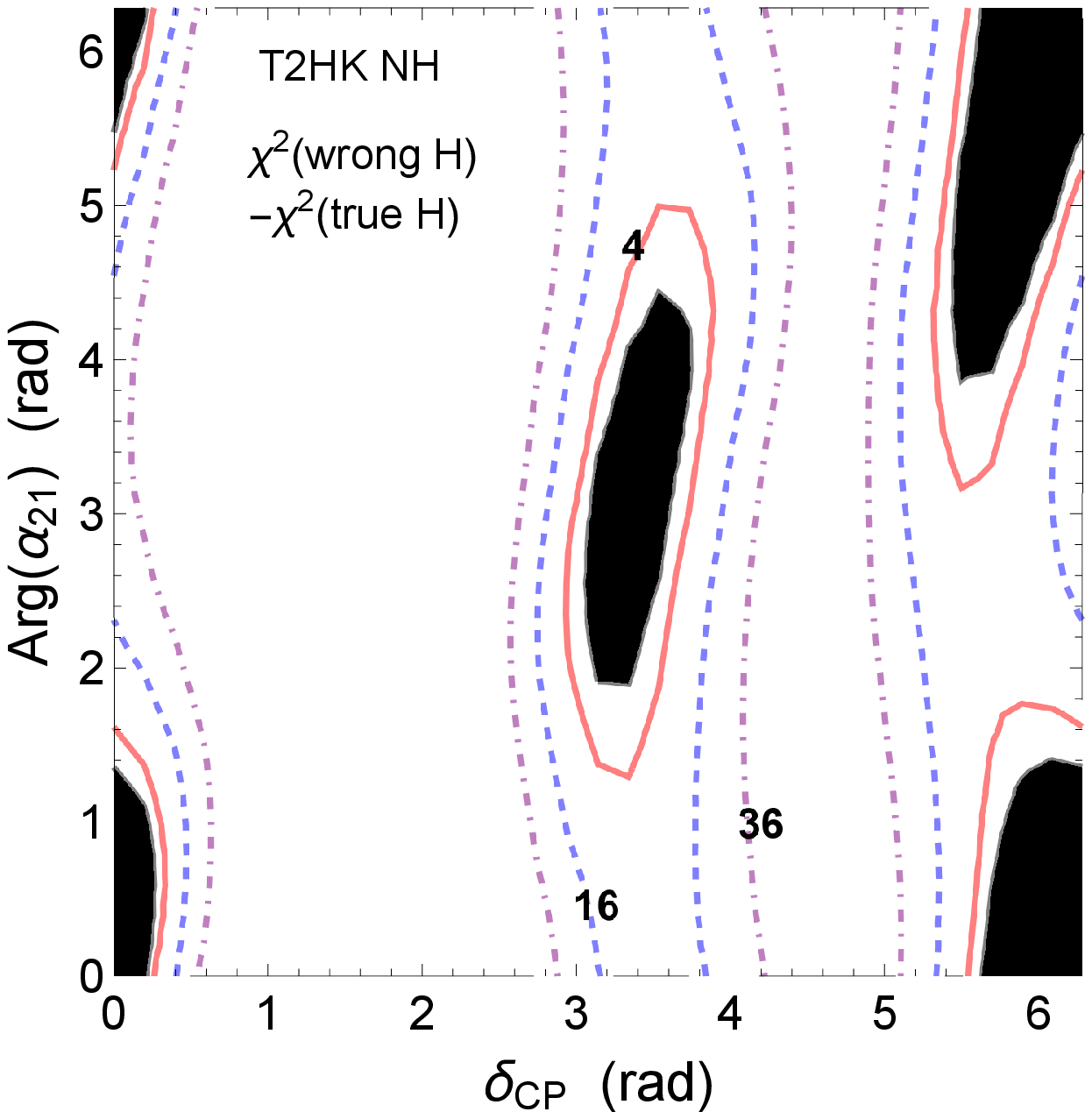} 
    \includegraphics[width=80mm]{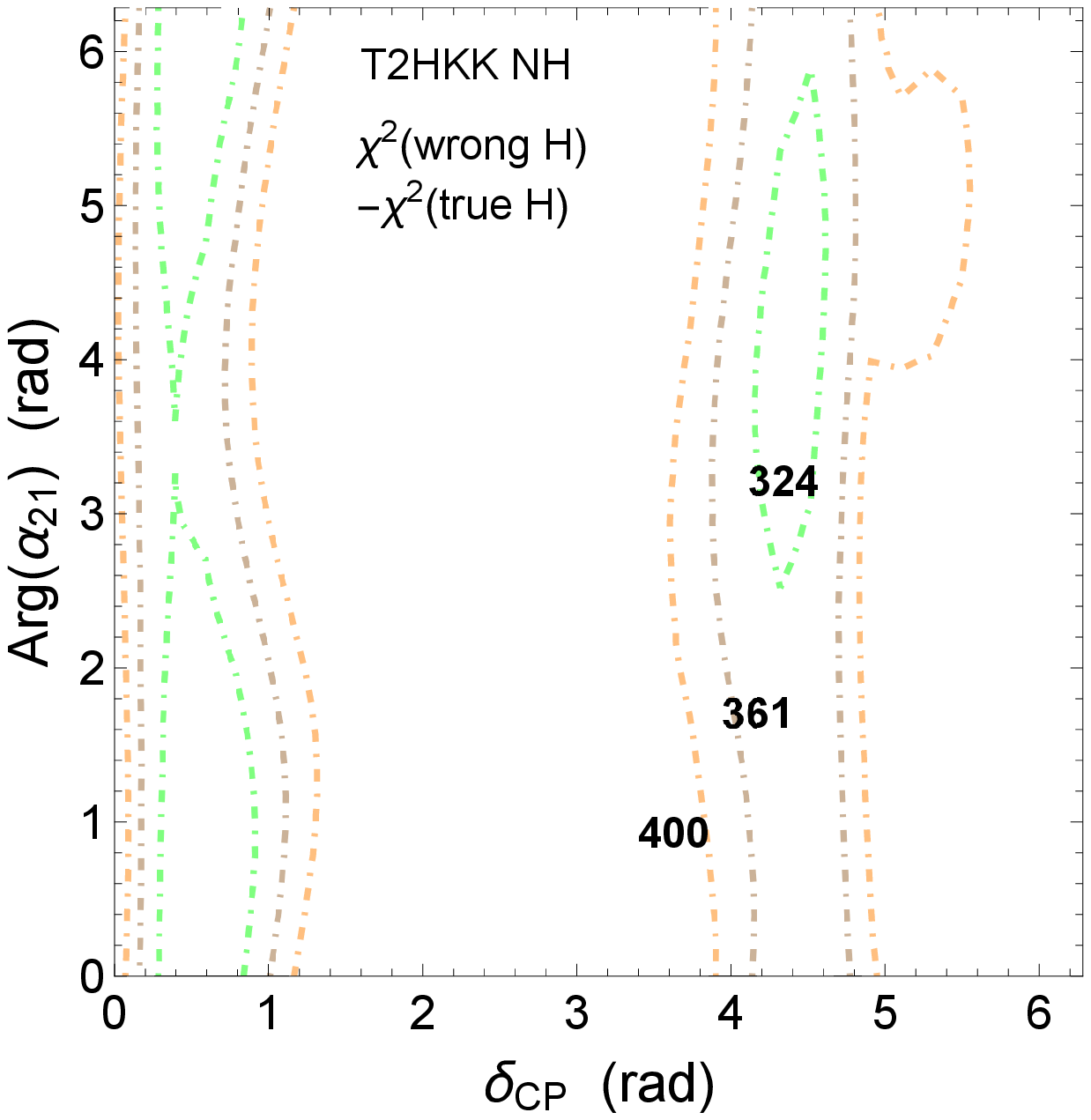}
    \\
    \includegraphics[width=80mm]{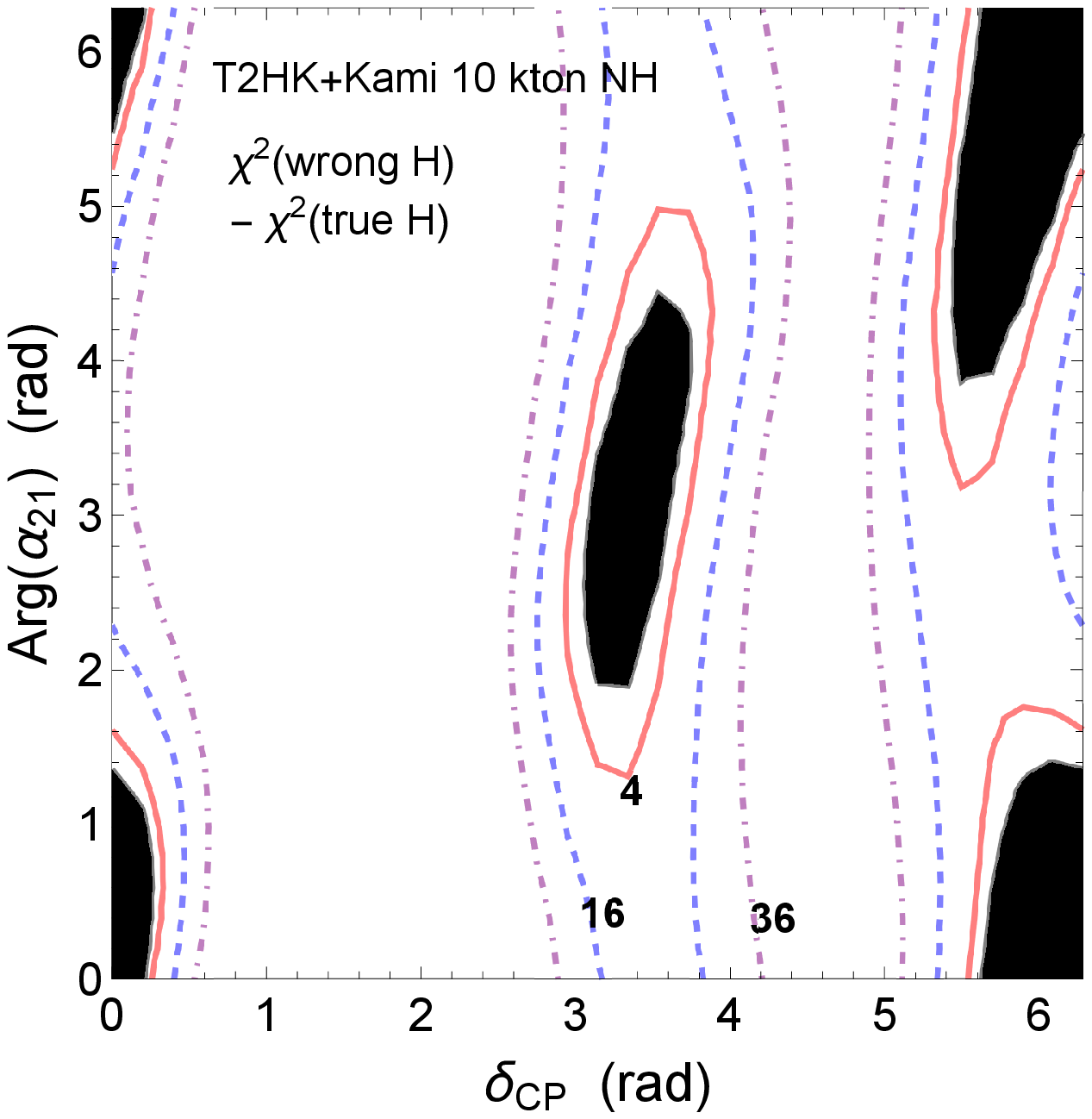}
    \includegraphics[width=80mm]{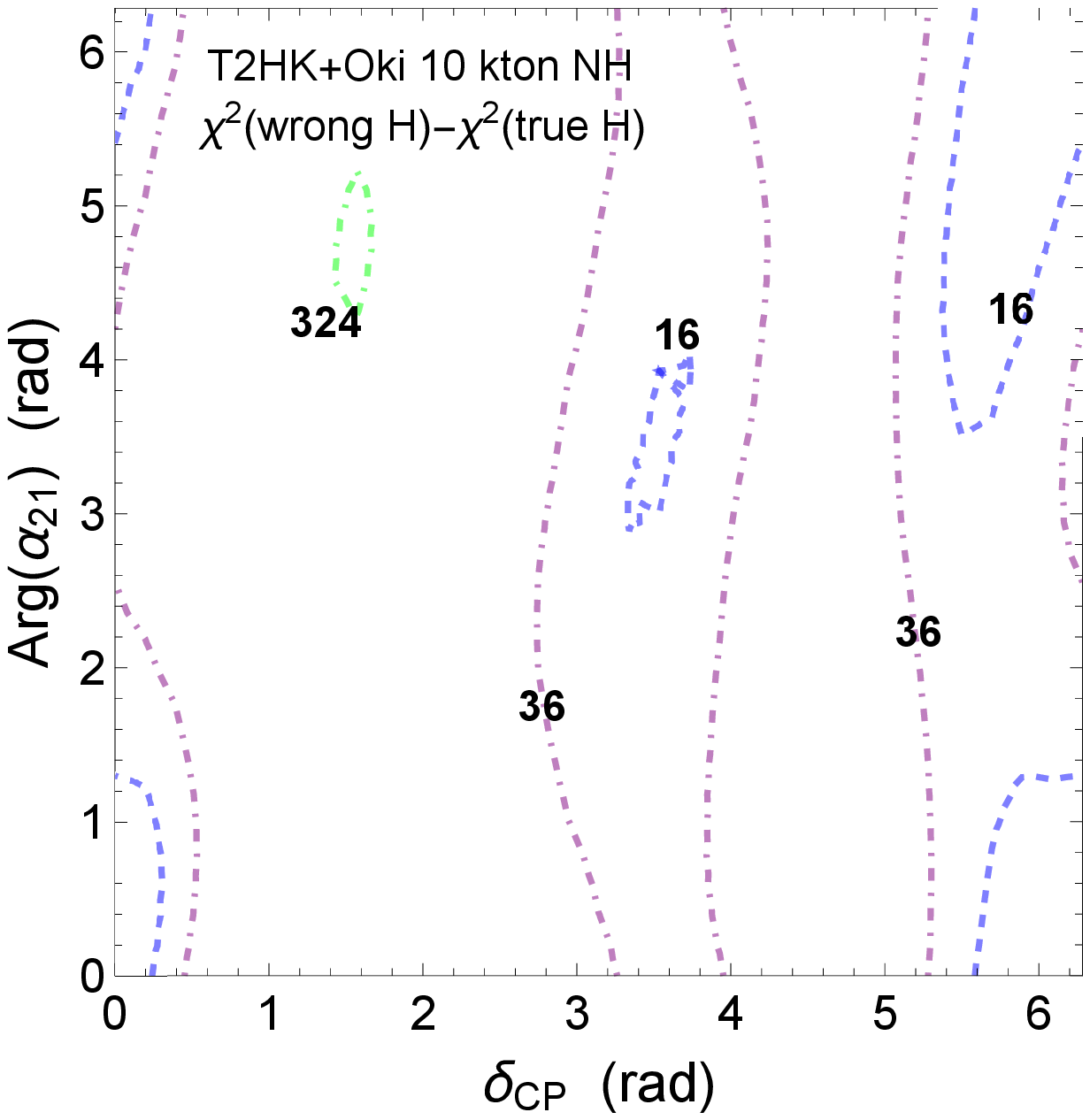}
    \caption{
    The difference between the minima (with respect to $\delta_{CP}$ only) of $\chi^2(\Pi)$ for the wrong and true mass hierarchies,
     $\{\min\chi^2({\rm wrong \ H})-\min\chi^2({\rm true \ H})\}$.
    The benchmark with Eq.~(\ref{ournu}) is assumed, and the true mass hierarchy is normal.
    The upper-left, upper-right and lower-right subplots correspond to the T2HK, the T2HKK, 
     and the plan of the T2HK plus a 10~kton water Cerenkov detector at Oki, respectively.
    For comparative study, we show in the lower-left a subplot for a plan of the T2HK plus a 10~kton water Cerenkov detector at Kamioka.
    $\{\min\chi^2({\rm wrong \ H})-\min\chi^2({\rm true \ H})\}=4,\,16,\,36$ on the red solid, blue dashed, and purple dot-dashed contours, respectively, and 
     $\{\min\chi^2({\rm wrong \ H})-\min\chi^2({\rm true \ H})\}=18^2,\,19^2,\,20^2$ on the green, brown, and orange dot-dashed contours, respectively.   
     The black-filled regions are where the wrong mass hierarchy is favored over the true mass hierarchy, \textit{i.e.},
 $\min\chi^2({\rm wrong \ H})-\min\chi^2({\rm true \ H})<0$.
     }
    \label{nhmass}
  \end{center}
\end{figure}

We study the impact of heavy neutrino mixing on the mass hierarchy measurement
 performed by fitting data without incorporating heavy neutrino mixing.
For this purpose, we evaluate the difference between the minimum (with respect to $\delta_{CP}$ only) of $\chi^2(\Pi)$ for the wrong hierarchy and that for the true mass hierarchy,
 which corresponds to the square of the significance of rejecting the wrong mass hierarchy.
We show in Figure~\ref{nhmass} contour plots of the difference,
\begin{align}
&\min\chi^2({\rm wrong \ H})-\min\chi^2({\rm true \ H}) 
\nonumber \\
&\equiv \min_{\delta_{CP}}\chi^2(\delta_{CP}, \ {\rm wrong} \ {\rm sgn}(\Delta m^2_{32})) - \min_{\delta_{CP}}\chi^2(\delta_{CP}, \ {\rm true} \ {\rm sgn}(\Delta m^2_{32})),
\end{align}
 on the same plane when the true mass hierarchy is normal.
The black-filled regions in the left subplots are where the wrong mass hierarchy is favored over the true mass hierarchy, \textit{i.e.},
 $\min\chi^2({\rm wrong \ H})-\min\chi^2({\rm true \ H})<0$.

\begin{figure}[H]
  \begin{center}
    \includegraphics[width=80mm]{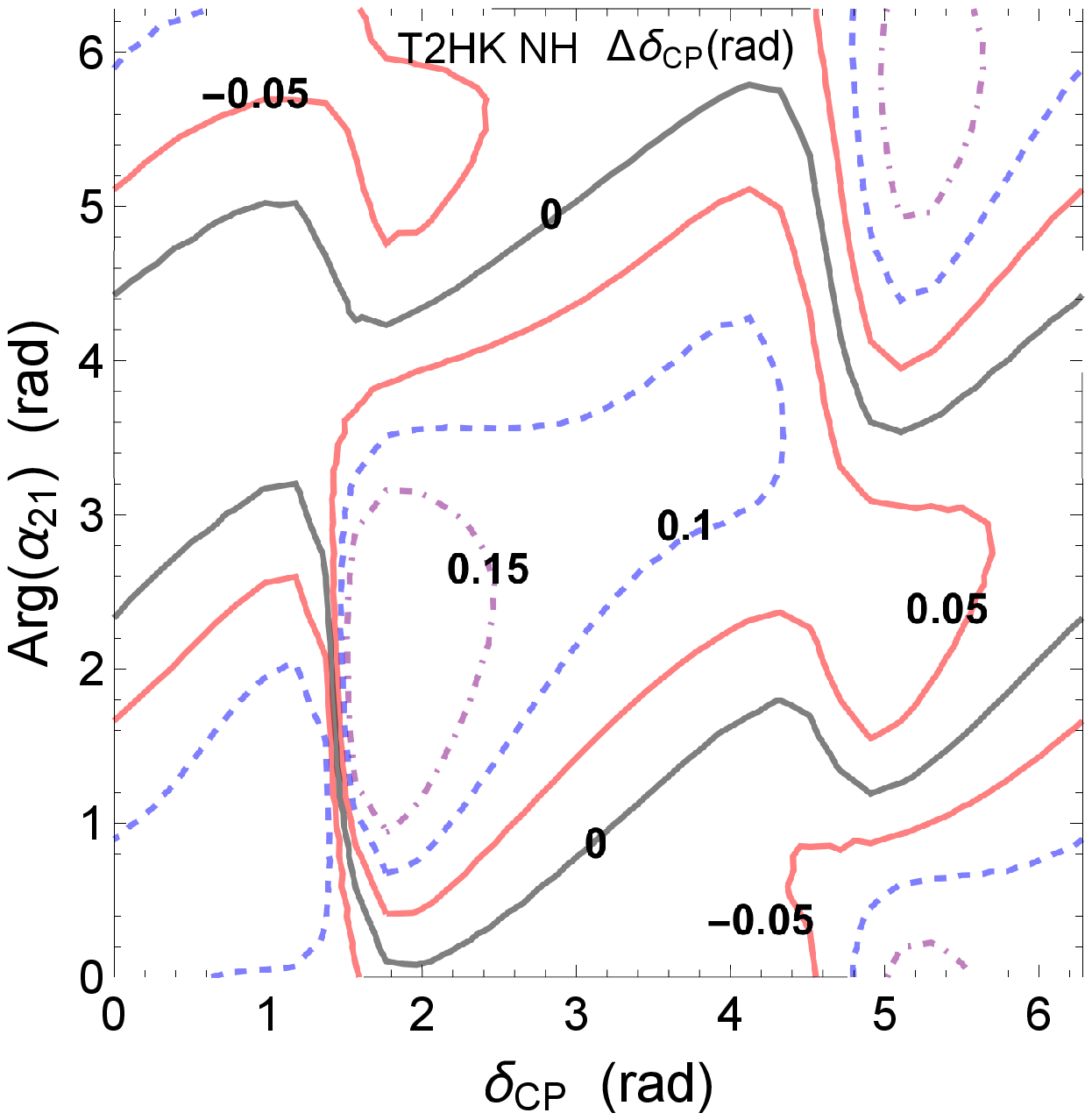}    
    \includegraphics[width=80mm]{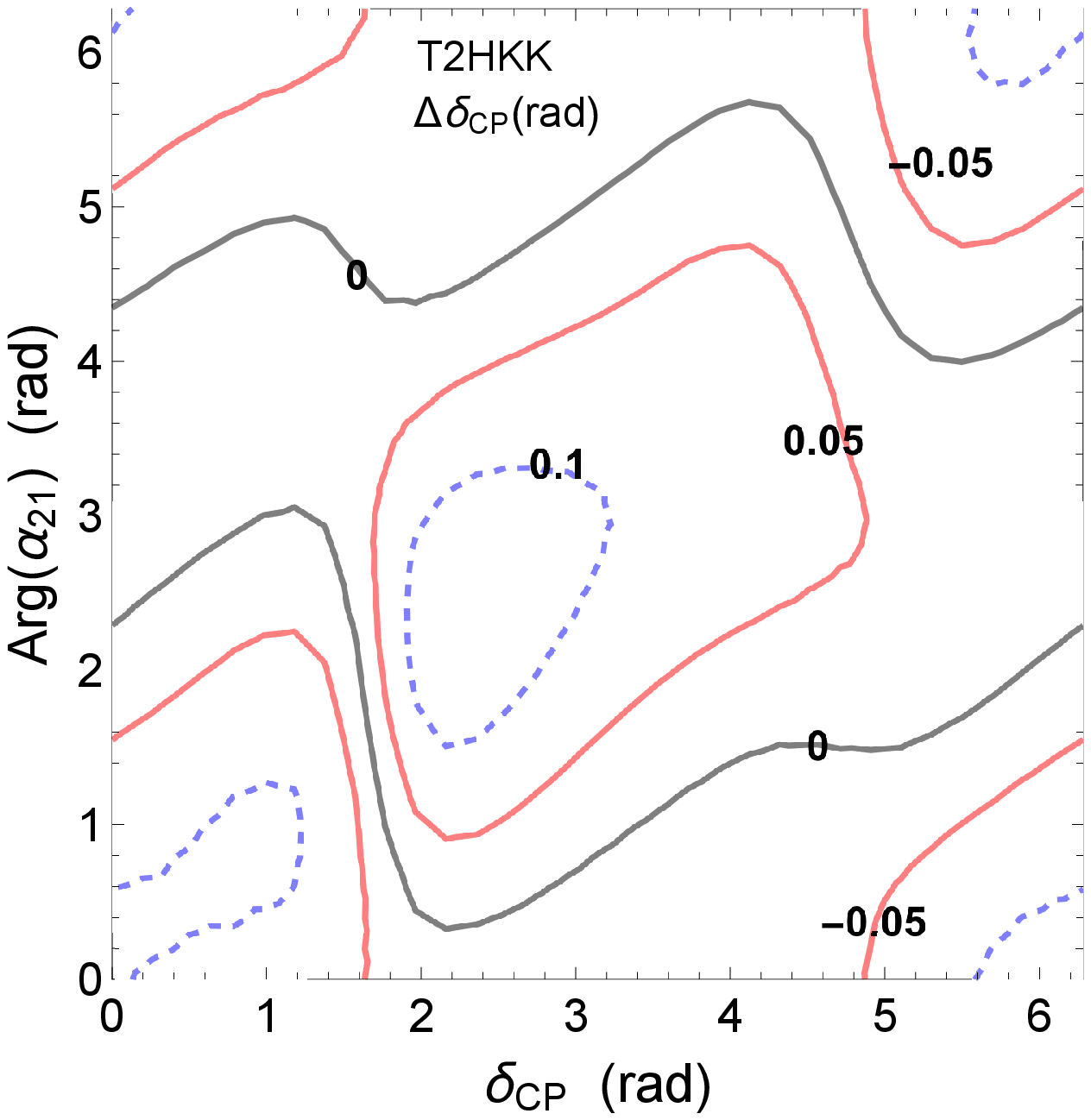}
    \\
    \includegraphics[width=80mm]{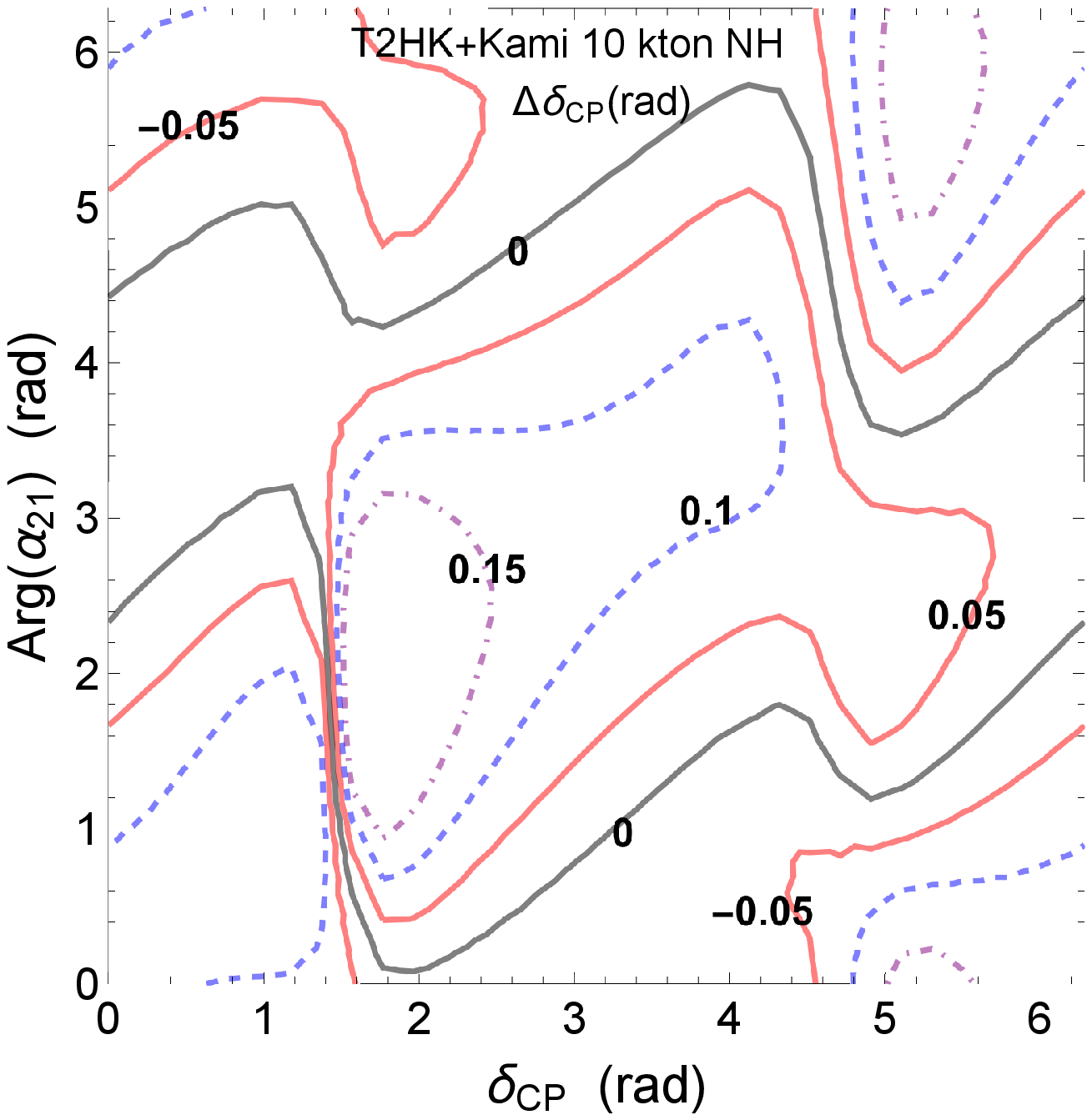}
    \includegraphics[width=80mm]{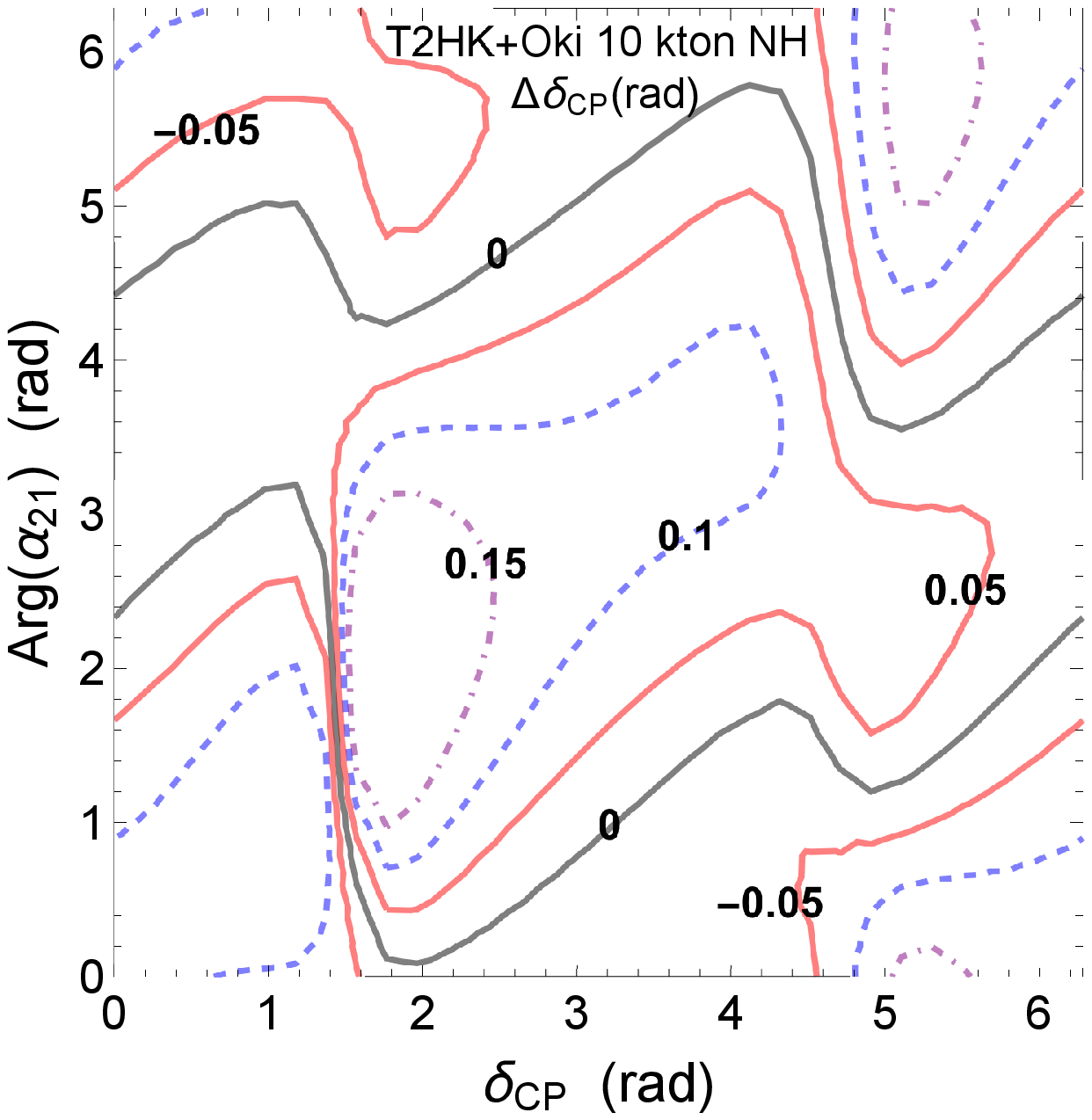}
    \caption{
    The difference between the true value of $\delta_{CP}$, and the value that minimizes $\chi^2(\Pi)$ of Eq.~(\ref{chi22})
     with true mass hierarchy.
    The benchmark with Eq.~(\ref{ournu}) is assumed, and the true mass hierarchy is normal.
    The upper-left, upper-right and lower-right subplots correspond to the T2HK, the T2HKK, 
     and the plan of the T2HK plus a 10~kton water Cerenkov detector at Oki, respectively.
    For comparative study, we show in the lower-left a subplot for a plan of the T2HK plus a 10~kton water Cerenkov detector at Kamioka.
   $\vert\Delta\delta_{CP}\vert=0, \ 0.05{\rm rad}, \ 0.1{\rm rad}$, and $0.15{\rm rad}$ on the black solid, red sold, blue dashed, and purple dot-dashed contours, respectively.     }
    \label{nhdch}
  \end{center}
\end{figure}

We study the impact of heavy neutrino mixing on the $\delta_{CP}$ measurement performed by fitting data without incorporating heavy neutrino mixing.
To assess the impact, we compute the difference between the true value of $\delta_{CP}$, and the value that minimizes $\chi^2(\Pi)$ of Eq.~(\ref{chi22}) with the true mass hierarchy.
We show in Figure~\ref{nhdch} contour plots of the difference,
\begin{align}
\Delta\delta_{CP} &\equiv (\delta_{CP} \ {\rm that \ minimizes} \ \chi^2(\delta_{CP}, \ {\rm true} \ {\rm sgn}(\Delta m^2_{32})) - ({\rm true} \ \delta_{CP})
\label{ddcp}
\end{align}
 on the same plane when the true mass hierarchy is normal.

\begin{figure}[H]
  \begin{center}
    \includegraphics[width=80mm]{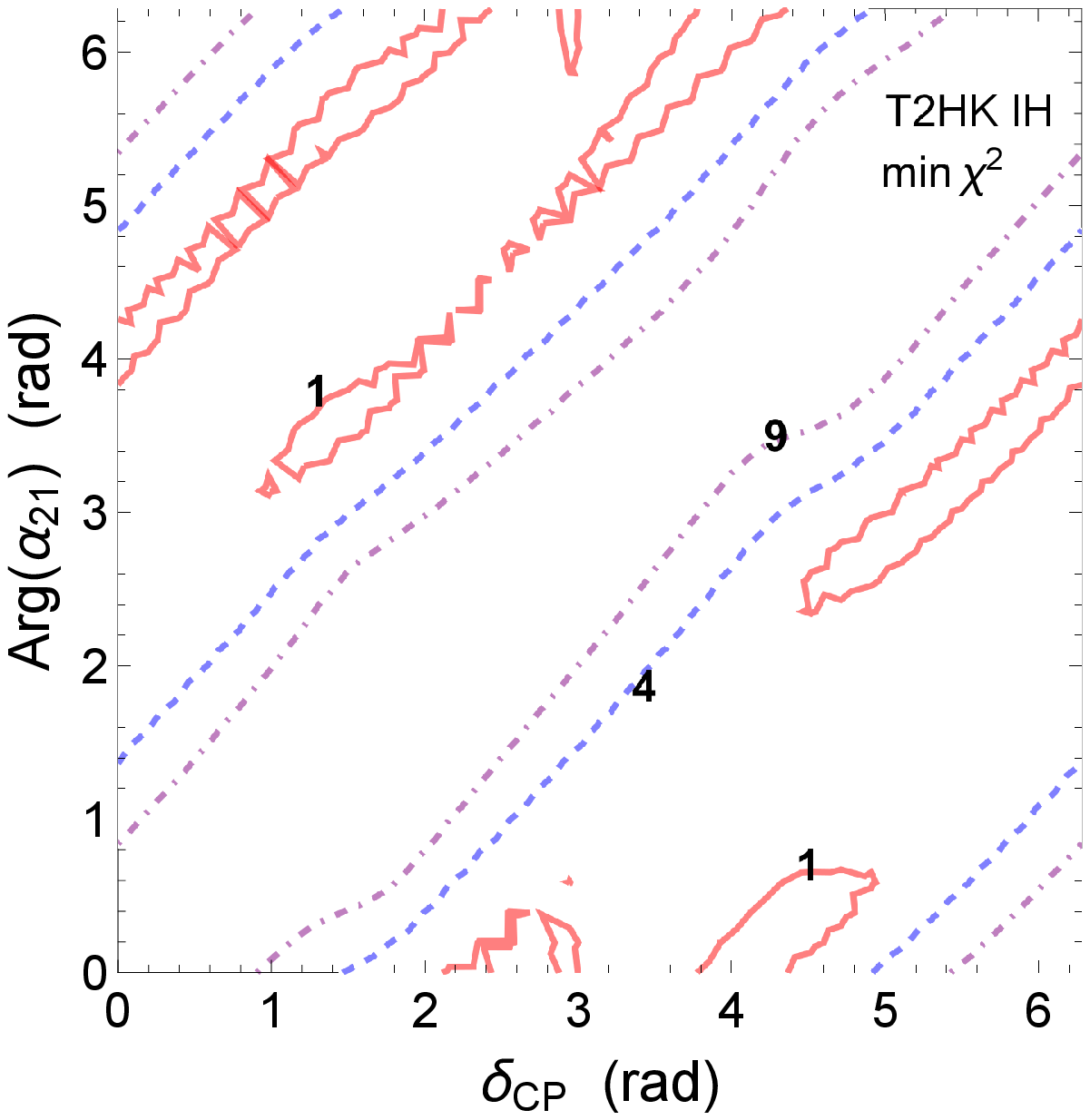}    
    \includegraphics[width=80mm]{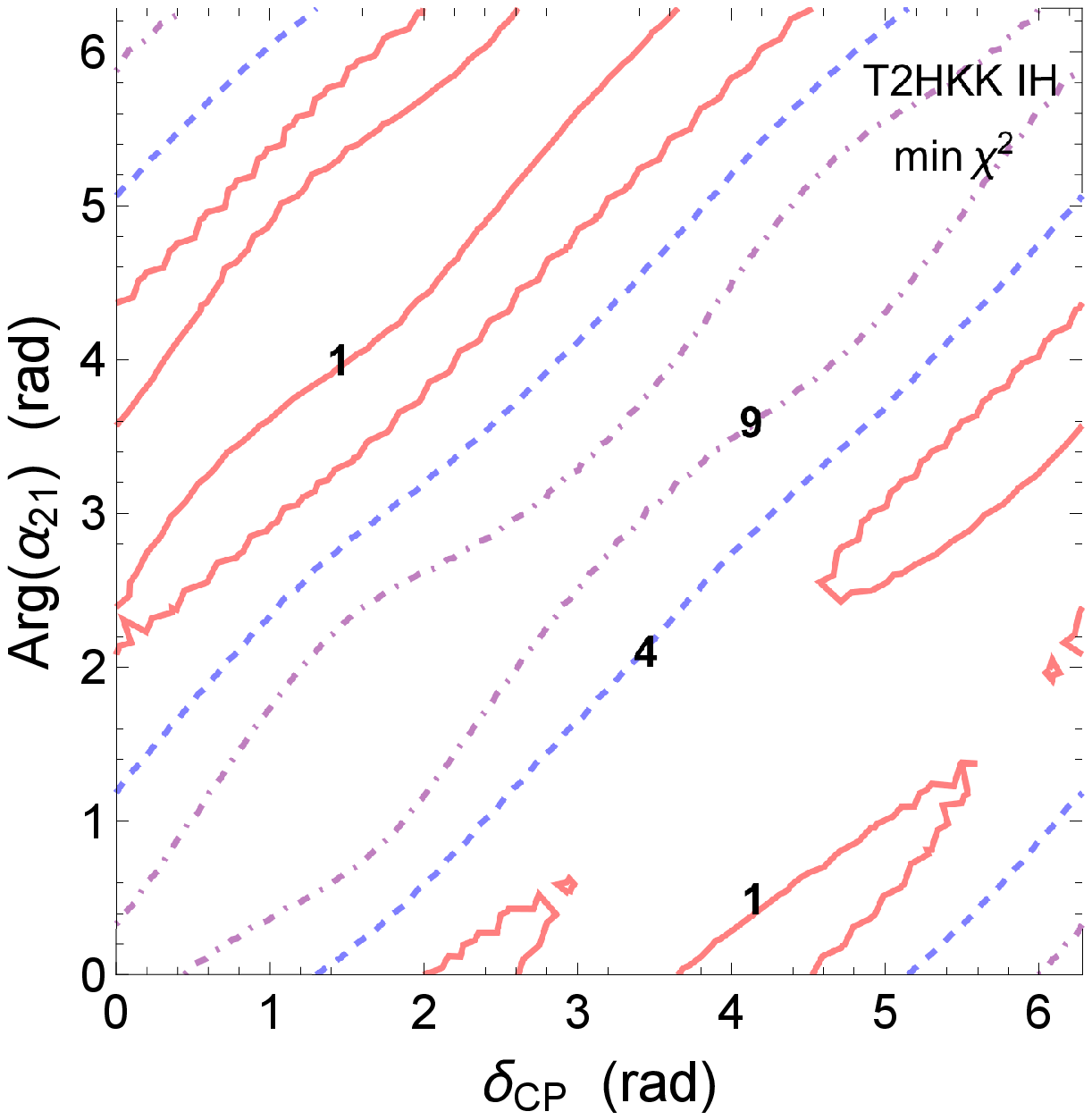}
    \\
    \includegraphics[width=80mm]{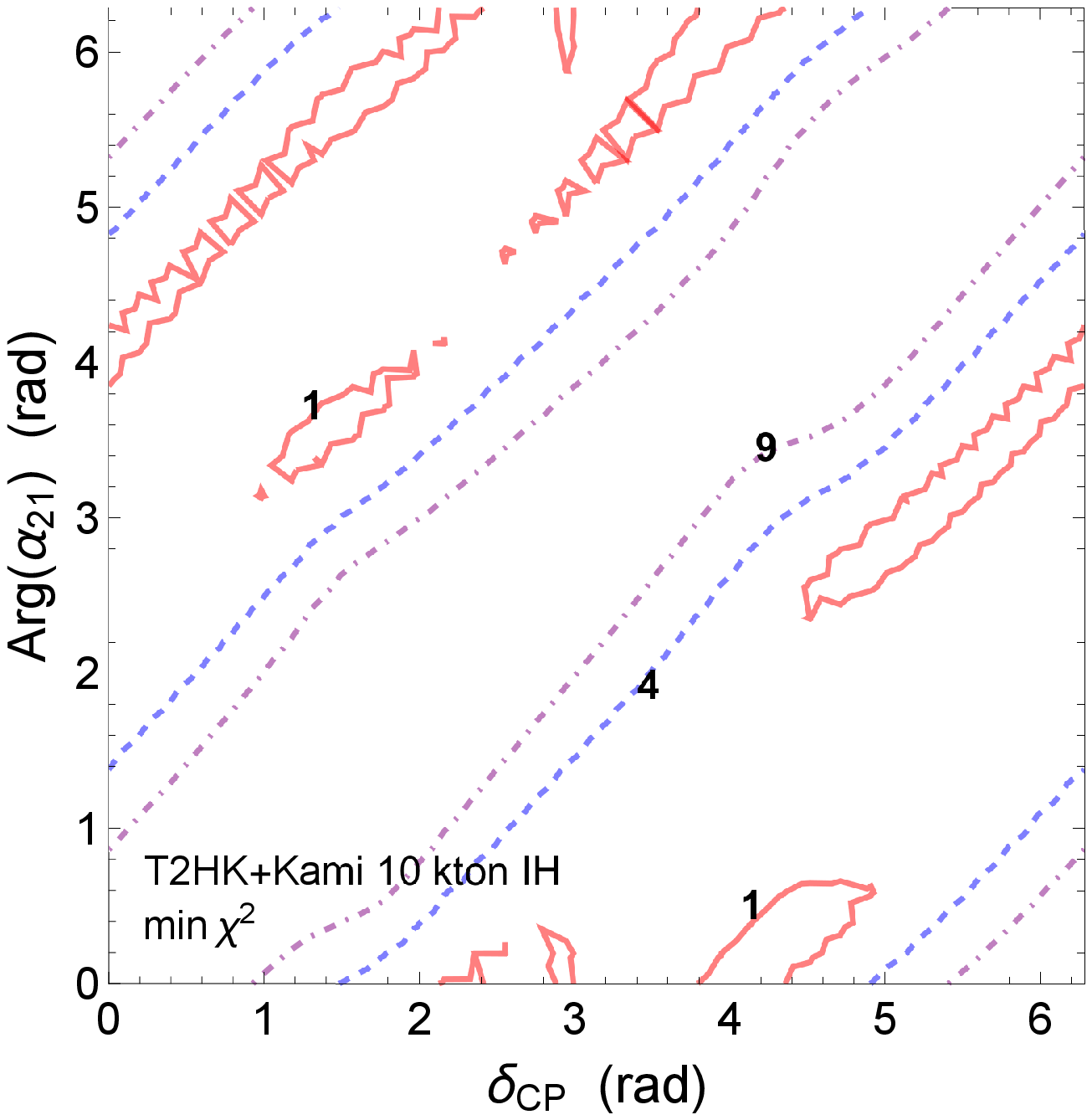}
    \includegraphics[width=80mm]{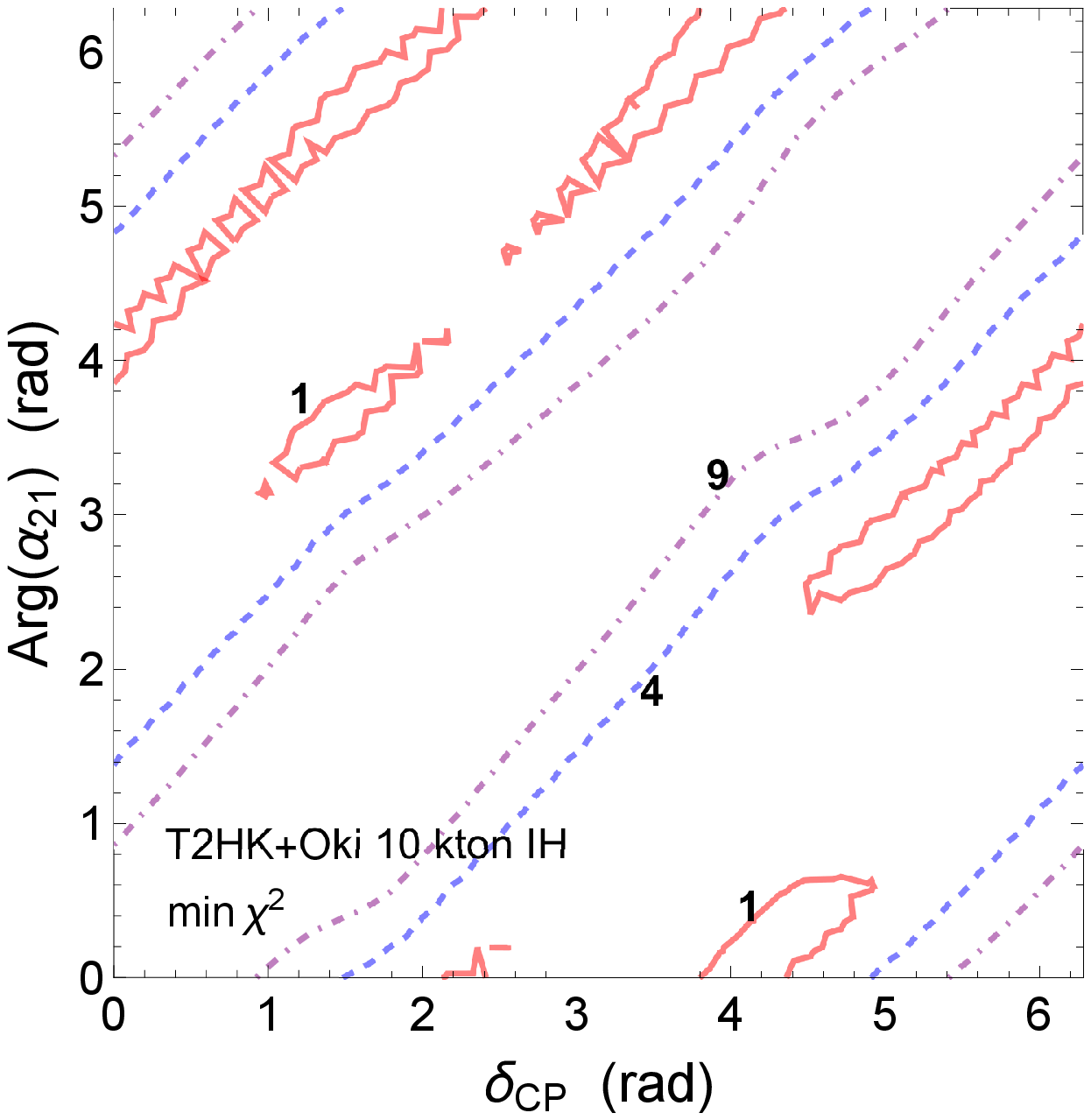}
    \caption{
    The minimum of $\chi^2(\Pi)$ Eq.~(\ref{chi22}), $\min \chi^2$.
    The benchmark with Eq.~(\ref{ournu}) is assumed, and the true mass hierarchy is inverted.
    The upper-left, upper-right and lower-right subplots correspond to the T2HK, the T2HKK, 
     and the plan of the T2HK plus a 10~kton water Cerenkov detector at Oki, respectively.
    For comparative study, we show in the lower-left a subplot for a plan of the T2HK plus a 10~kton water Cerenkov detector at Kamioka.
    $\min \chi^2=1,\,4,\,9$ on the red solid, blue dashed, and purple dot-dashed contours, respectively.
    }
    \label{ihch}
  \end{center}
\end{figure}

\begin{figure}[H]
  \begin{center}    
    \includegraphics[width=80mm]{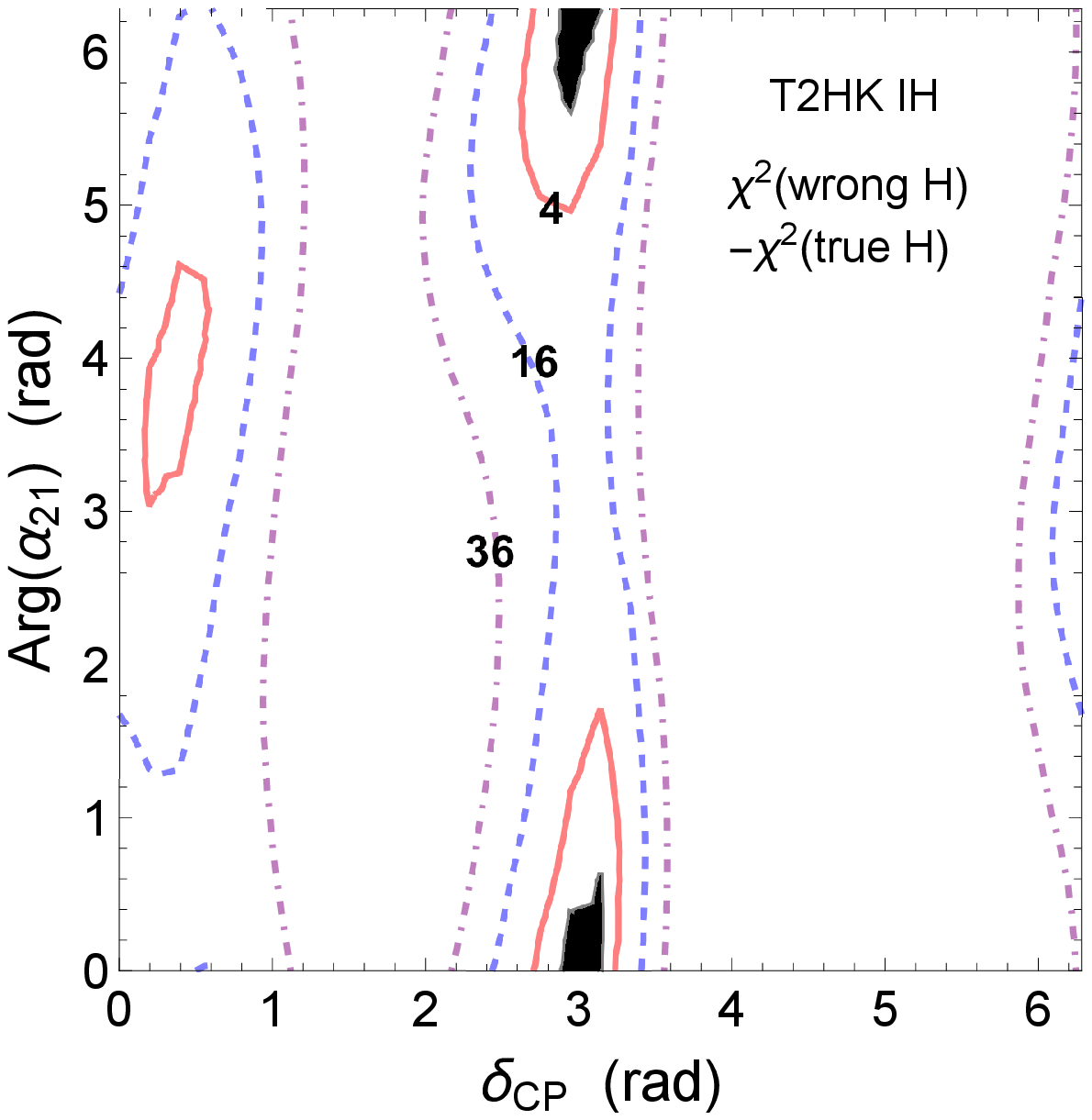}
    \includegraphics[width=80mm]{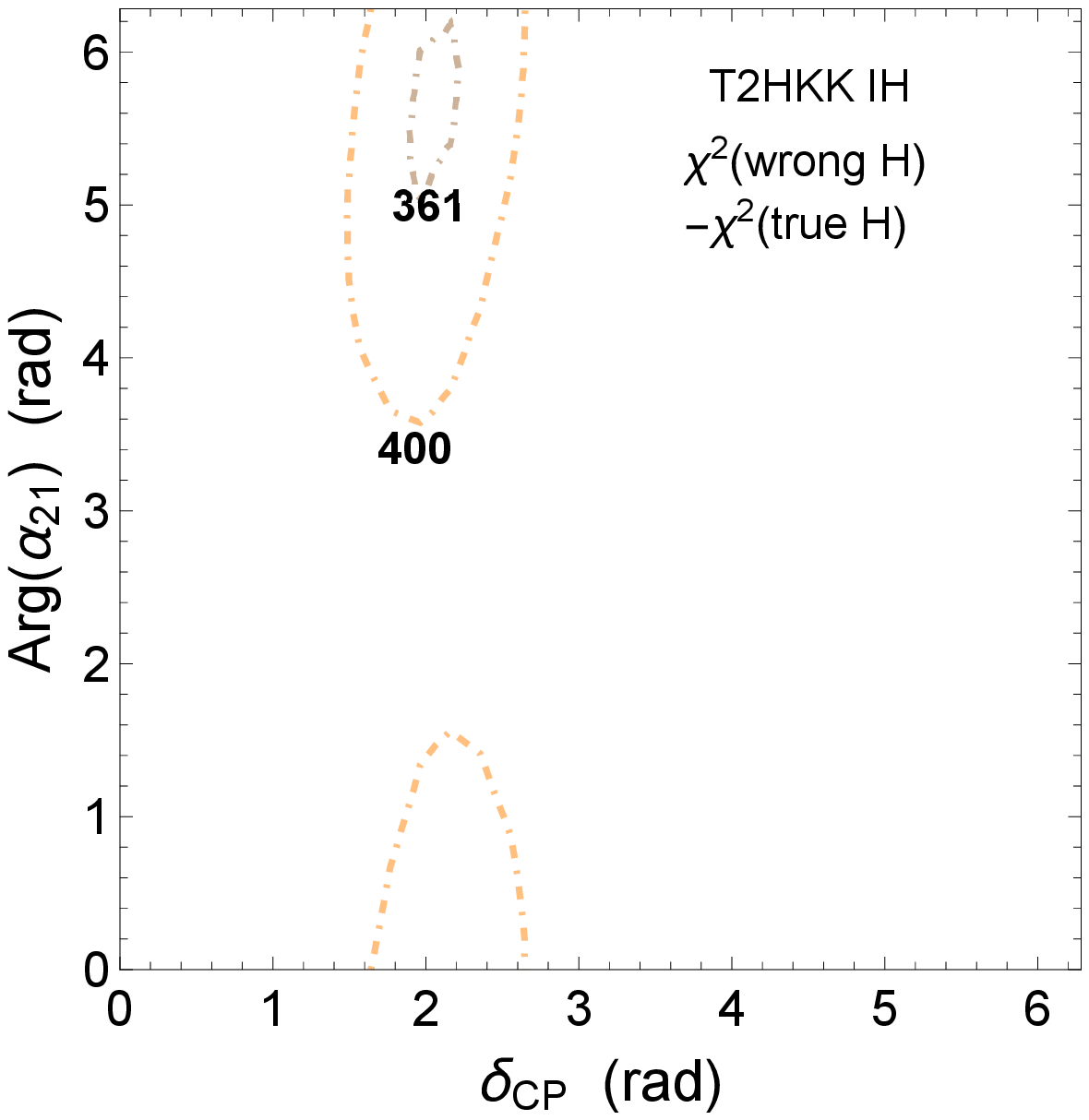}
    \\
    \includegraphics[width=80mm]{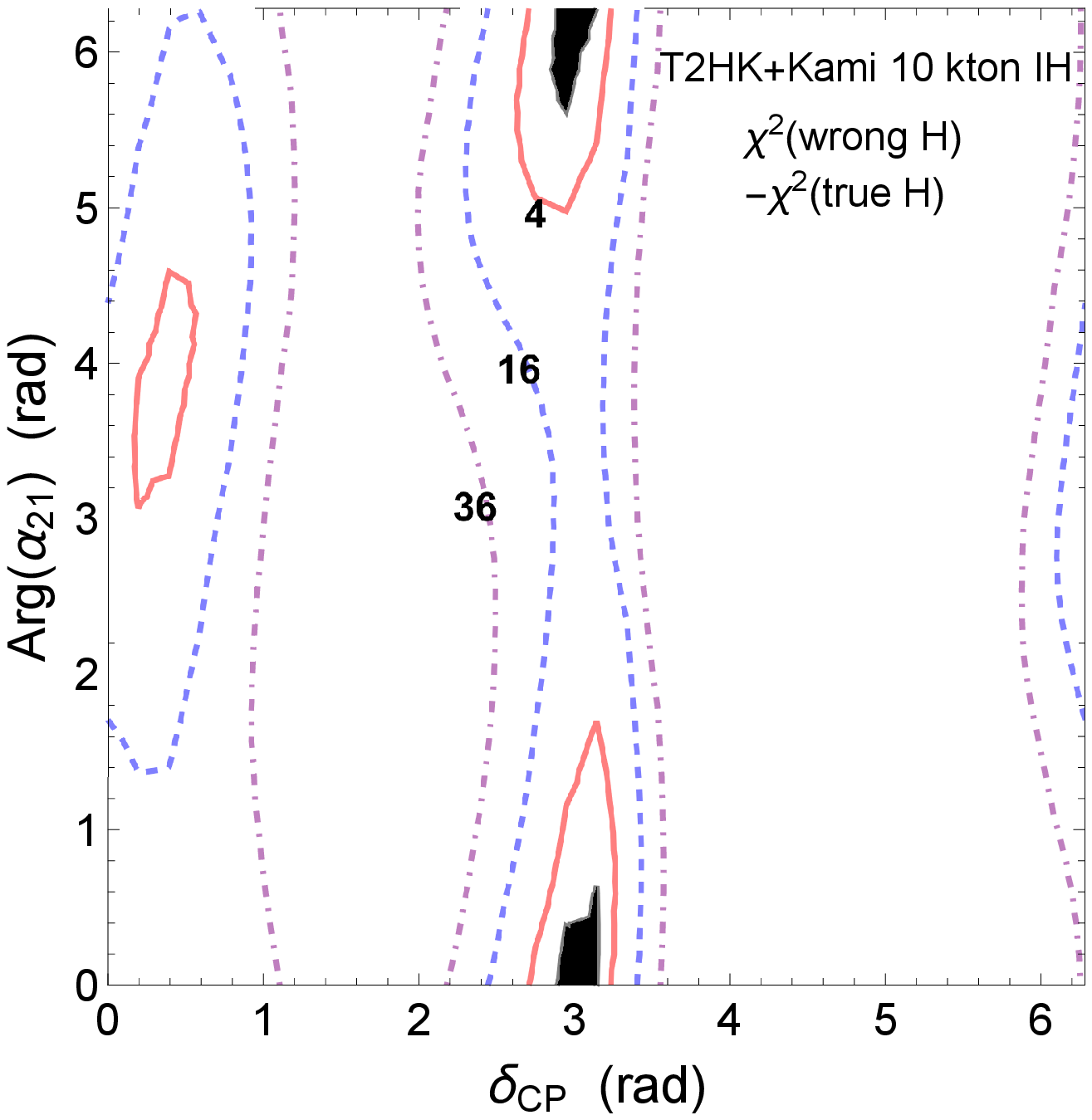}
    \includegraphics[width=80mm]{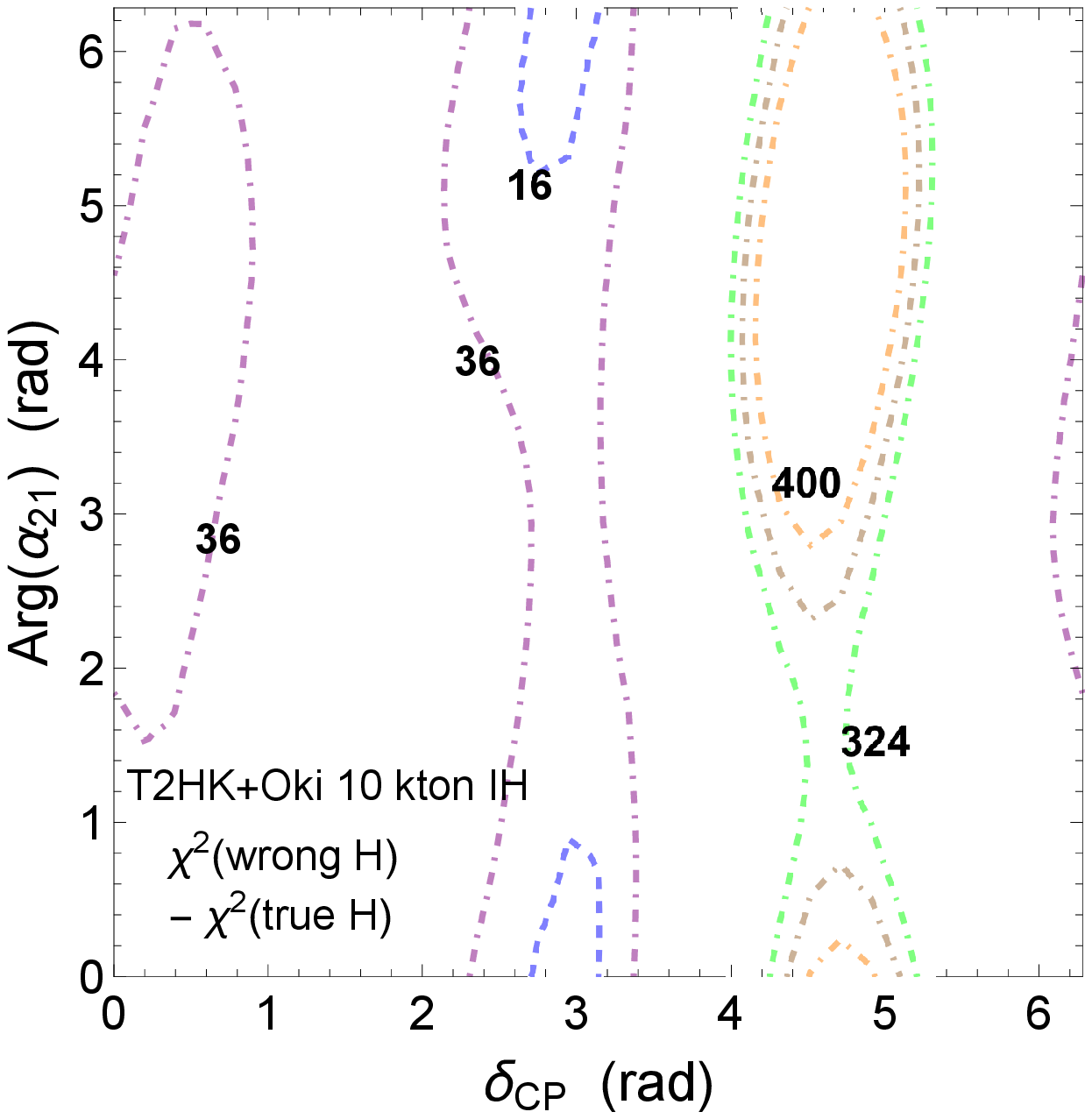}
    \caption{
    The difference between the minima (with respect to $\delta_{CP}$ only) of $\chi^2(\Pi)$ for the wrong and true mass hierarchy,
     $\{\min\chi^2({\rm wrong \ H})-\min\chi^2({\rm true \ H})\}$, which is the square of the significance of rejecting the wrong hierarchy.
    The benchmark with Eq.~(\ref{ournu}) is assumed, and the true mass hierarchy is inverted.
    The upper-left, upper-right and lower-right subplots correspond to the T2HK, the T2HKK, 
     and the plan of the T2HK plus a 10~kton water Cerenkov detector at Oki, respectively.
    For comparative study, we show in the lower-left a subplot for a plan of the T2HK plus a 10~kton water Cerenkov detector at Kamioka.
    $\{\min\chi^2({\rm wrong \ H})-\min\chi^2({\rm true \ H})\}=4,\,16,\,36$ on the red solid, blue dashed, and purple dot-dashed contours, respectively, and 
     $\{\min\chi^2({\rm wrong \ H})-\min\chi^2({\rm true \ H})\}=18^2,\,19^2,\,20^2$ on the green, brown, and orange dot-dashed contours, respectively.   
     The black-filled regions are where the wrong mass hierarchy is favored over the true mass hierarchy, \textit{i.e.},
 $\min\chi^2({\rm wrong \ H})-\min\chi^2({\rm true \ H})<0$.
     }
    \label{ihmass}
  \end{center}
\end{figure}

\begin{figure}[H]
  \begin{center}
    \includegraphics[width=80mm]{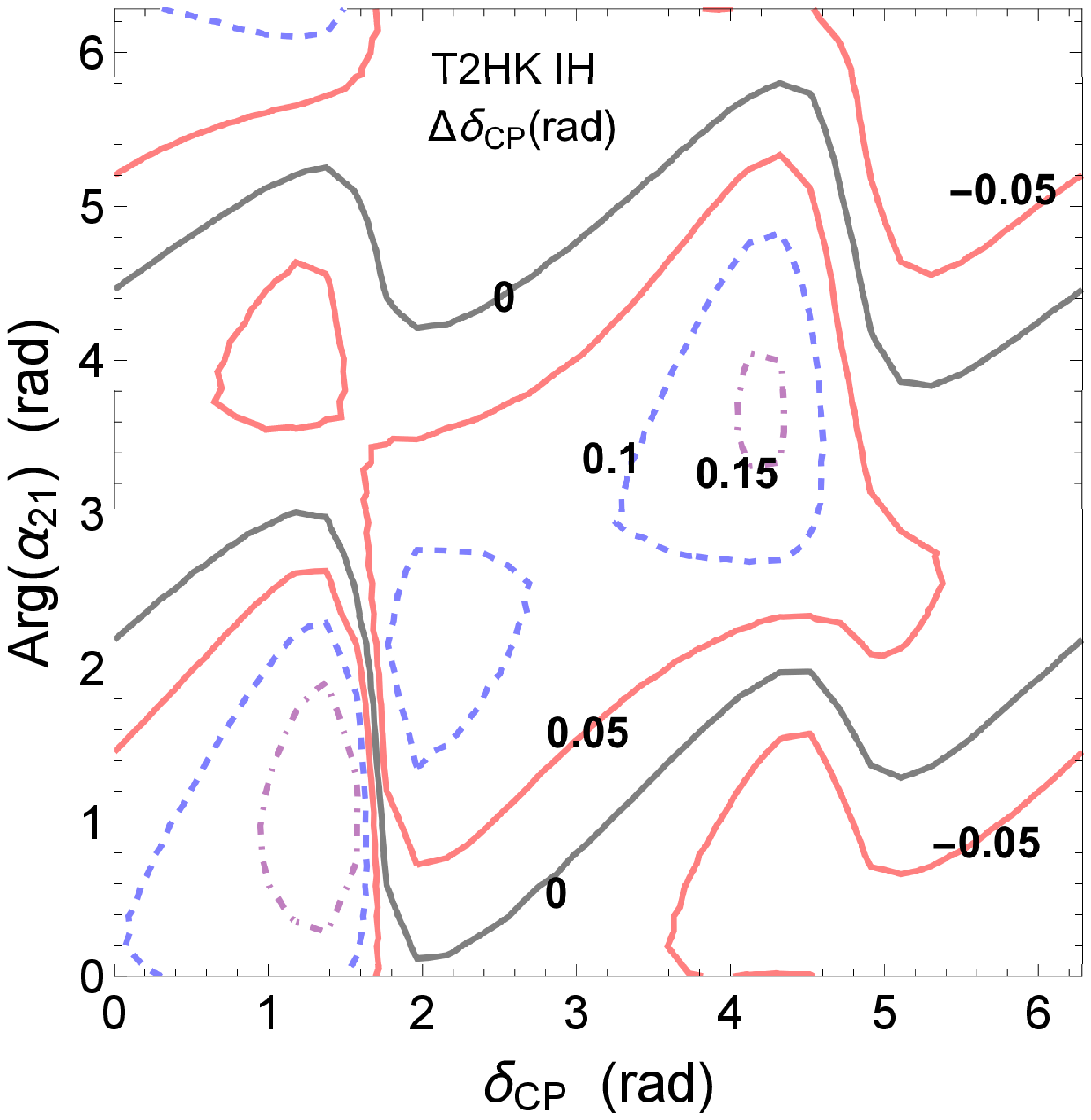}   
    \includegraphics[width=80mm]{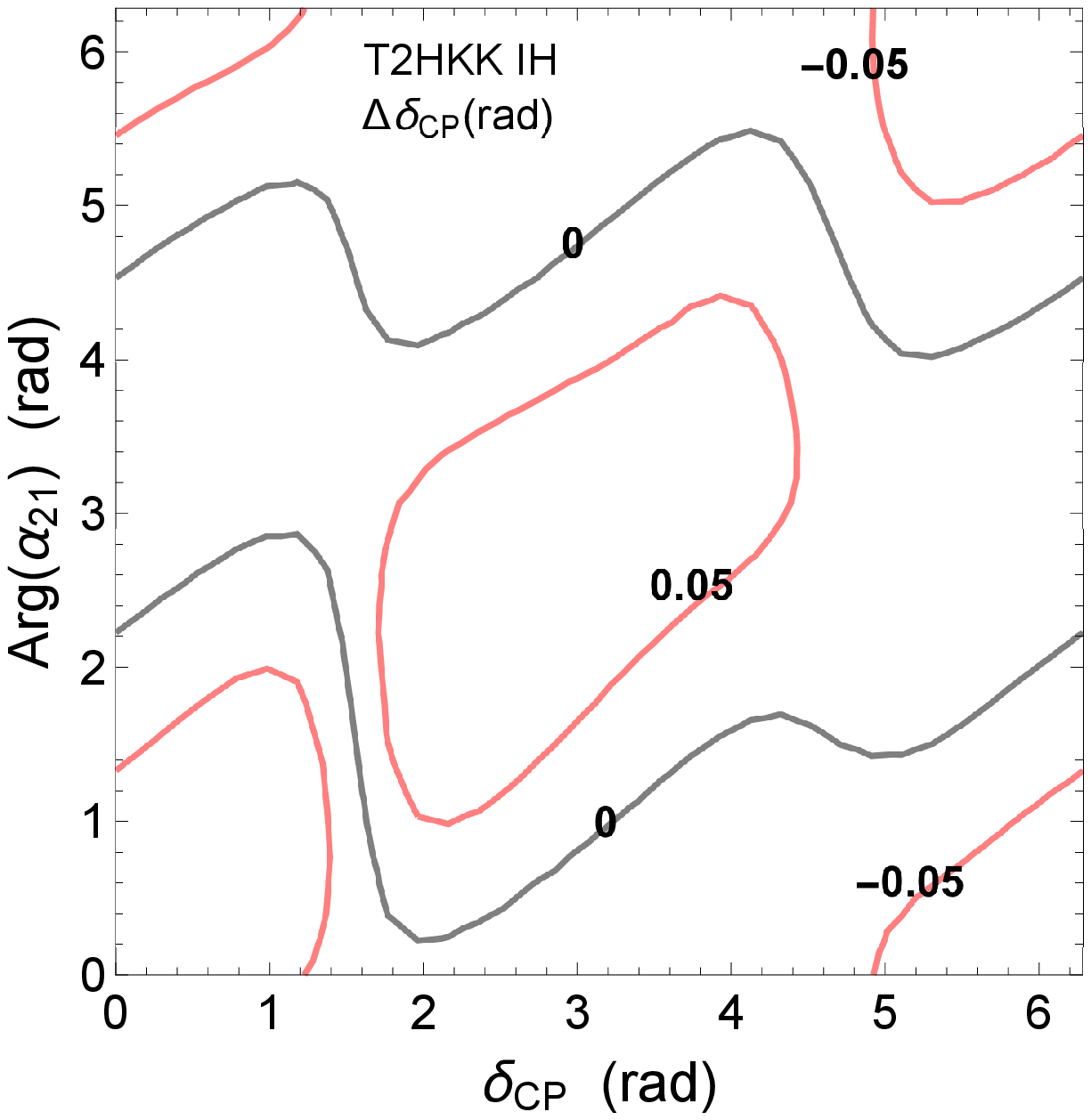}
    \\
    \includegraphics[width=80mm]{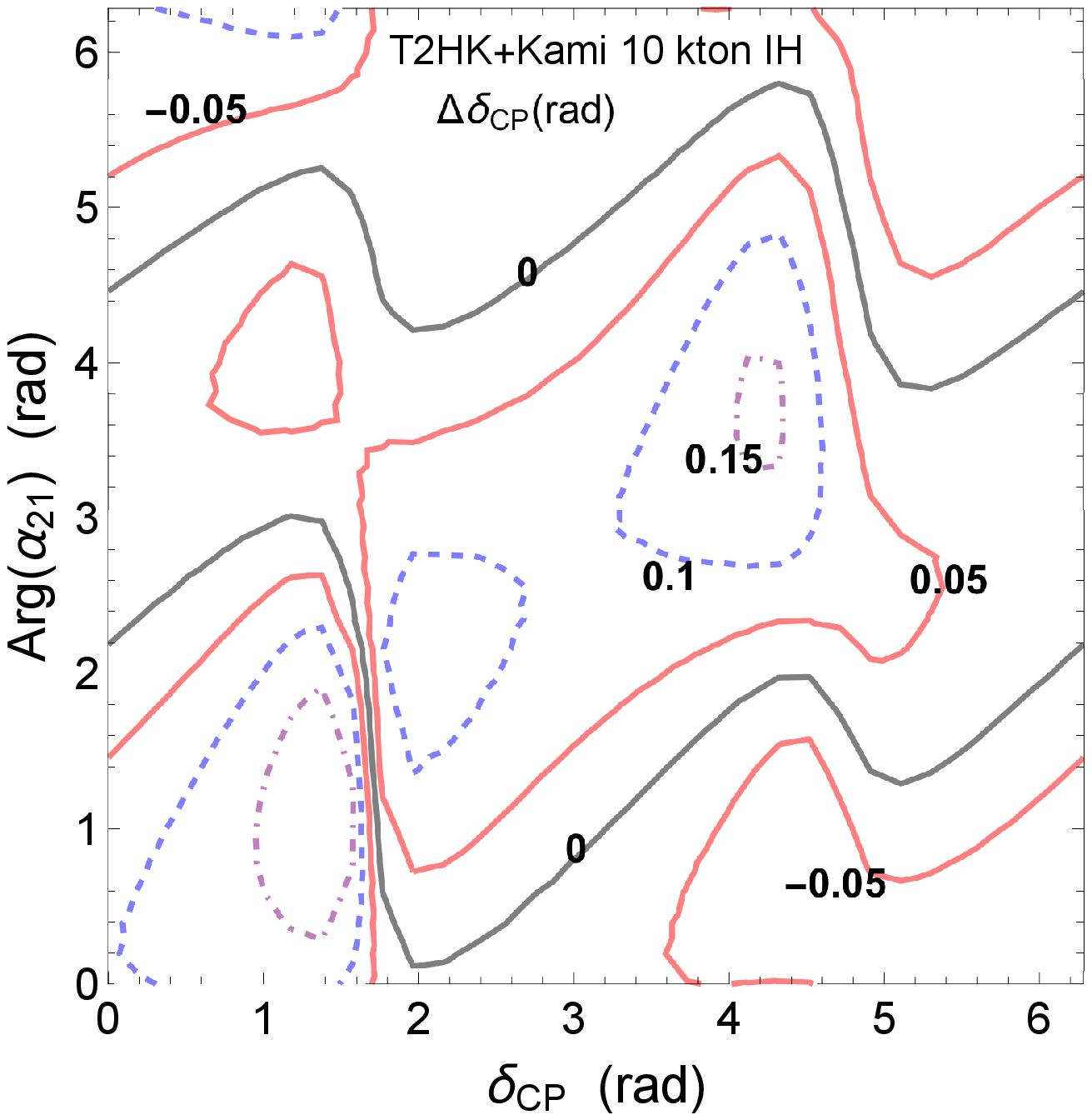}
    \includegraphics[width=80mm]{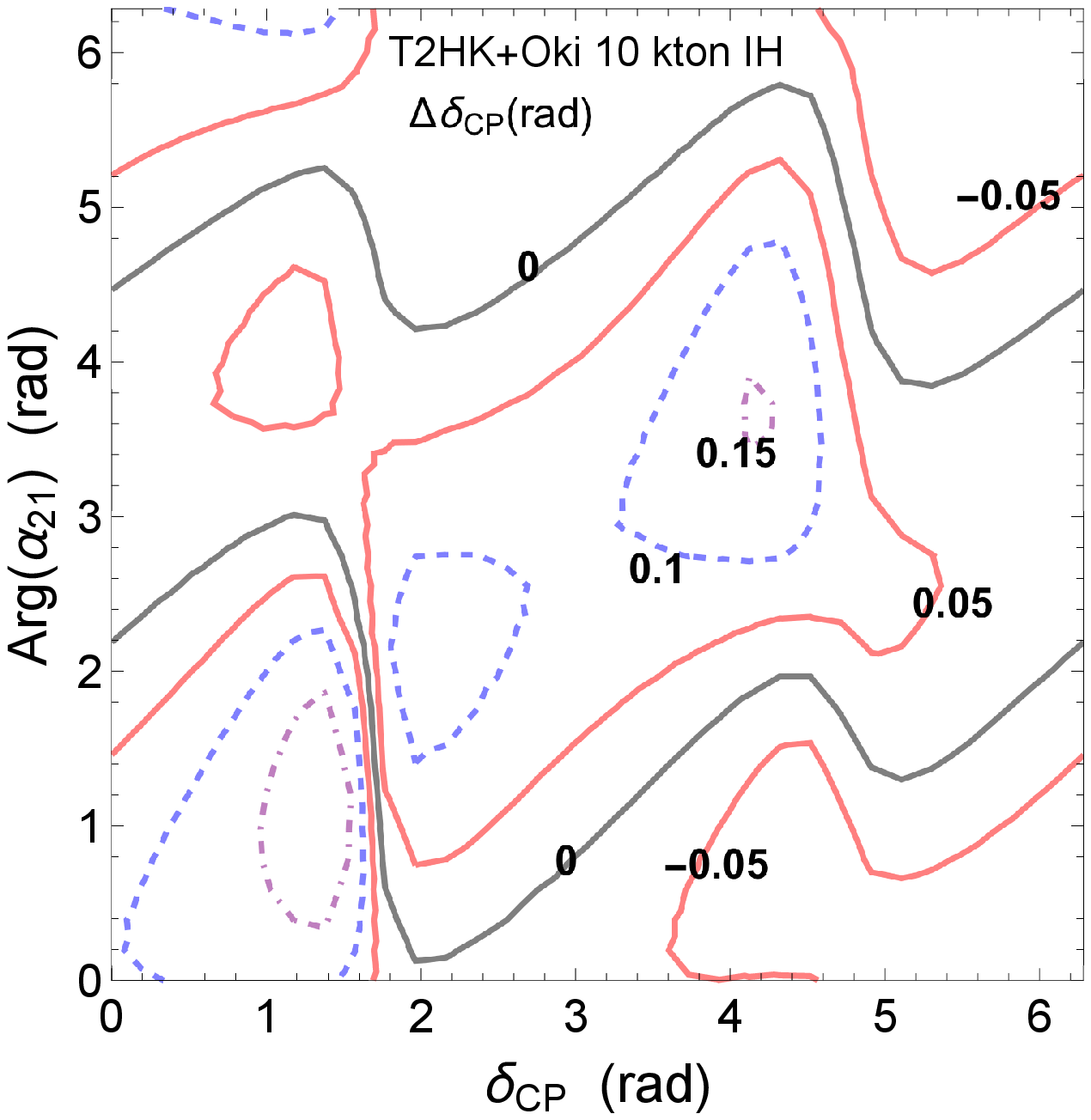}
    \caption{
       The difference between the true value of $\delta_{CP}$, and the value that minimizes $\chi^2(\Pi)$ of Eq.~(\ref{chi22})
    with the true mass hierarchy.
    The benchmark with Eq.~(\ref{ournu}) is assumed, and the true mass hierarchy is inverted.
    The upper-left, upper-right and lower-right subplots correspond to the T2HK, the T2HKK, 
     and the plan of the T2HK plus a 10~kton water Cerenkov detector at Oki, respectively.
    For comparative study, we show in the lower-left a subplot for a plan of the T2HK plus a 10~kton water Cerenkov detector at Kamioka.
   $\vert\Delta\delta_{CP}\vert=0, \ 0.05{\rm rad}, \ 0.1{\rm rad}$, and $0.15{\rm rad}$ on the black solid, red sold, blue dashed, and purple dot-dashed contours, respectively.    
    }
    \label{ihdch}
  \end{center}
\end{figure}

Figures~\ref{ihch},~\ref{ihmass},~\ref{ihdch} are the corresponding figures when the true mass hierarchy is inverted.

The following observations are made:
\\

\noindent
(A) 
In Figures~\ref{nhch},\ref{ihch}, we see that the significance of heavy neutrino mixing can be above $3\sigma$ for Arg($\alpha_{21})\sim\delta_{CP}$,
 but decreases considerably for Arg($\alpha_{21})\simeq\delta_{CP}\pm\pi/2$ in all the experiments,
 confirming that the sensitivity deteriorates due to suppression of interference between the standard oscillation amplitude and non-unitary mixing $\alpha_{21}$ at the first oscillation peak, as can be read from Eqs.~(\ref{interference1}),~(\ref{interference2}).
The regions of depleted sensitivity is slightly off the lines Arg($\alpha_{21})=\delta_{CP}\pm\pi/2$, because of the influence of the second, subleading interference term in Eqs.~(\ref{interference1}),~(\ref{interference2}).
We also find that the significance does not ameliorate in the T2HKK and in the extension of the T2HK with a 10~kton detector at Oki.
Rather, the T2HKK shows inferior performance due to the loss of statistics, which is seen in Table~\ref{numbers}.
\\

\begin{figure}[H]
  \begin{center}
    \includegraphics[width=120mm]{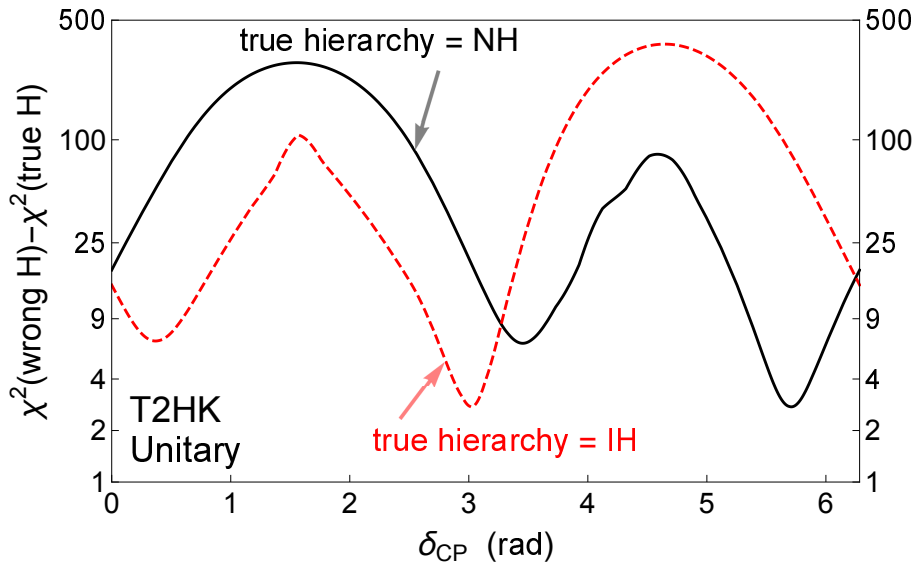}   
    \caption{
    The difference between the minima (with respect to $\delta_{CP}$ only) of $\chi^2(\Pi)$ for the wrong and true mass hierarchy,
     $\{\min\chi^2({\rm wrong \ H})-\min\chi^2({\rm true \ H})\}$, when the neutrino mixing is unitary (no heavy neutrino mixing is present), in the T2HK.
     The results are presented as functions of the true value of $\delta_{CP}$.
     The black solid line corresponds to the case when the true hierarchy is normal, and the red dashed line to the case when the true hierarchy is inverted.
        }
    \label{masshieunitary}
  \end{center}
\end{figure}

\noindent
(B)
In Figures~\ref{nhmass},\ref{ihmass},  we observe that in the T2HK, the wrong mass hierarchy is favored 
 when $(\delta_{CP}, \, {\rm Arg}(\alpha_{21}))\sim(0, \, 0), \, (\pi, \, \pi)$ holds in the normal hierarchy case, and when
 $(\delta_{CP}, \, {\rm Arg}(\alpha_{21}))\sim(\pi, \, 0)$ holds in the inverted hierarchy case.
To estimate the impact of non-unitary mixing on the mass hierarchy measurement in the T2HK, we show in Figure~\ref{masshieunitary}
 the values of $\{\min\chi^2({\rm wrong \ H})-\min\chi^2({\rm true \ H})\}$ when non-unitarity is absent, \textit{i.e.} $N_{NU}=I$,
 calculated with the same procedure as Figures~\ref{nhmass},~\ref{ihmass}.
\noindent
Comparison of Figures~\ref{nhmass},\ref{ihmass} with Figure~\ref{masshieunitary} indicates that for $\delta_{CP}\sim0, \pi$,
 the T2HK has limited sensitivity to the mass hierarchy in the unitary case,
 and the presence of non-unitary mixing may aggravate it to the extent that the wrong mass hierarchy is favored.
Such impact of non-unitary mixing is because $\alpha_{21}$ cancels matter effects in the neutrino transition probability Eq.~(\ref{amp1}) when $(\delta_{CP}, \, {\rm Arg}(\alpha_{21}))\sim(0, \, 0), \, (\pi, \, \pi)$ holds in the normal hierarchy case, and when
 $(\delta_{CP}, \, {\rm Arg}(\alpha_{21}))\sim(\pi, \, 0)$ holds in the inverted hierarchy case.
Although $\alpha_{21}$ and matter effects add positively in the antineutrino transition probability Eq.~(\ref{amp2}) in the same cases,
 it does not change the situation because neutrino flux is larger than antineutrino flux in our setup (see Table~\ref{flux} and Figure~\ref{fluxfig}).

The T2HKK maintains strong sensitivity to the mass hierarchy in the presence of heavy neutrino mixing.
Also, adding a 10~kton detector at Oki to the T2HK drastically ameliorates sensitivity to the mass hierarchy, in spite of the small size of the Oki detector.
The advantage of the T2HKK and T2HK+Oki is because measurements with different matter effects resolve degeneracy between non-unitary mixing $\alpha_{21}$ and matter effects.

We stress that even a small 10~kton detector at Oki negates the impact of heavy neutrino mixing on the mass hierarchy measurement and allows correct measurement.
\\

\begin{figure}[H]
  \begin{center}
    \includegraphics[width=80mm]{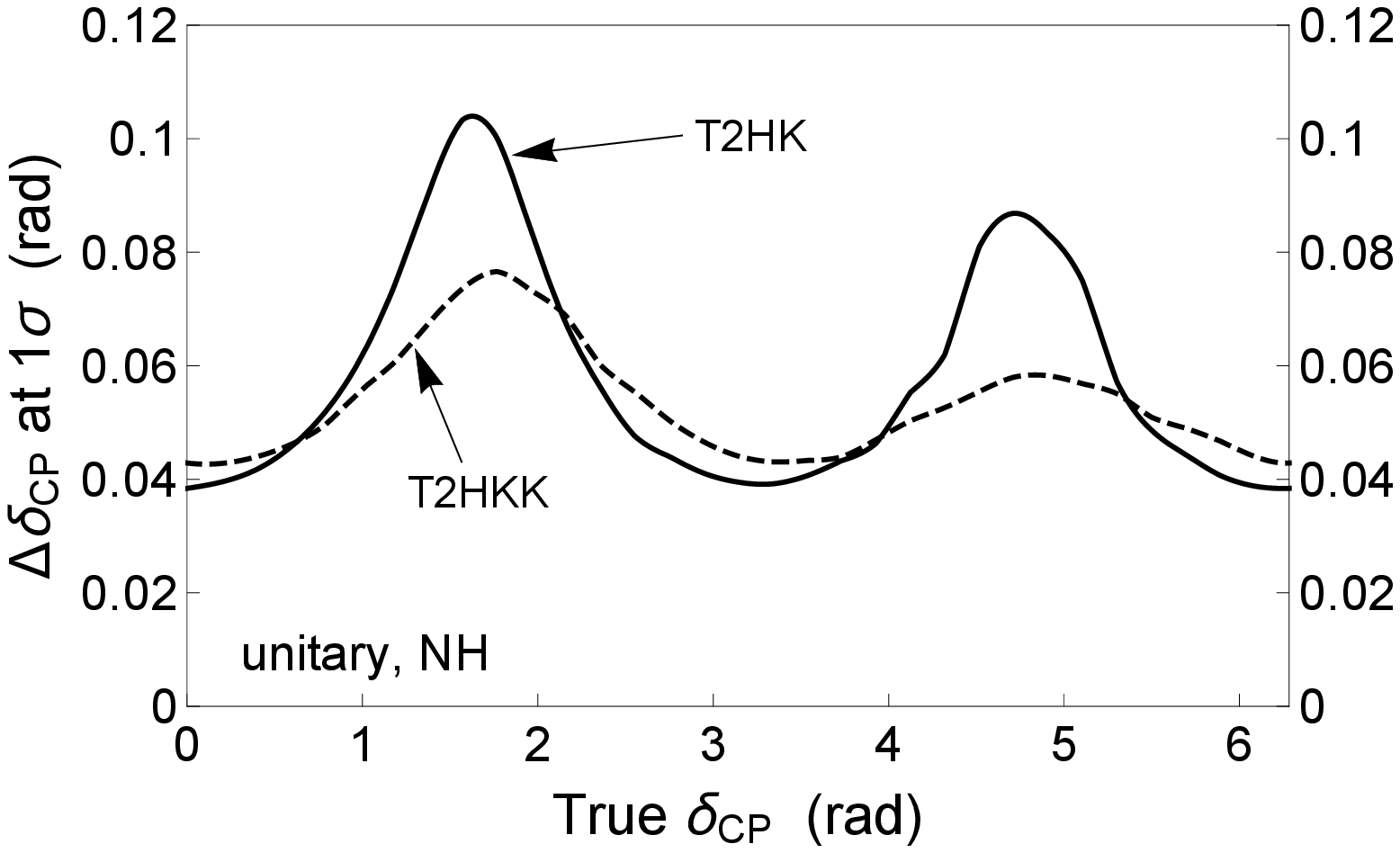}   
    \includegraphics[width=80mm]{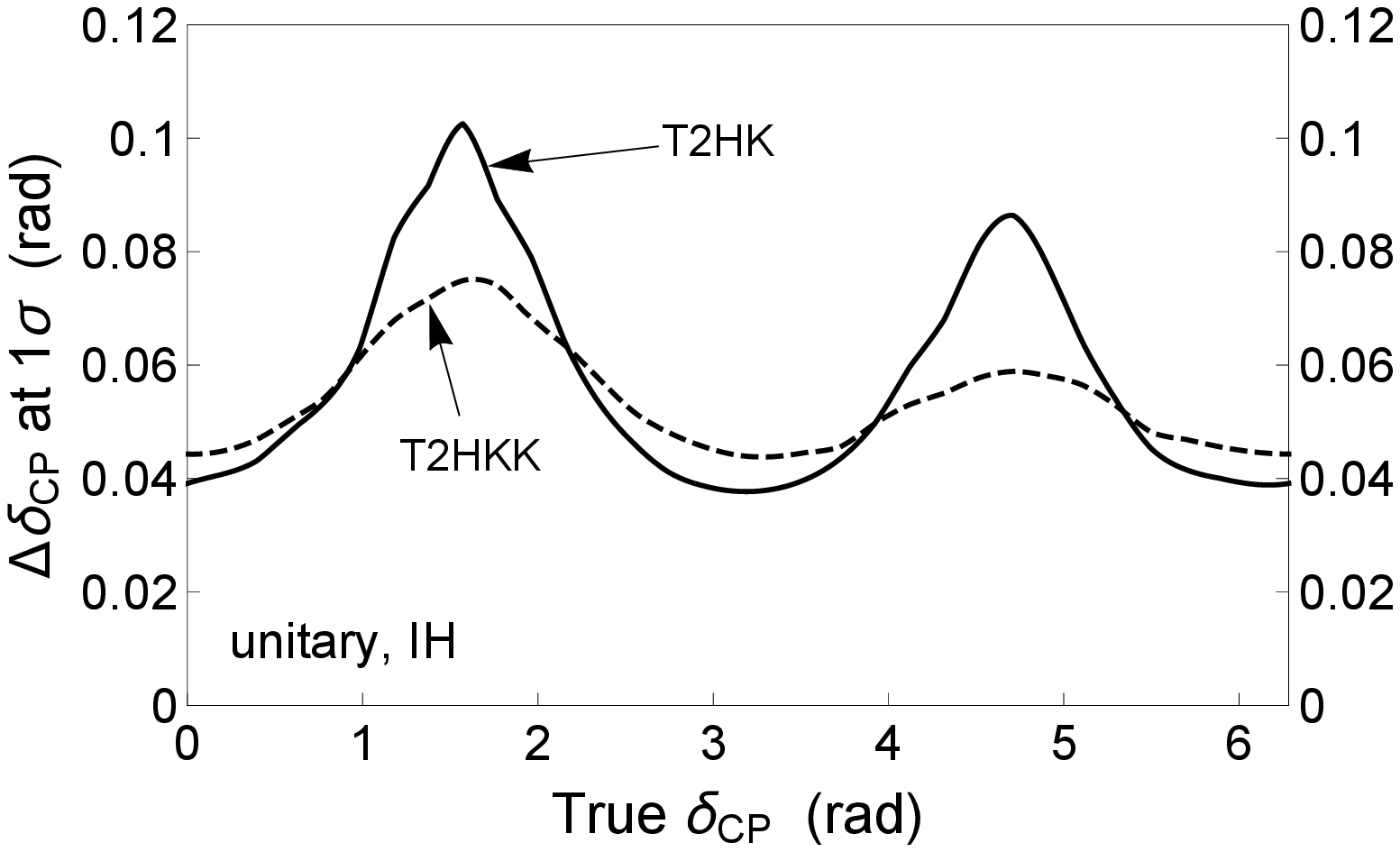}   
    \caption{
    The resolution of the $\delta_{CP}$ measurement when non-unitarity is absent,
     quantified as the difference between $\delta_{CP}$ that gives $\chi^2=1$ and the true $\delta_{CP}$ Eq.~(\ref{ddcp1sigma}).  
     The results are presented as functions of the true $\delta_{CP}$.
     The left and right plots correspond to the cases when the true mass hierarchy is normal and inverted, respectively.
     In each plot, the solid line indicates the sensitivity in the T2HK, and the dashed line the sensitivity in the T2HKK.}
    \label{deltaCPunitary}
  \end{center}
\end{figure}

\noindent
(C)
In Figures~\ref{nhdch},\ref{ihdch}, we find that the value of $\delta_{CP}$ that minimizes $\chi^2$ under the assumption of standard unitary oscillation can deviate from the true $\delta_{CP}$ by more than 0.15~radian in the T2HK, and by more than 0.1~radian in the T2HKK.
The result should be compared to the resolution of $\delta_{CP}$.
To this end, we present in Figure~\ref{deltaCPunitary} the difference between $\delta_{CP}$ that gives $\chi^2=1$ and the true $\delta_{CP}$
 when non-unitarity is absent ($N_{NU}=I$),
\begin{align} 
(\Delta\delta_{CP} \ {\rm at} \ 1\sigma) &\equiv 
(\delta_{CP} \ {\rm that \ gives} \ \chi^2=1) - ({\rm true} \ \delta_{CP}),
\label{ddcp1sigma}
\end{align}
 which quantifies $1\sigma$ resolution of the $\delta_{CP}$ measurement in the standard oscillation case.
Eq.~(\ref{ddcp1sigma})  is calculated through a simulation performed with the same procedure.
The $1\sigma$ resolution for $\delta_{CP}$ is better than 0.1~radian in the T2HK, and better than 0.08~radian in the T2HKK,
 which is of the same order as the deviation of $\delta_{CP}$ due to non-unitarity.
This indicates that non-unitarity can have a non-negligible effect on the measurement of $\delta_{CP}$,
 and hence, if a hint of non-unitarity has been discovered, it must be incorporated in the data fitting in order to measure $\delta_{CP}$ correctly.

The deviation due to non-unitarity is enhanced when $({\rm true} \ \delta_{CP})\sim{\rm Arg}(\alpha_{21})$ holds.
This is because interference between the part of the standard oscillation amplitude proportional to $e^{i\delta_{CP}}$
 and the amplitude involving non-unitary mixing $\alpha_{21}$ is maximized,
 and so is the influence of $\alpha_{21}$ on the value of $\delta_{CP}$ that minimizes $\chi^2$.

The deviation is smaller in the T2HKK than in the T2HK.
This is due to the fact that the energy dependence of the part of the standard oscillation amplitude proportional to 
 $e^{i\delta_{CP}}$, which helps distinguish the standard and non-unitary amplitudes, is more accurately measured with a longer baseline.
Hence, the $\delta_{CP}$ measurement is less affected by non-unitarity in the T2HKK.
\\

\section{Summary}

We have studied the discovery potential for the mixing of heavy isospin-singlet neutrinos
 in the Tokai-to-Hyper-Kamiokande (T2HK), the Tokai-to-Hyper-Kamiokande-to-Korea (T2HKK), and a plan of adding a small detector at Oki Islands to the T2HK,
 and further examined the feasibility of measuring the mass hierarchy and the standard $CP$-violating phase $\delta_{CP}$ in the presence of heavy neutrino mixing by fitting data without assuming heavy neutrino mixing.
The mixing of heavy neutrinos is parametrized with a non-unitary mixing matrix for active flavors.
A benchmark that maximizes the non-unitary mixing and is consistent with the current experimental and theoretical bounds is employed to estimate the largest possible significance of heavy neutrino mixing and its impact on the mass hierarchy and $\delta_{CP}$ measurement.

Through a simulation, we have revealed that the significance of heavy neutrino mixing can be above $3\sigma$ in all the experiments 
 when the standard phase $\delta_{CP}$ and a new $CP$-violating phase originating from heavy neutrino mixing, Arg($\alpha_{21}$), satisfy
 Arg($\alpha_{21})\simeq\delta_{CP}$.
The significance decreases considerably for Arg($\alpha_{21})\simeq\delta_{CP}\pm\pi/2$, due to suppression of interference between the standard oscillation amplitude and a non-unitary mixing term.

In the T2HK, the mass hierarchy measurement is so affected by heavy neutrino mixing that the wrong mass hierarchy is favored for $(\delta_{CP}, \, {\rm Arg}(\alpha_{21}))\sim(0, \, 0), \, (\pi, \, \pi)$ in the normal hierarchy case, and for $(\delta_{CP}, \, {\rm Arg}(\alpha_{21}))\sim(\pi, \, 0)$ in the inverted hierarchy case.
This is because a non-unitary mixing term possibly cancels matter effects.
In contrast, the T2HKK and the extension of the T2HK with a 10~kton detector at Oki show strong sensitivity to the mass hierarchy even with heavy neutrino mixing, 
 because measurements with different baseline lengths at Kamioka and in Korea or at Oki resolve degeneracy between a non-unitary mixing term and matter effects.
We thus conclude that (i) the mass hierarchy measurement in the T2HKK is highly stable against any effects of heavy neutrino mixing,
 and that (ii) although the mass hierarchy measurement in the T2HK is easily affected, the addition of a small detector at Oki negates effects of heavy neutrino mixing.

In the presence of non-unitarity, the value of $\delta_{CP}$ measured without assuming non-unitarity deviates from the true $\delta_{CP}$.
The deviation is enhanced when $\delta_{CP}$ and ${\rm Arg}(\alpha_{21})$ are similar,
 because in such a case, the part of the standard oscillation amplitude proportional to $e^{i\delta_{CP}}$ interferes maximally with a non-unitary mixing term at oscillation peaks.
The deviation can be of the same order as $1\sigma$ resolution of the $\delta_{CP}$ measurement,
 which suggests that if a hint of non-unitary mixing has been discovered, one must employ an appropriate ansatz including the non-unitary mixing terms
 to fit neutrino oscillation data and measure $\delta_{CP}$ correctly.
\\

\section*{Acknowledgement}

The authors would like to thank Kaoru Hagiwara (KEK), Hiroyuki Ishida (NCTS) and Yuya Yamaguchi (Hokkaido University).
This work is partially supported by Scientific Grants by the Ministry of Education, Culture, Sports, Science and Technology of Japan (Nos. 24540272, 26247038, 15H01037, 16H00871, and 16H02189).
\\

\section*{Appendix A}

\begin{figure}[H]
  \begin{center}
    \includegraphics[width=80mm]{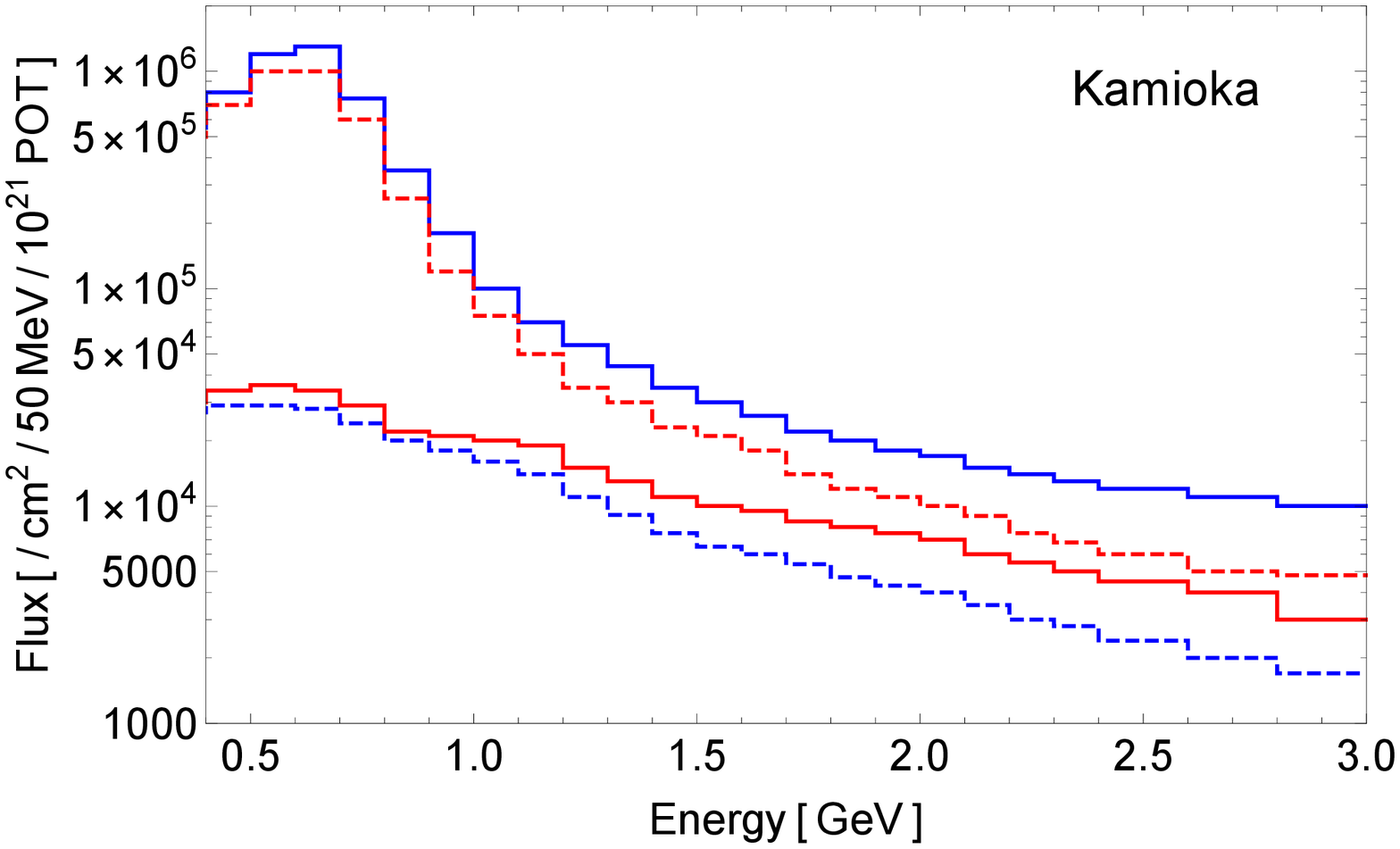}
    \\
    \includegraphics[width=80mm]{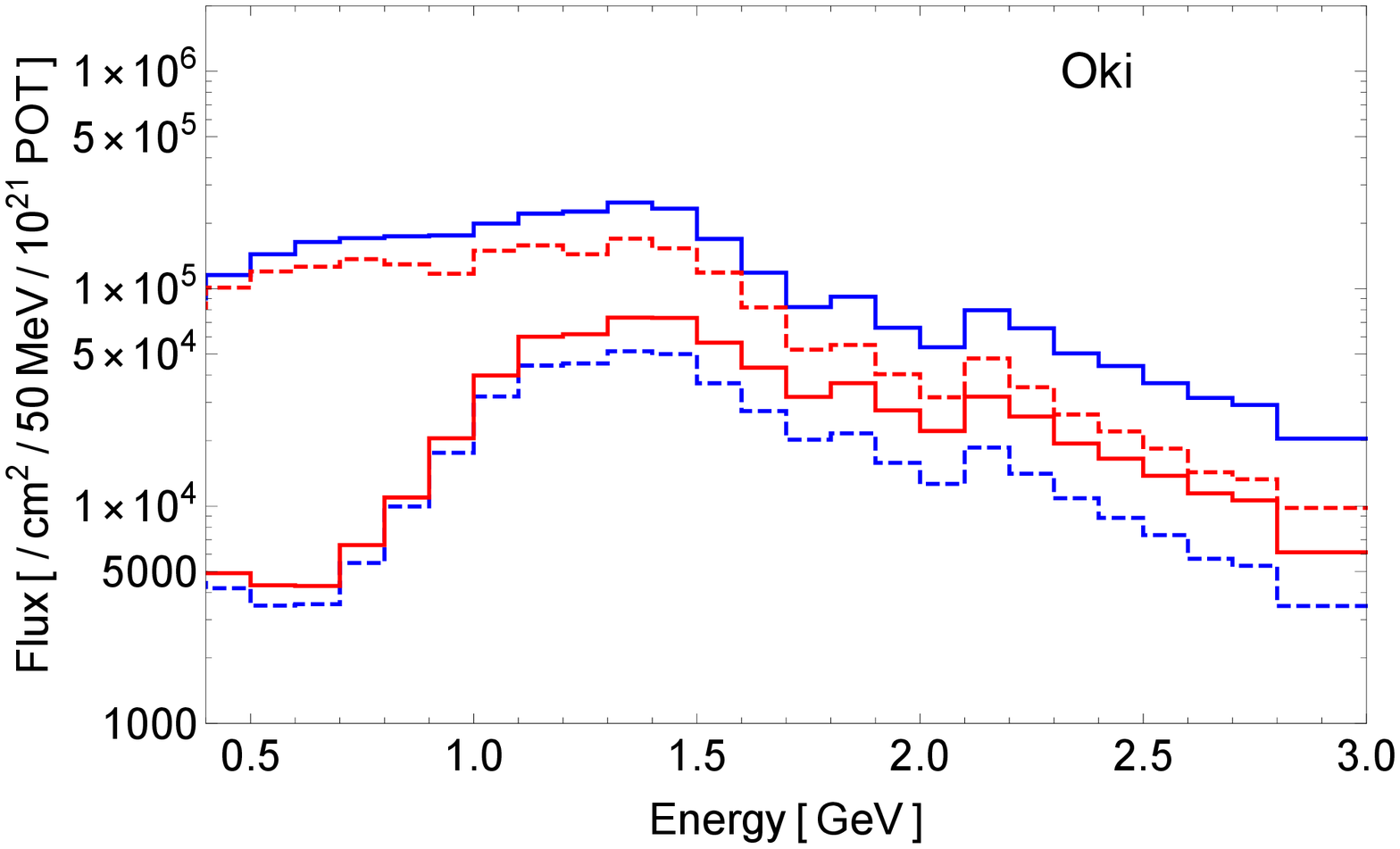}
    \includegraphics[width=80mm]{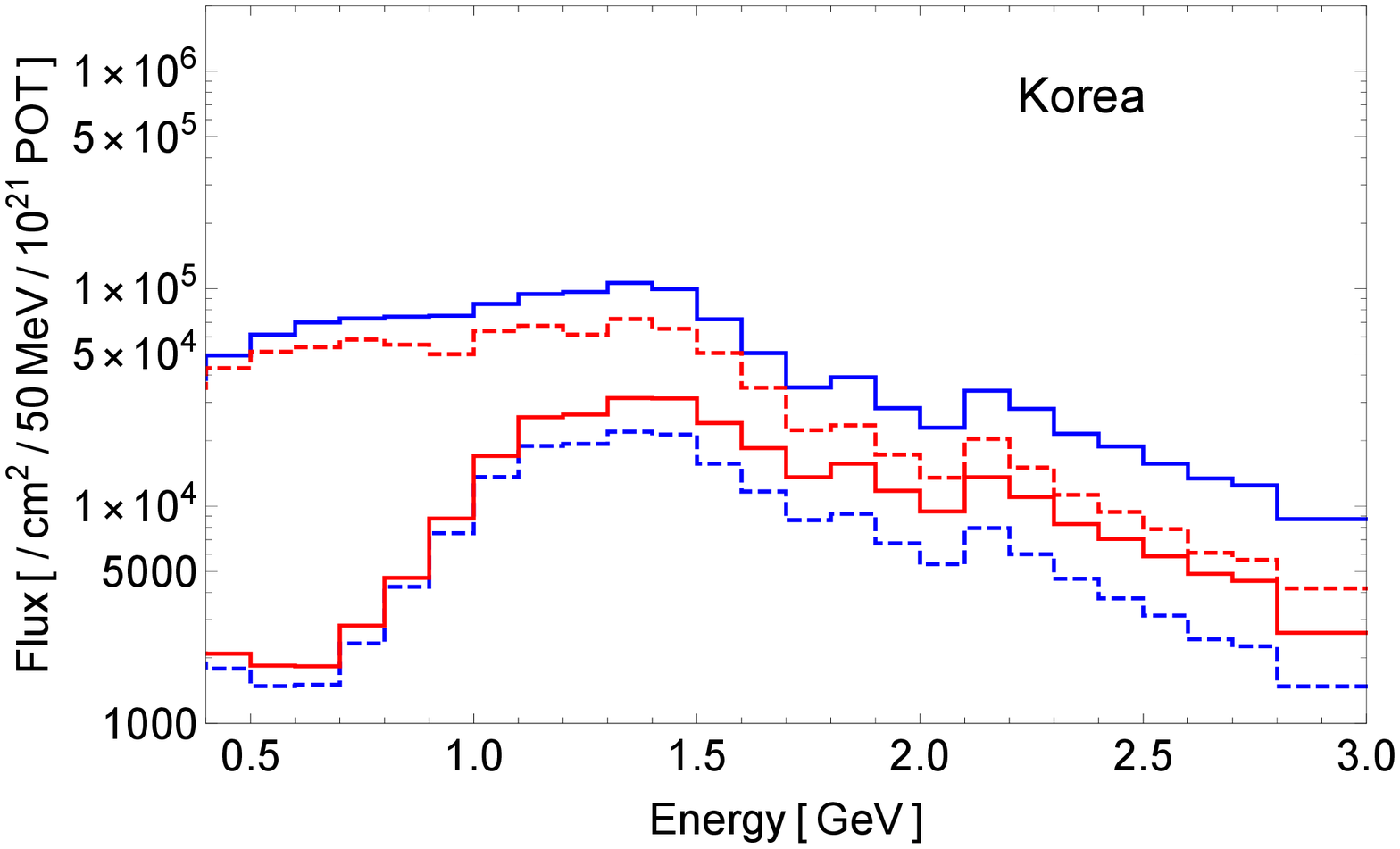}
    \caption{
    Flux of $\nu_\mu$ and $\bar{\nu}_\mu$ in neutrino-focusing and antineutrino-focusing beams detected at
     a water Cerenkov detector at Kamioka, Oki and in Korea if the neutrino oscillation were absent.
    The upper plot, the lower-left plot and the lower-right plot respectively correspond to Kamioka, Oki and Korea.
    In each plot, the blue lines correspond to a neutrino-focusing beam and the red lines correspond to an antineutrino-focusing beam.
    The solid lines denote $\nu_\mu$ flux and the dashed lines denote $\bar{\nu}_\mu$ flux.
    }
    \label{fluxfig}
  \end{center}
\end{figure}

In Figure~\ref{fluxfig}, we show the flux of $\nu_\mu$ and $\bar{\nu}_\mu$ in neutrino-focusing and antineutrino-focusing beams from J-PARC, detected at
 a water Cerenkov detector at Kamioka, Oki and in Korea if the neutrino oscillation were absent.
The baseline length and beam off-axis angle assumed are given in Table~\ref{exppara}.
In each plot, the blue lines correspond to a neutrino-focusing beam and the red lines correspond to an antineutrino-focusing beam.
The solid lines denote $\nu_\mu$ flux and the dashed lines denote $\bar{\nu}_\mu$ flux.

\section*{Appendix B}

\begin{figure}[H]
  \begin{center}
    \includegraphics[width=100mm]{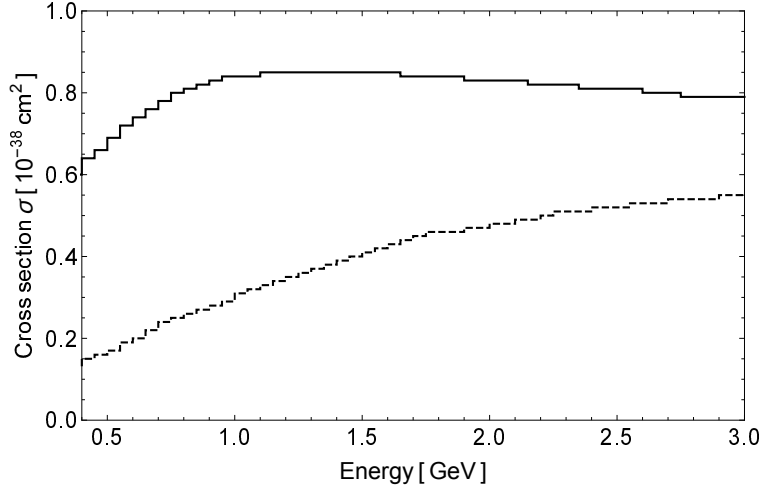}
    \caption{
    The cross section for charged current quasi-elastic scatterings
     $\nu_\ell n \to \ell^- p$ and $\bar{\nu}_\ell p \to \ell^+ n$ ($\ell=e, \, \mu$).
    The solid line corresponds to the cross section for $\nu_\ell$, and the dashed line to that for $\bar{\nu}_\ell$.
    }
    \label{crosssectionfig}
  \end{center}
\end{figure}

In Figure~\ref{crosssectionfig}, we show the cross sections for charged current quasi-elastic scatterings
 $\nu_\ell n \to \ell^- p$ and $\bar{\nu}_\ell p \to \ell^+ n$ ($\ell=e, \, \mu$).
The solid line corresponds to the cross section for $\nu_\ell$, and the dashed line to that for $\bar{\nu}_\ell$.

\section*{Appendix C}

\begin{figure}[H]
  \begin{center}
    \includegraphics[width=80mm]{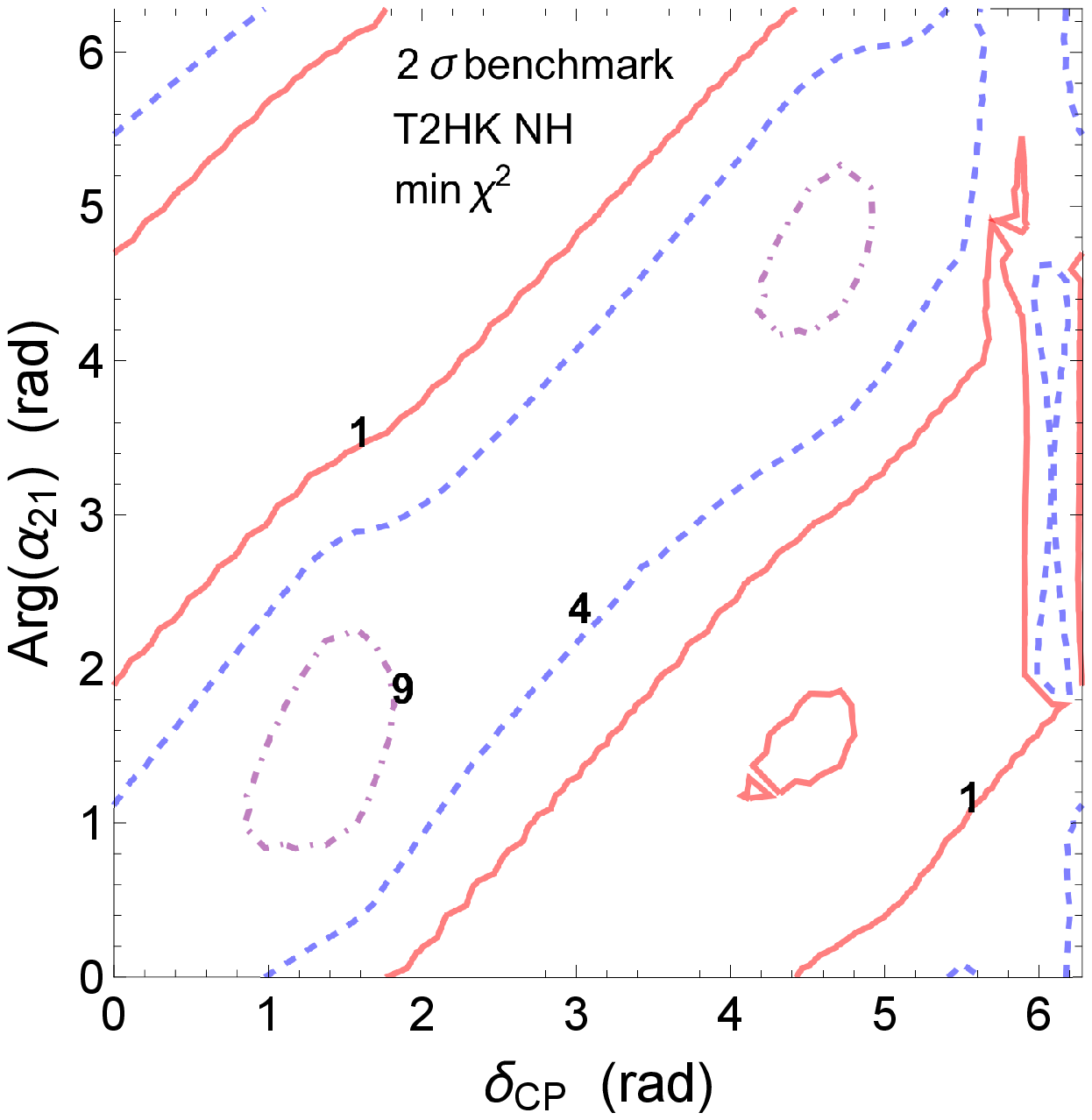} 
    \includegraphics[width=80mm]{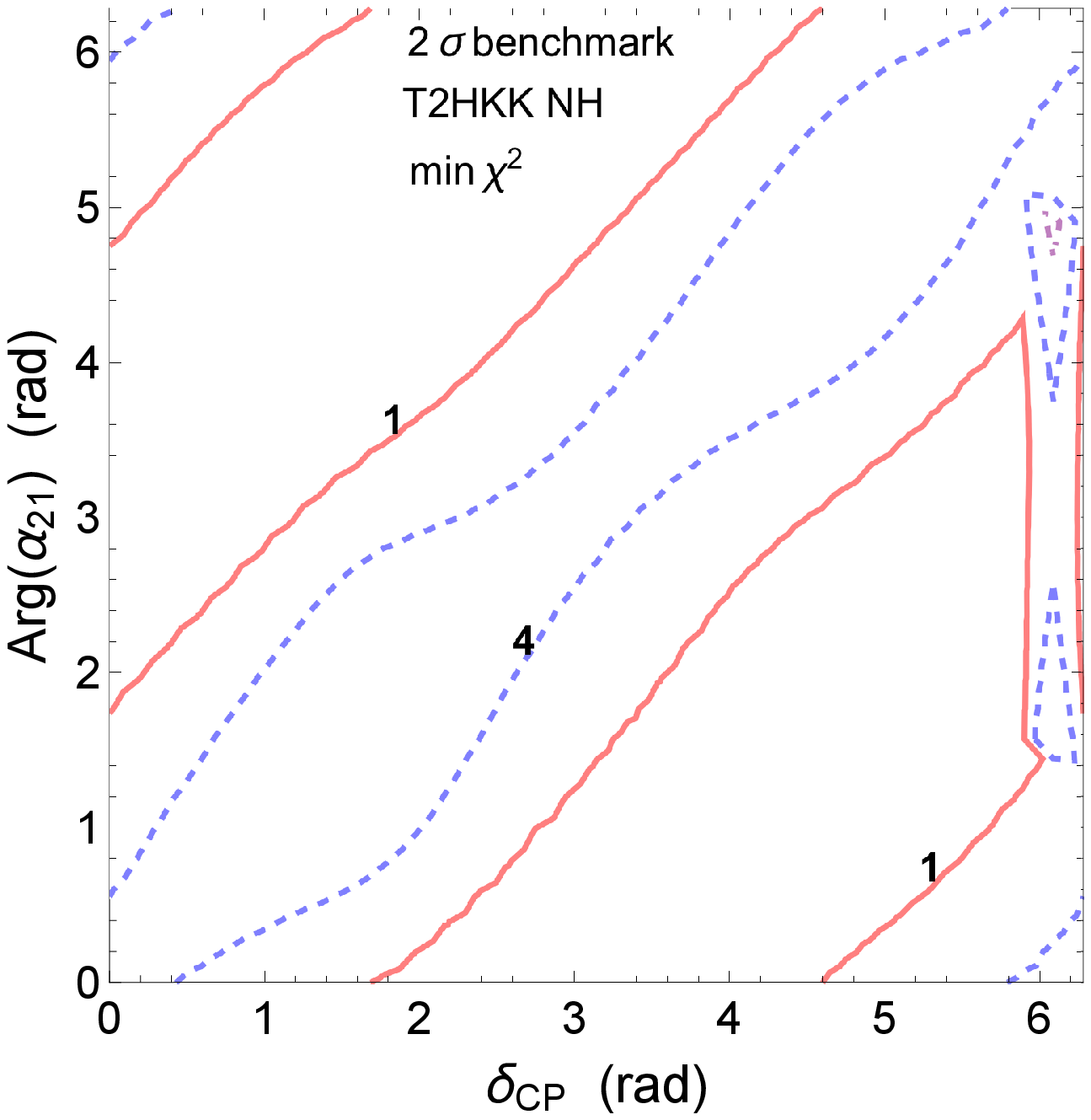}
    \\
    \includegraphics[width=80mm]{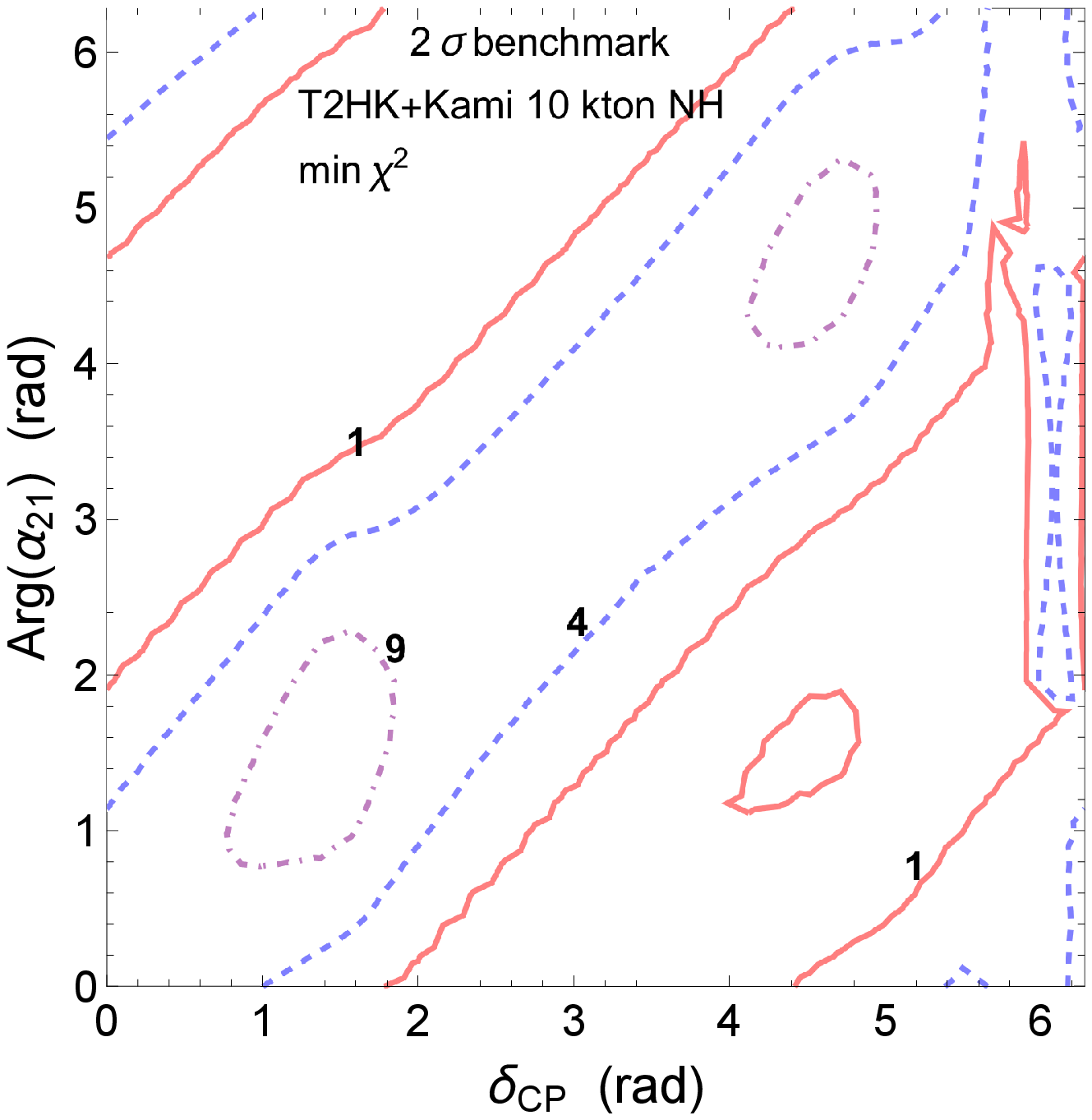}
    \includegraphics[width=80mm]{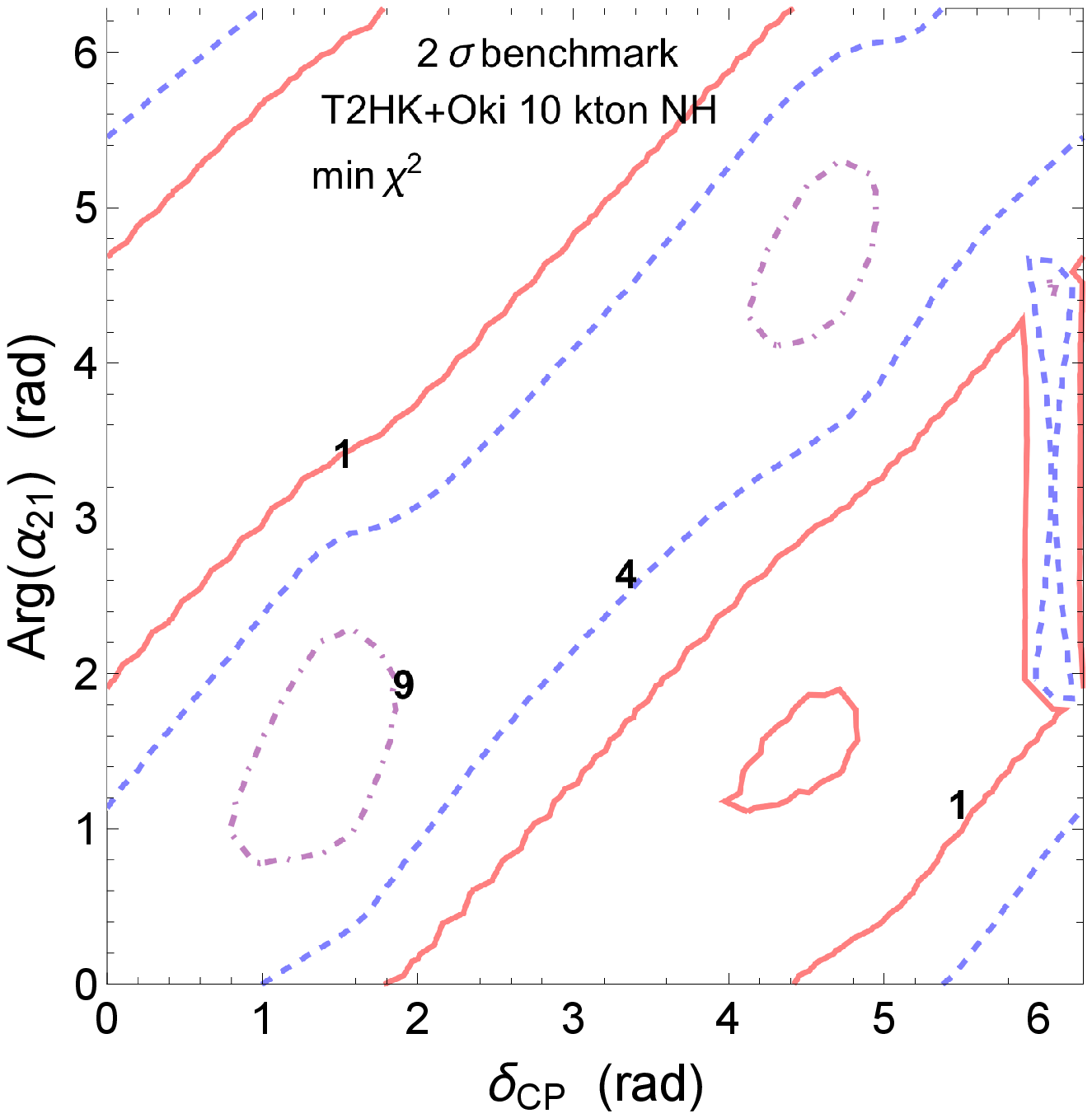}
    \caption{
    The minimum of $\chi^2(\Pi)$ Eq.~(\ref{chi22}), $\min \chi^2$,
     on the plane of $\delta_{CP}$ and Arg($\alpha_{21}$).
    The benchmark with Eq.~(\ref{altnu}) is assumed, and the true mass hierarchy is normal.
    The upper-left, upper-right and lower-right subplots correspond to the T2HK, the T2HKK, 
     and the plan of the T2HK plus a 10~kton water Cerenkov detector at Oki, respectively.
    For comparative study, we show in the lower-left a subplot for a plan of the T2HK plus a 10~kton water Cerenkov detector at Kamioka.
    $\min \chi^2=1,\,4,\,9$ on the red solid, blue dashed, and purple dot-dashed contours, respectively.
    }
    \label{nhch2s}
  \end{center}
\end{figure}

\begin{figure}[H]
  \begin{center}
    \includegraphics[width=80mm]{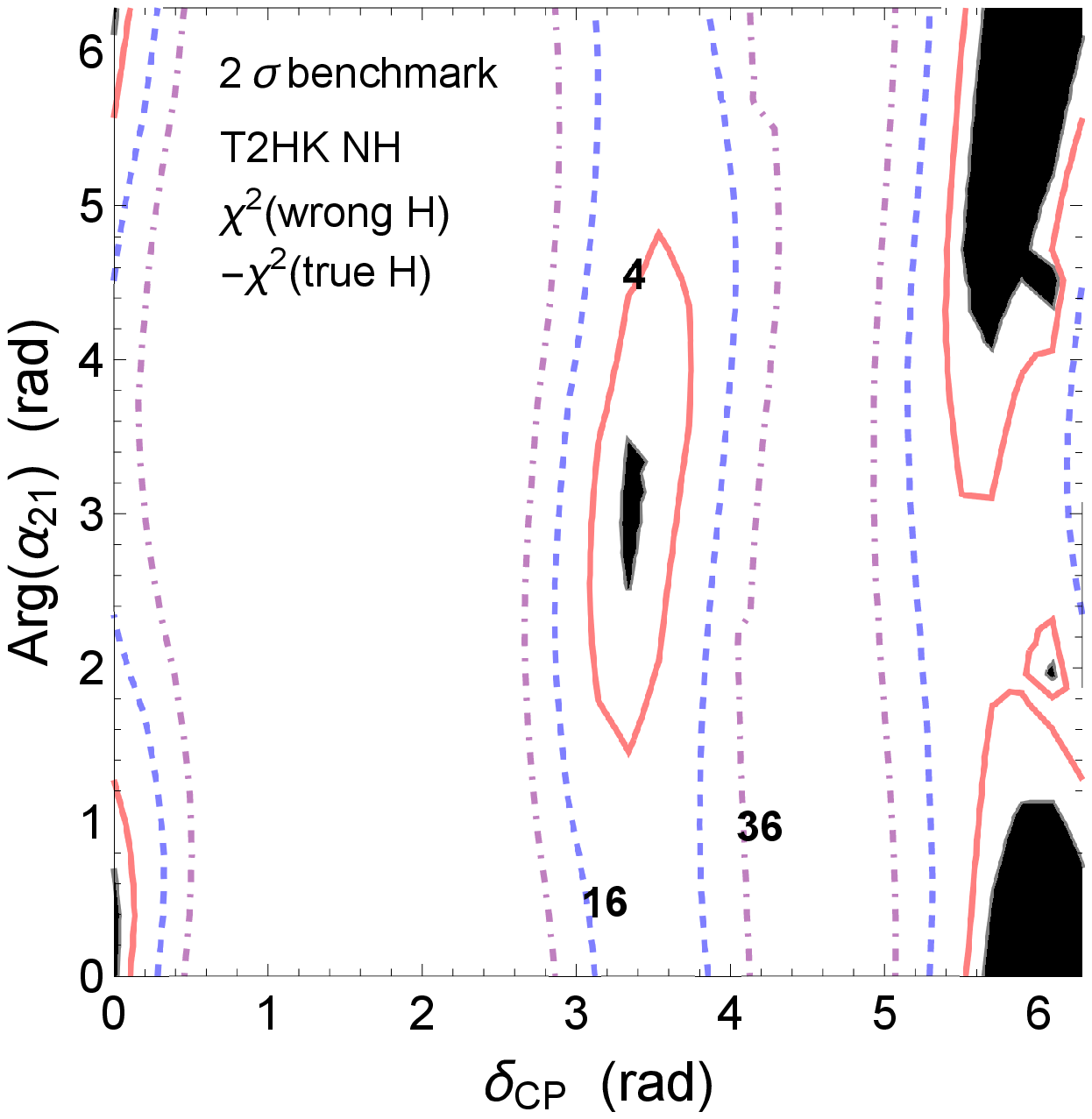} 
    \includegraphics[width=80mm]{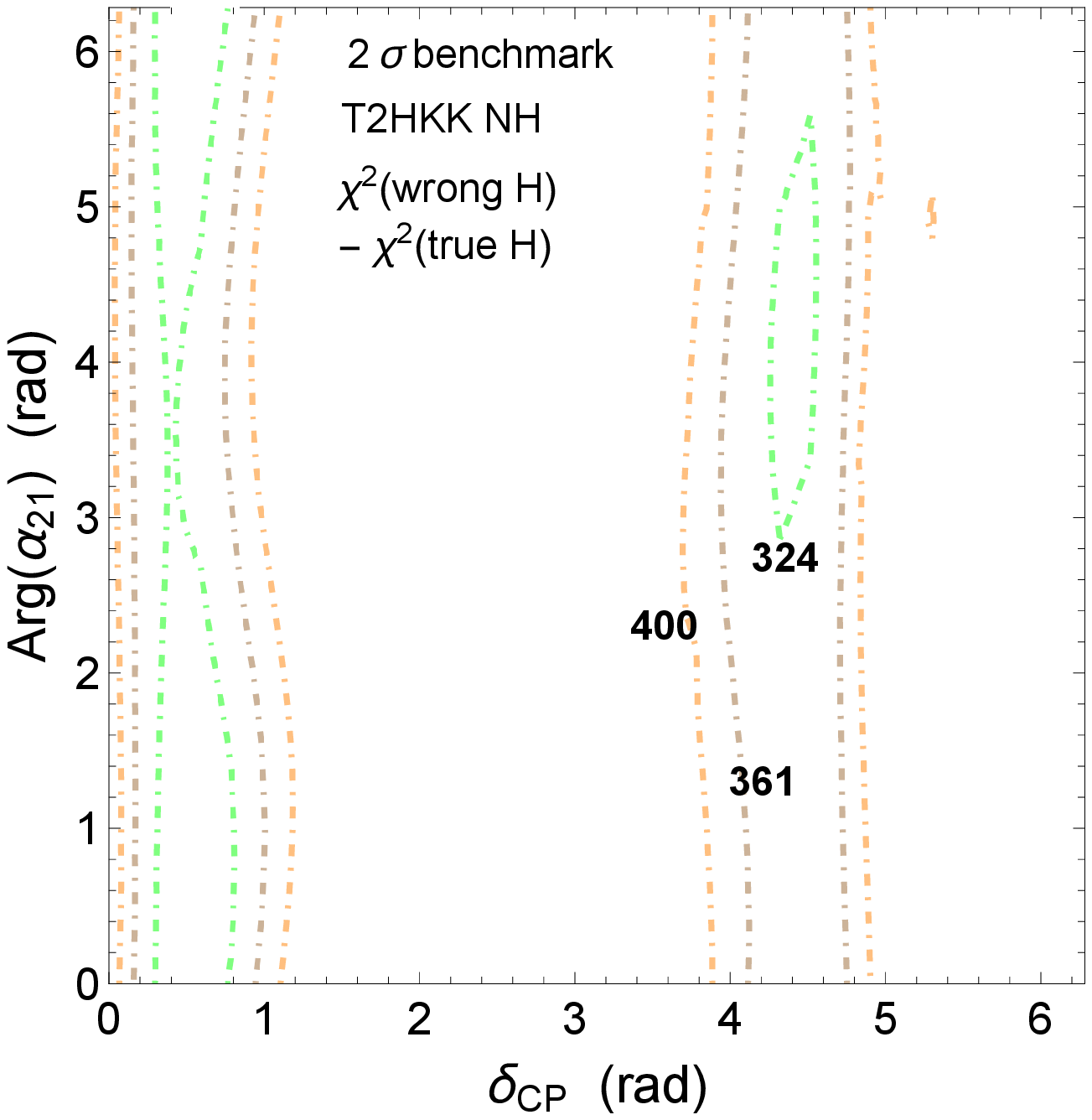}
    \\
    \includegraphics[width=80mm]{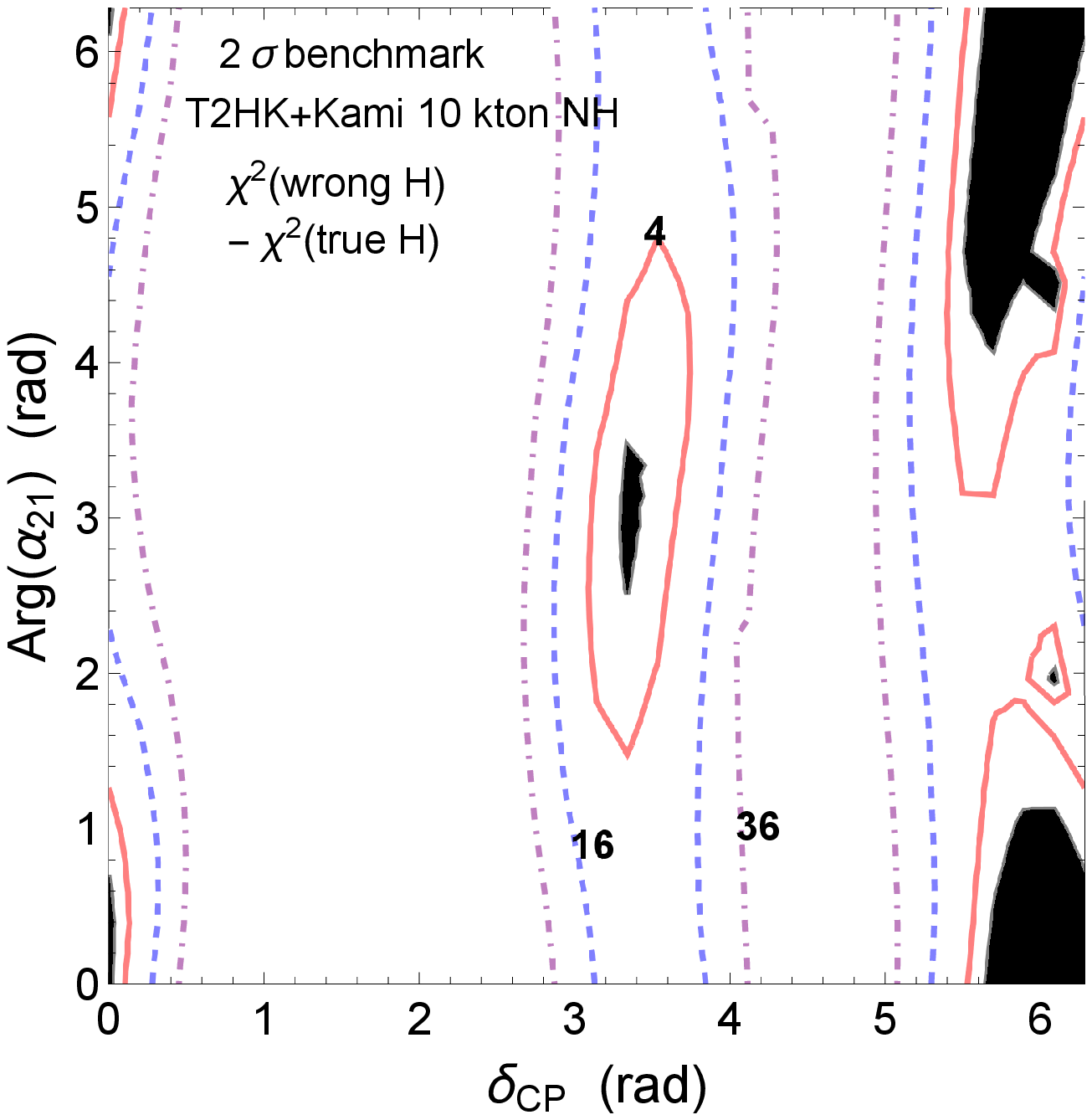}
    \includegraphics[width=80mm]{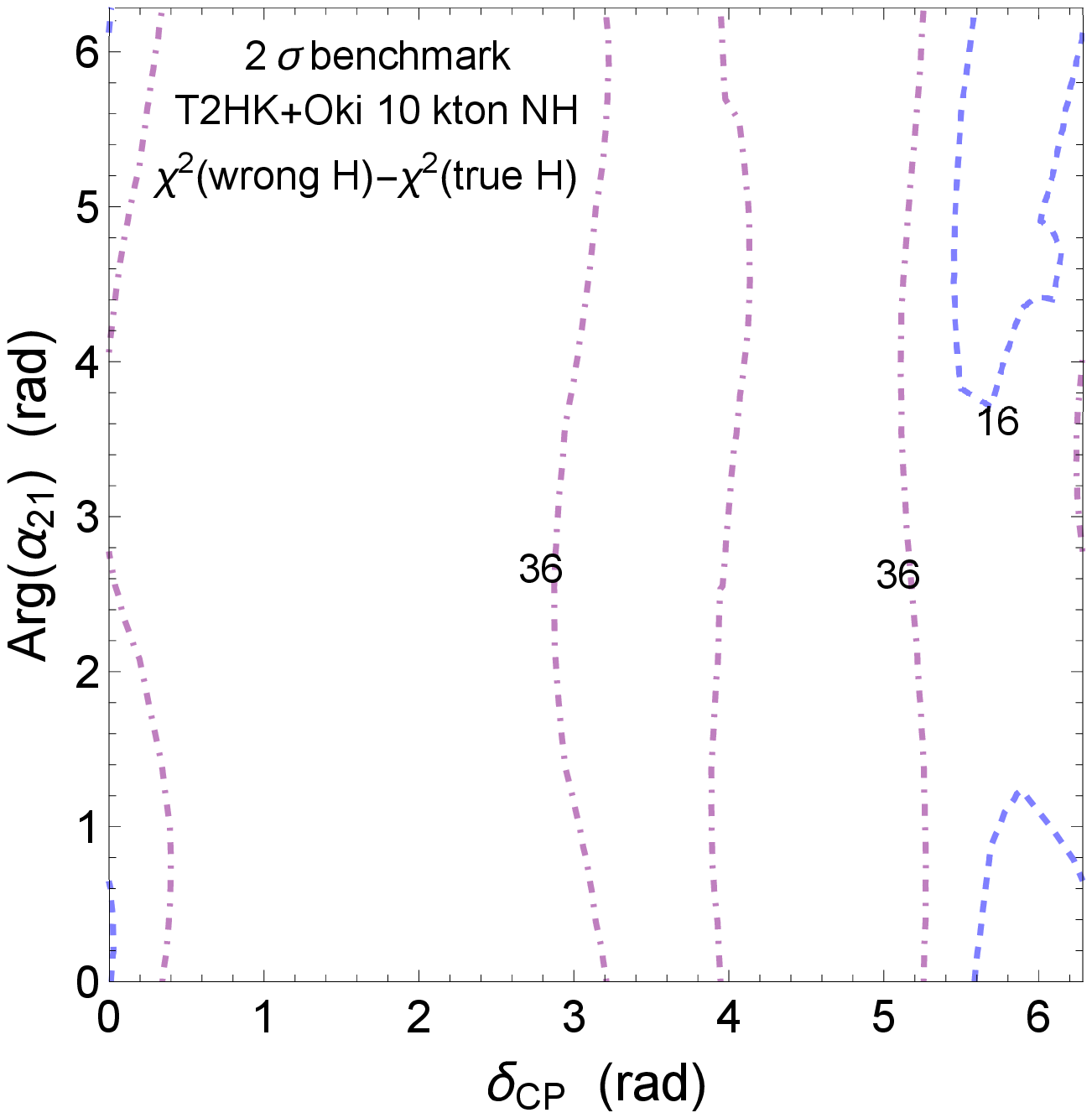}
    \caption{
    The difference between the minima (with respect to $\delta_{CP}$ only) of $\chi^2(\Pi)$ for the wrong and true mass hierarchy,
     $\{\min\chi^2({\rm wrong \ H})-\min\chi^2({\rm true \ H})\}$.
    The benchmark with Eq.~(\ref{altnu}) is assumed, and the true mass hierarchy is normal.
    The upper-left, upper-right and lower-right subplots correspond to the T2HK, the T2HKK, 
     and the plan of the T2HK plus a 10~kton water Cerenkov detector at Oki, respectively.
    For comparative study, we show in the lower-left a subplot for a plan of the T2HK plus a 10~kton water Cerenkov detector at Kamioka.
    $\{\min\chi^2({\rm wrong \ H})-\min\chi^2({\rm true \ H})\}=4,\,16,\,36$ on the red solid, blue dashed, and purple dot-dashed contours, respectively, and 
     $\{\min\chi^2({\rm wrong \ H})-\min\chi^2({\rm true \ H})\}=18^2,\,19^2,\,20^2$ on the green, brown, and orange dot-dashed contours, respectively.   
     The black-filled regions are where the wrong mass hierarchy is favored over the true mass hierarchy, \textit{i.e.},
 $\min\chi^2({\rm wrong \ H})-\min\chi^2({\rm true \ H})<0$.
     }
    \label{nhmass2s}
  \end{center}
\end{figure}

\begin{figure}[H]
  \begin{center}
    \includegraphics[width=80mm]{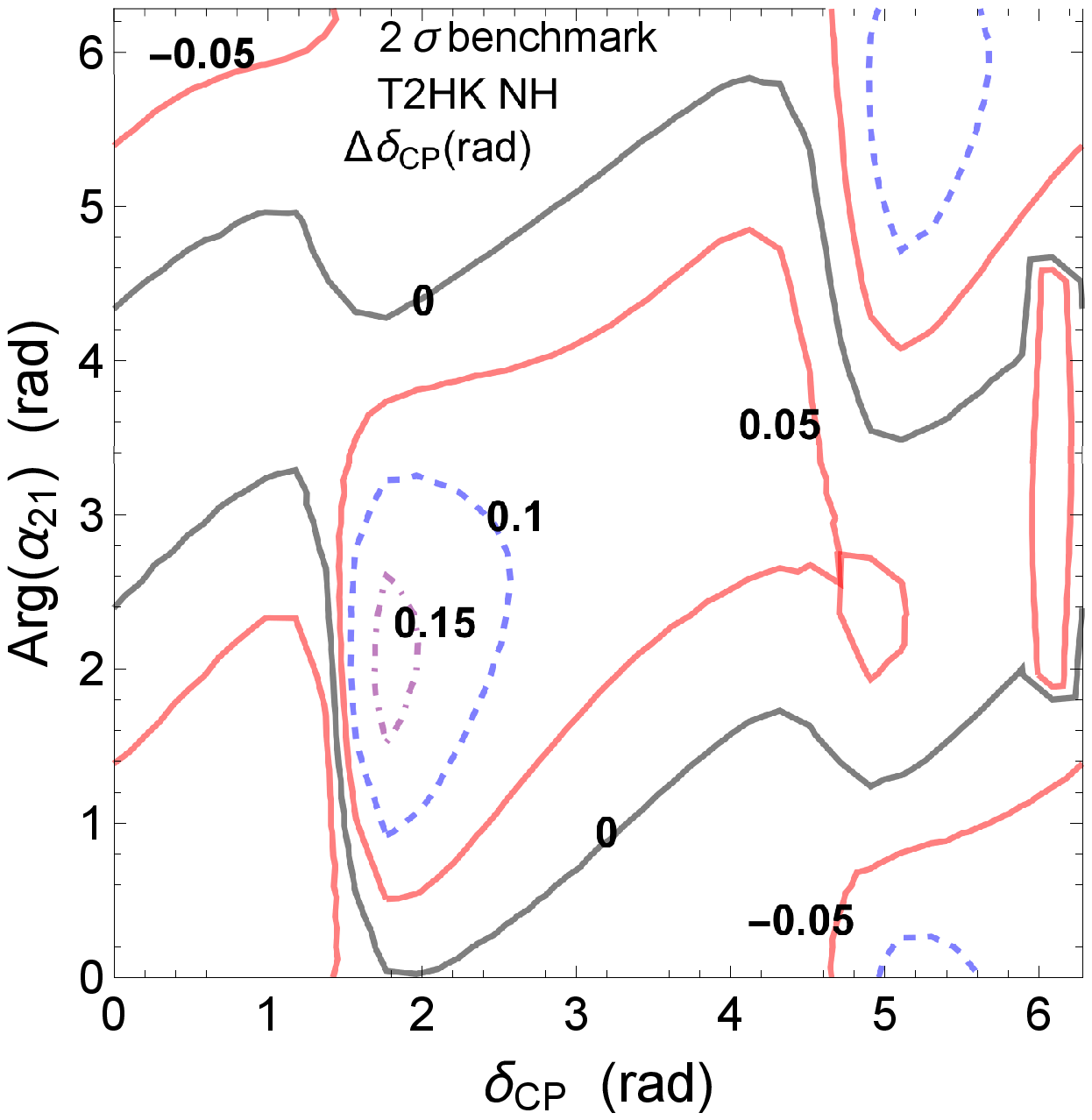}    
    \includegraphics[width=80mm]{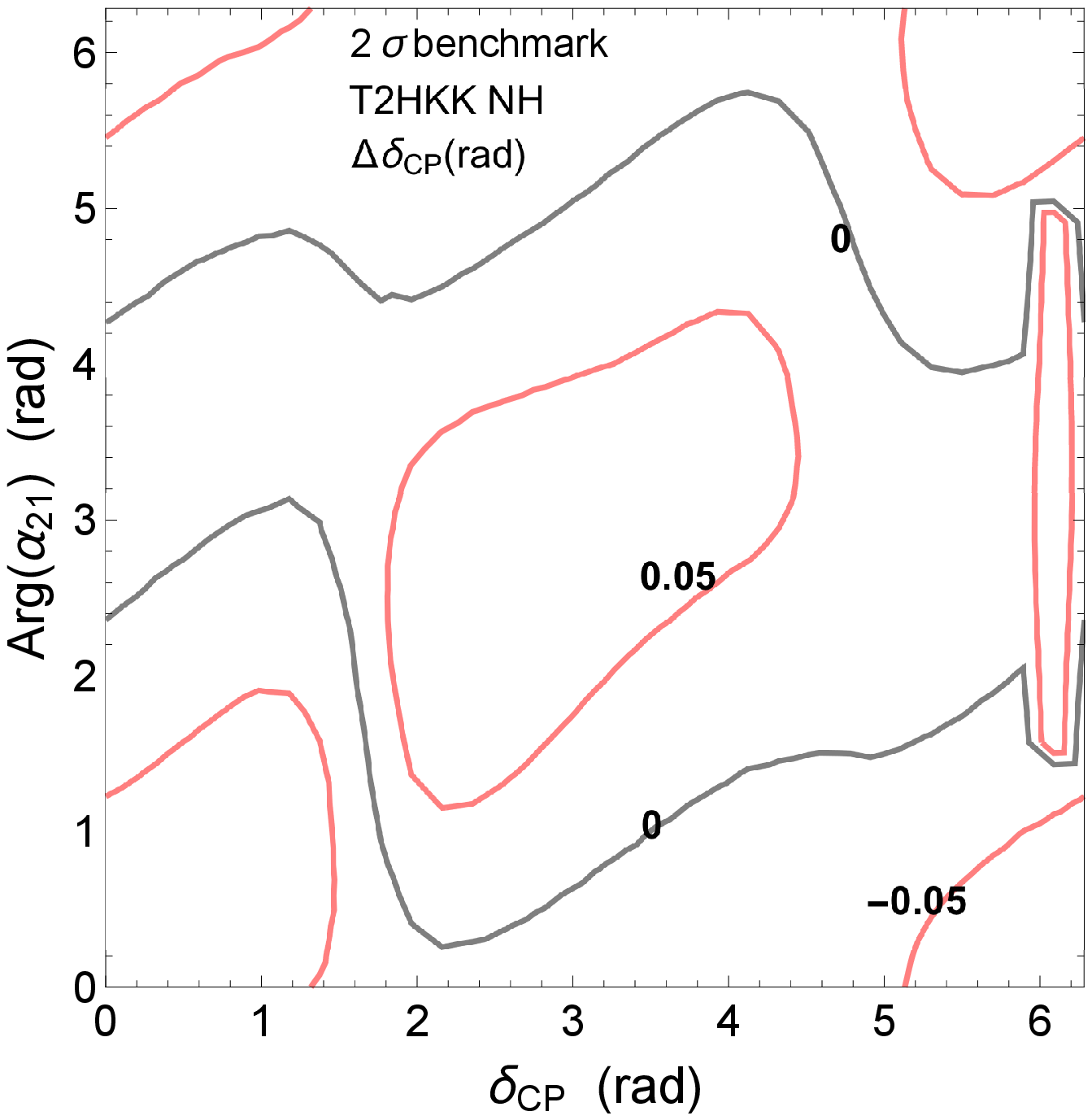}
    \\
    \includegraphics[width=80mm]{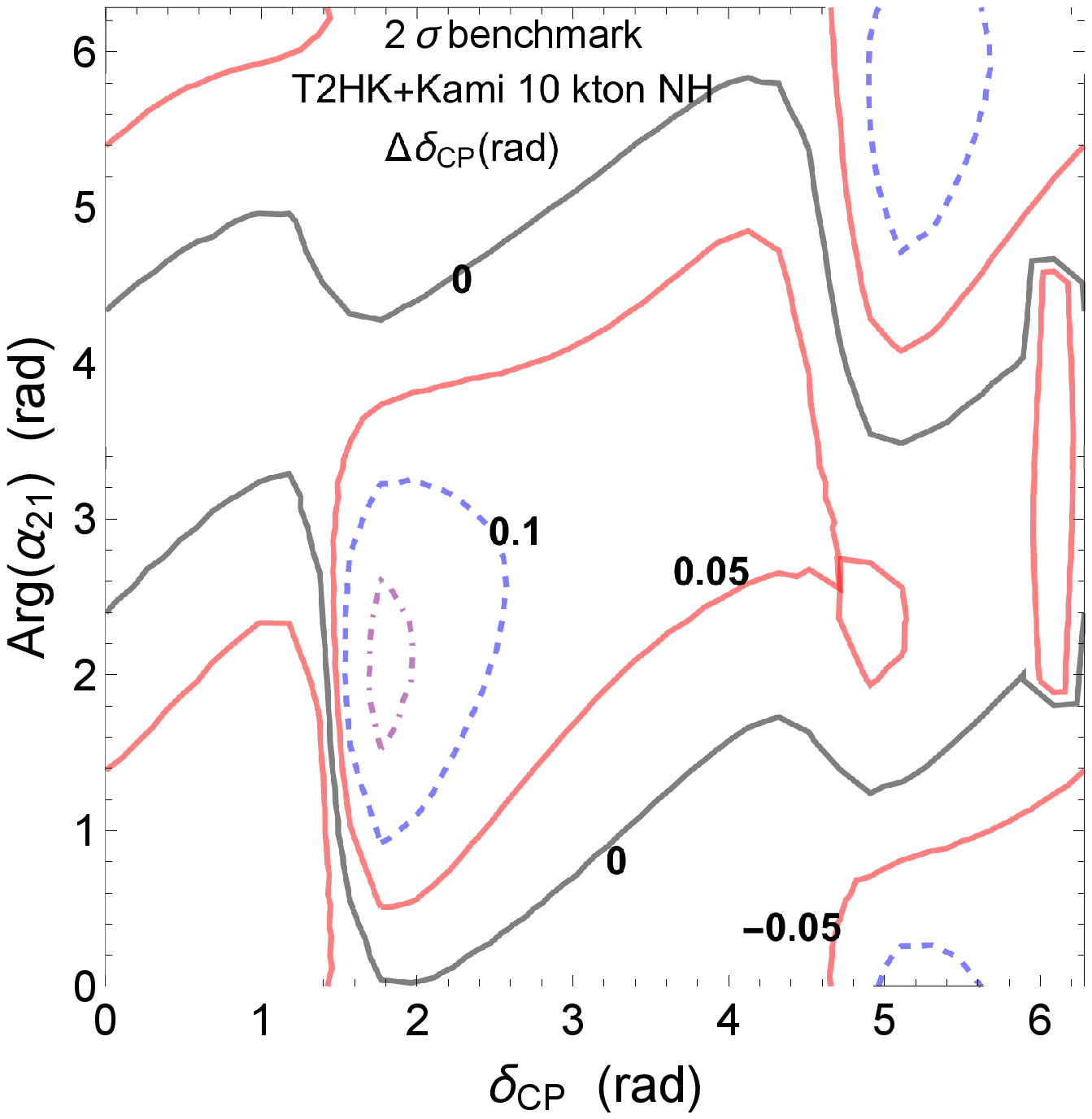}
    \includegraphics[width=80mm]{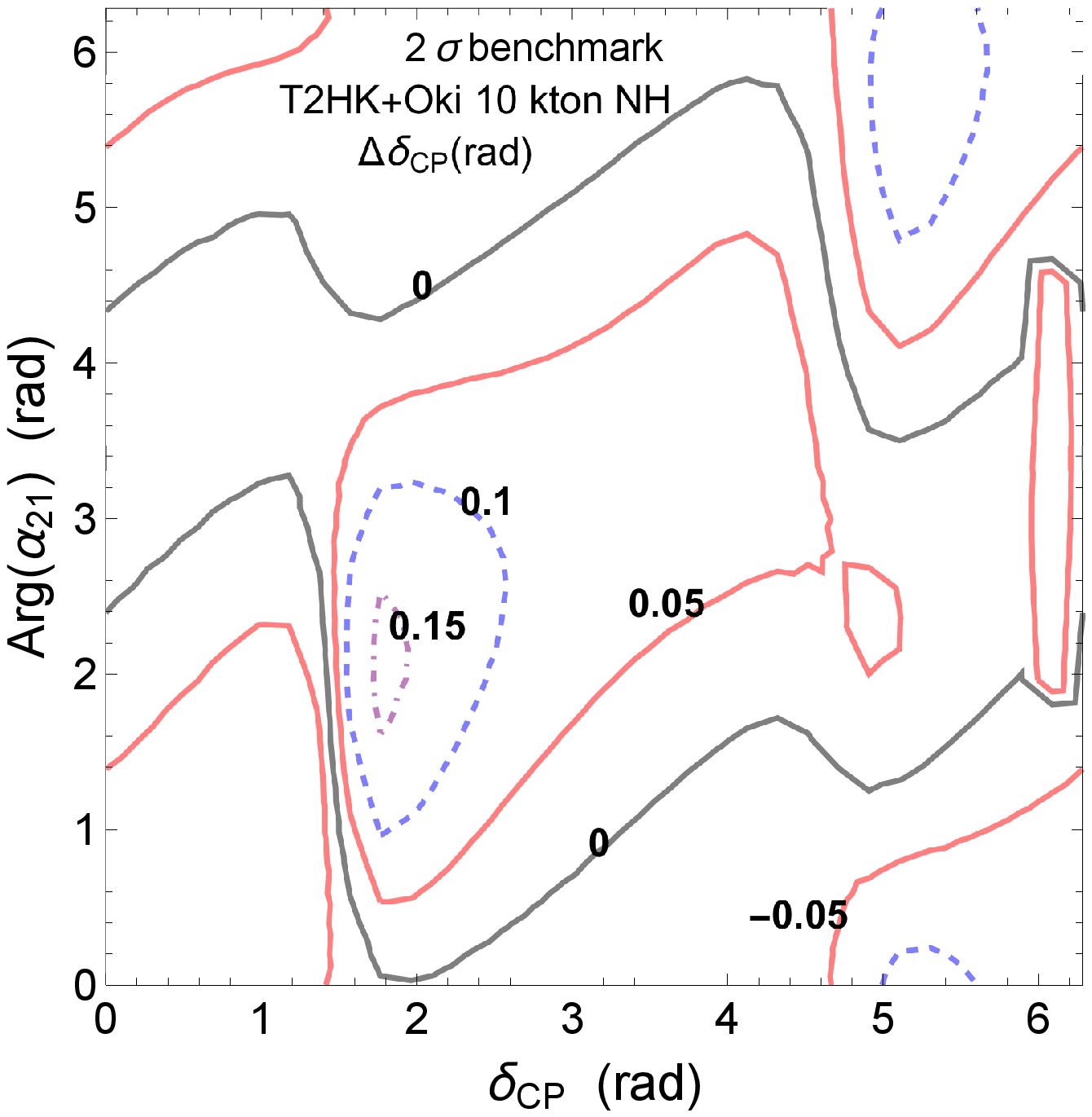}
    \caption{
       The difference between the true value of $\delta_{CP}$, and the value that minimizes $\chi^2(\Pi)$ of Eq.~(\ref{chi22})
    with the true mass hierarchy.
    The benchmark with Eq.~(\ref{altnu}) is assumed, and the true mass hierarchy is normal.
    The upper-left, upper-right and lower-right subplots correspond to the T2HK, the T2HKK, 
     and the plan of the T2HK plus a 10~kton water Cerenkov detector at Oki, respectively.
    For comparative study, we show in the lower-left a subplot for a plan of the T2HK plus a 10~kton water Cerenkov detector at Kamioka.
   $\vert\Delta\delta_{CP}\vert=0, \ 0.05{\rm rad}, \ 0.1{\rm rad}$, and $0.15{\rm rad}$ on the black solid, red sold, blue dashed, and purple dot-dashed contours, respectively.    
     }
    \label{nhdch2s}
  \end{center}
\end{figure}

\begin{figure}[H]
  \begin{center}
    \includegraphics[width=80mm]{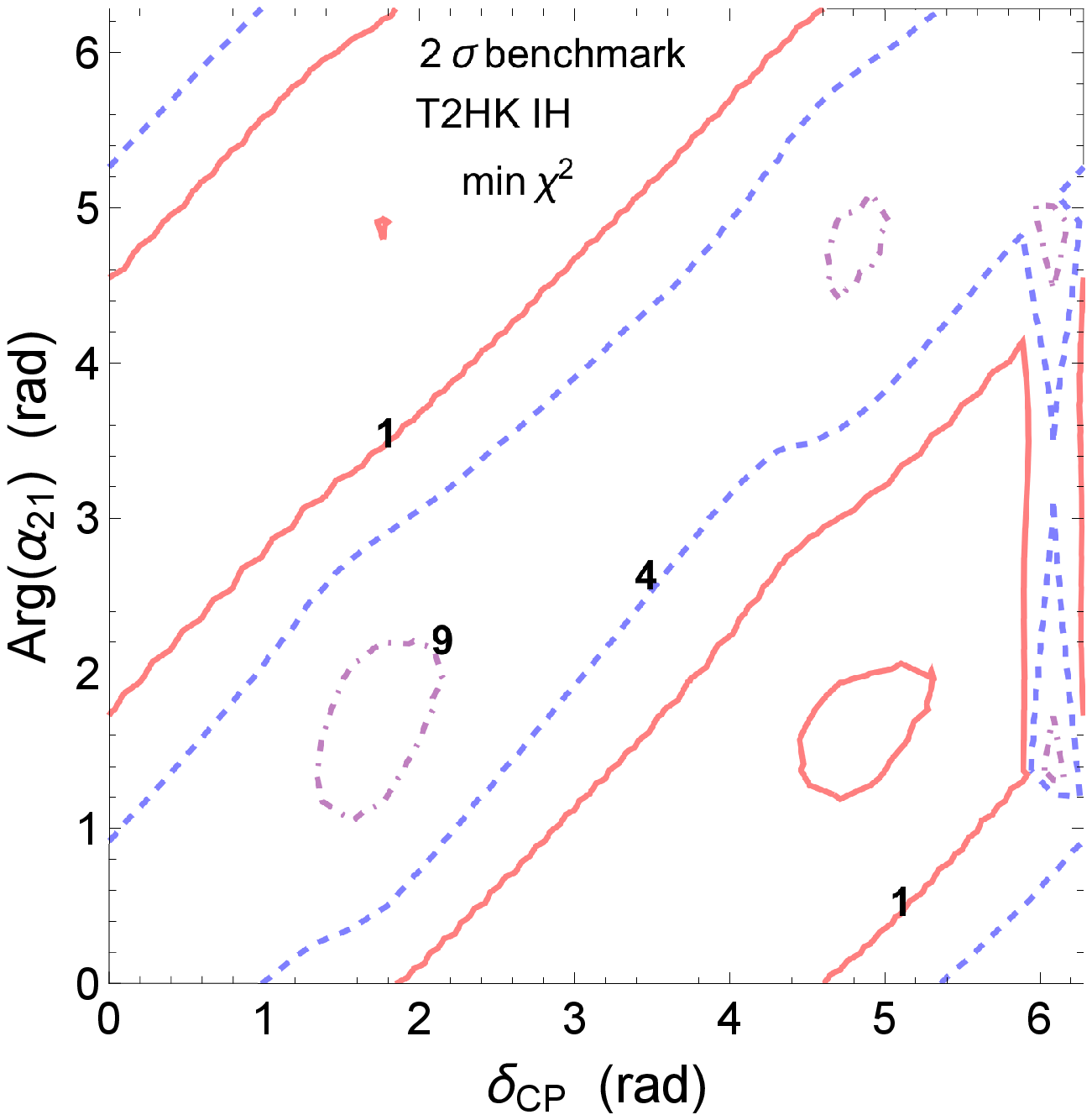}    
    \includegraphics[width=80mm]{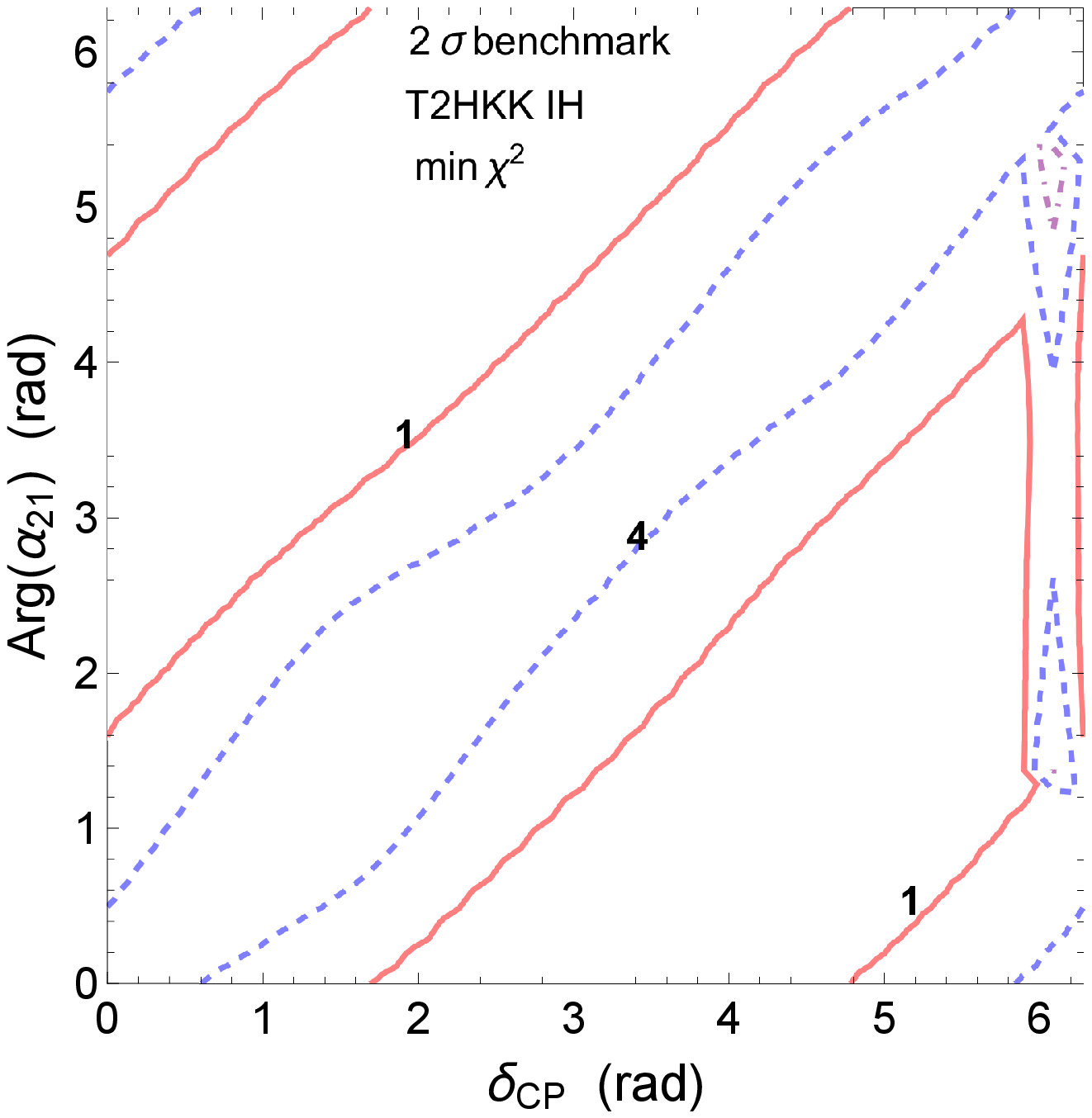}
    \\
    \includegraphics[width=80mm]{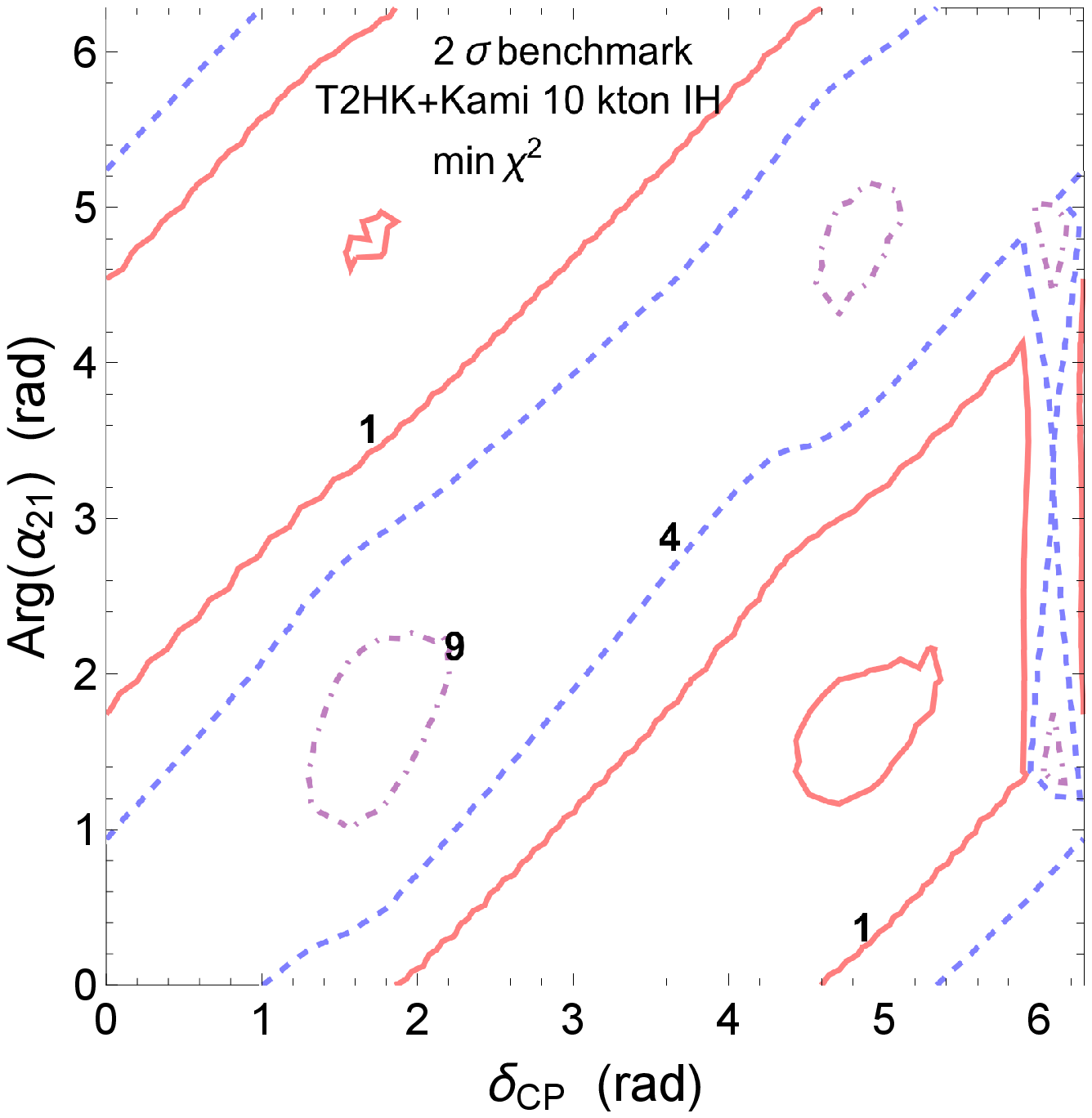}
    \includegraphics[width=80mm]{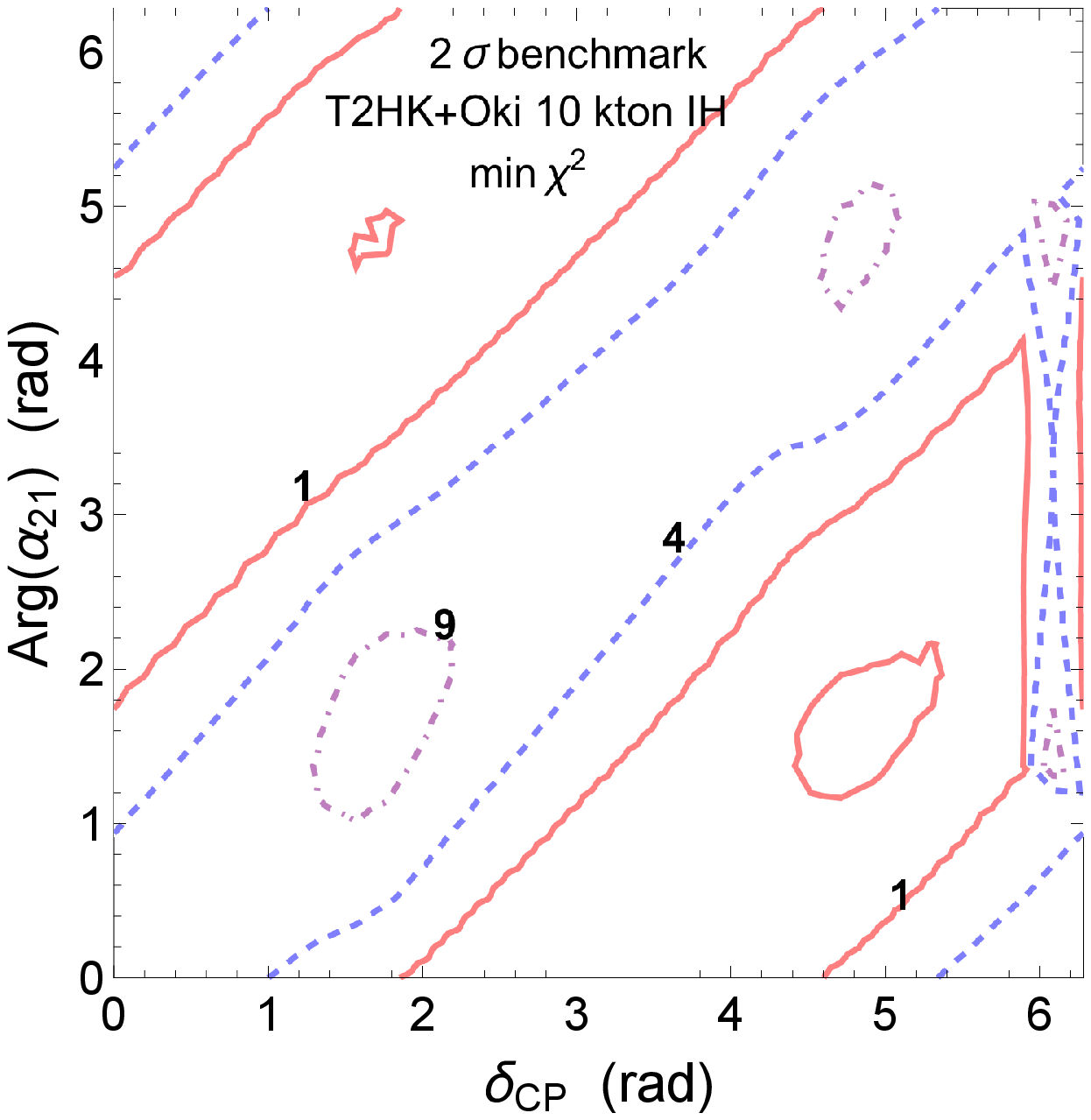}
    \caption{
    The minimum of $\chi^2(\Pi)$ Eq.~(\ref{chi22}), $\min \chi^2$.
    The benchmark with Eq.~(\ref{altnu}) is assumed, and the true mass hierarchy is inverted.
    The upper-left, upper-right and lower-right subplots correspond to the T2HK, the T2HKK, 
     and the plan of the T2HK plus a 10~kton water Cerenkov detector at Oki, respectively.
    For comparative study, we show in the lower-left a subplot for a plan of the T2HK plus a 10~kton water Cerenkov detector at Kamioka.
    $\min \chi^2=1,\,4,\,9$ on the red solid, blue dashed, and purple dot-dashed contours, respectively.
    }
    \label{ihch2s}
  \end{center}
\end{figure}

\begin{figure}[H]
  \begin{center}    
    \includegraphics[width=80mm]{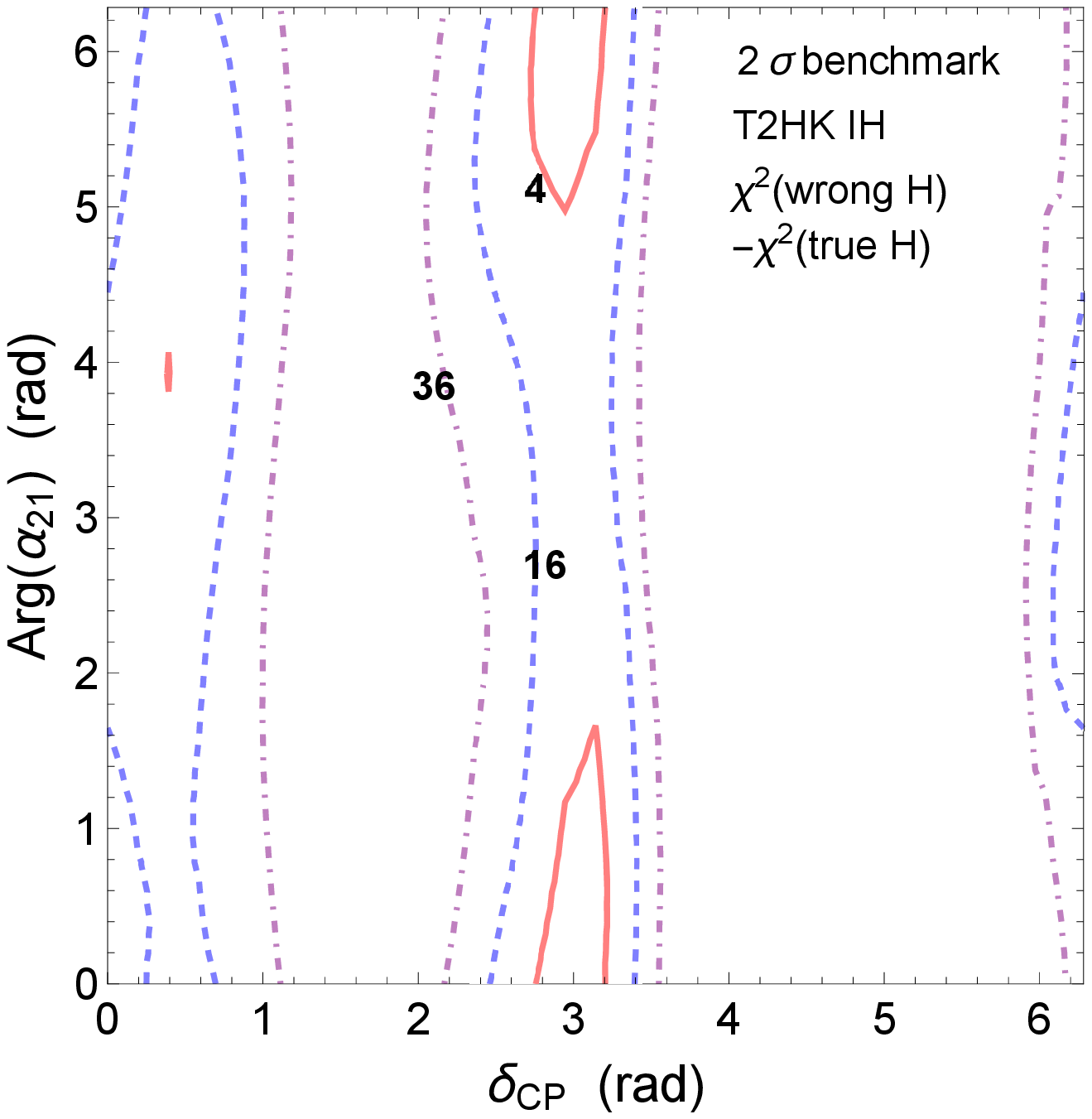}
    \includegraphics[width=80mm]{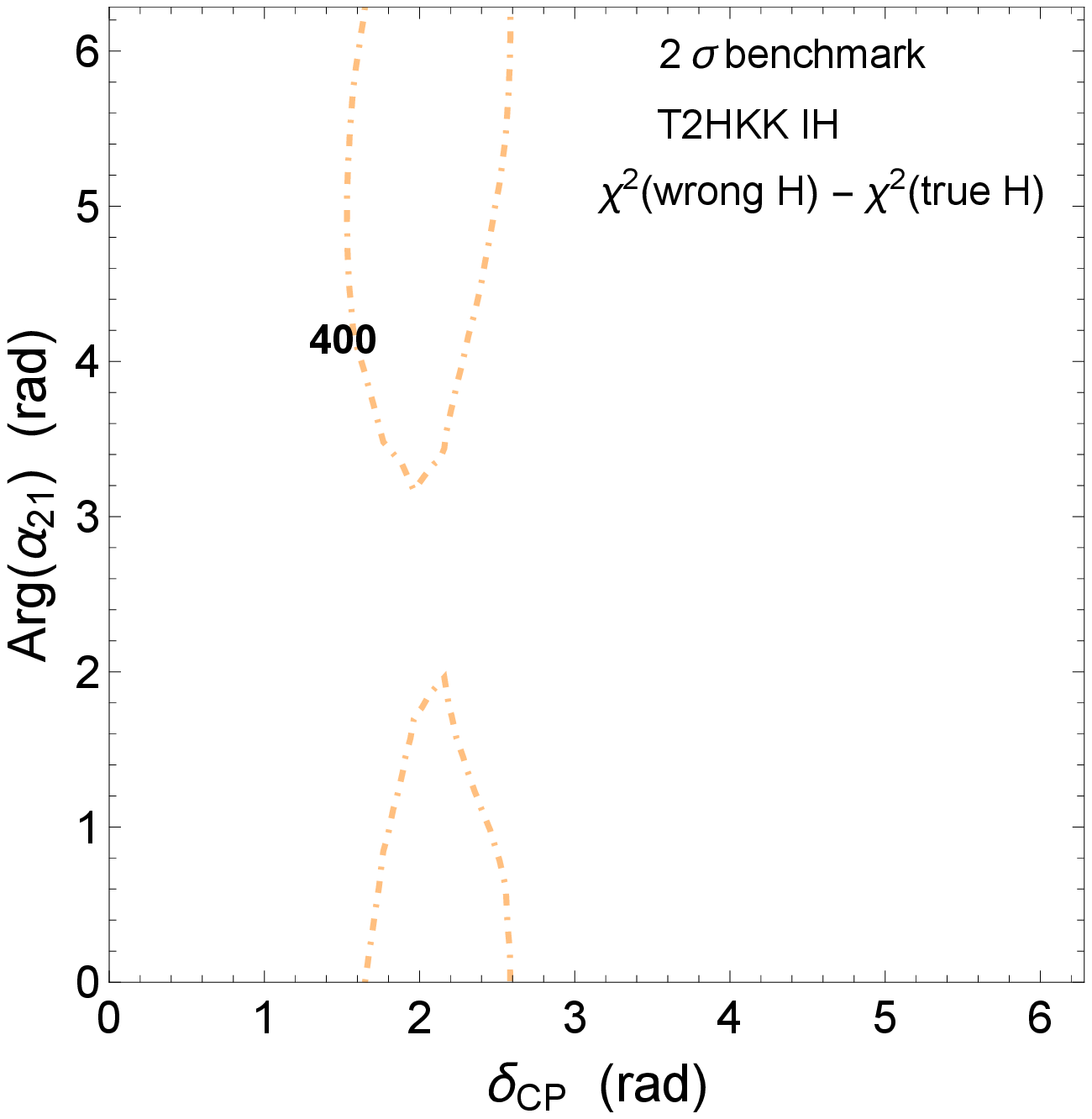}
    \\
    \includegraphics[width=80mm]{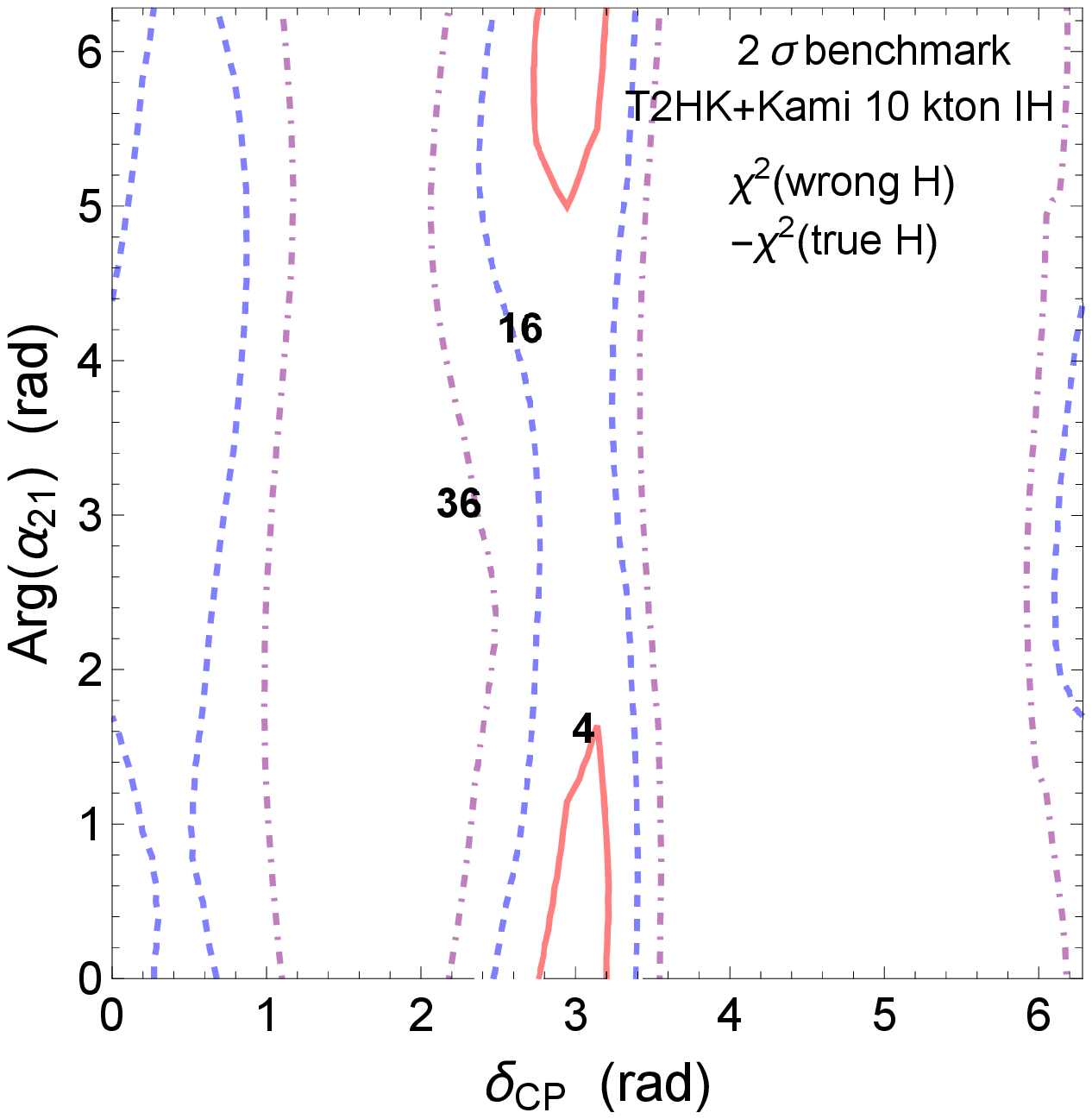}
    \includegraphics[width=80mm]{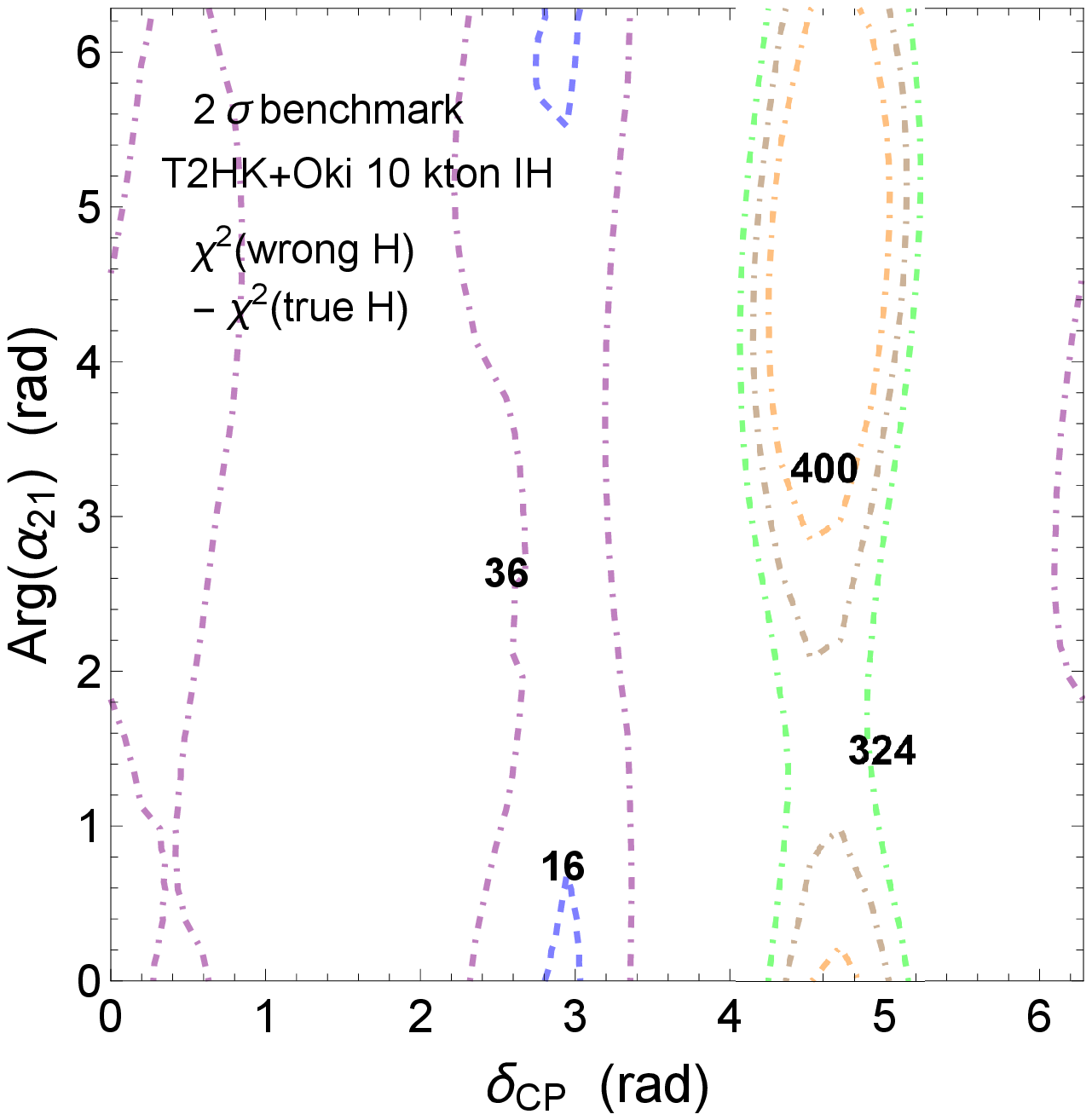}
    \caption{
    The difference between the minima (with respect to $\delta_{CP}$ only) of $\chi^2(\Pi)$ for the wrong and true mass hierarchy,
     $\{\min\chi^2({\rm wrong \ H})-\min\chi^2({\rm true \ H})\}$.
    The benchmark with Eq.~(\ref{altnu}) is assumed, and the true mass hierarchy is inverted.
    The upper-left, upper-right and lower-right subplots correspond to the T2HK, the T2HKK, 
     and the plan of the T2HK plus a 10~kton water Cerenkov detector at Oki, respectively.
    For comparative study, we show in the lower-left a subplot for a plan of the T2HK plus a 10~kton water Cerenkov detector at Kamioka.
    $\{\min\chi^2({\rm wrong \ H})-\min\chi^2({\rm true \ H})\}=4,\,16,\,36$ on the red solid, blue dashed, and purple dot-dashed contours, respectively, and 
     $\{\min\chi^2({\rm wrong \ H})-\min\chi^2({\rm true \ H})\}=18^2,\,19^2,\,20^2$ on the green, brown, and orange dot-dashed contours, respectively.   
     }
    \label{ihmass2s}
  \end{center}
\end{figure}

\begin{figure}[H]
  \begin{center}
    \includegraphics[width=80mm]{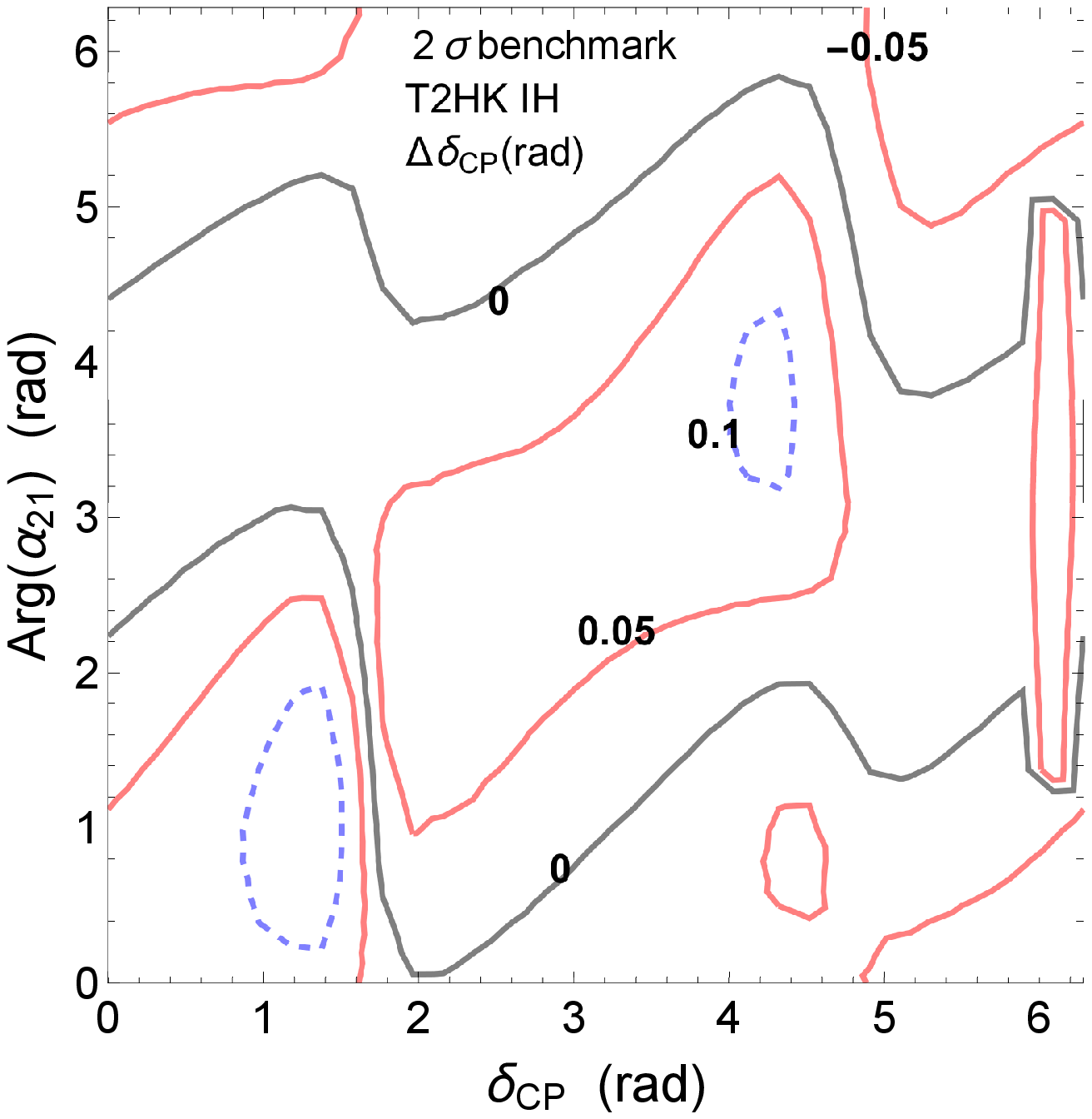}   
    \includegraphics[width=80mm]{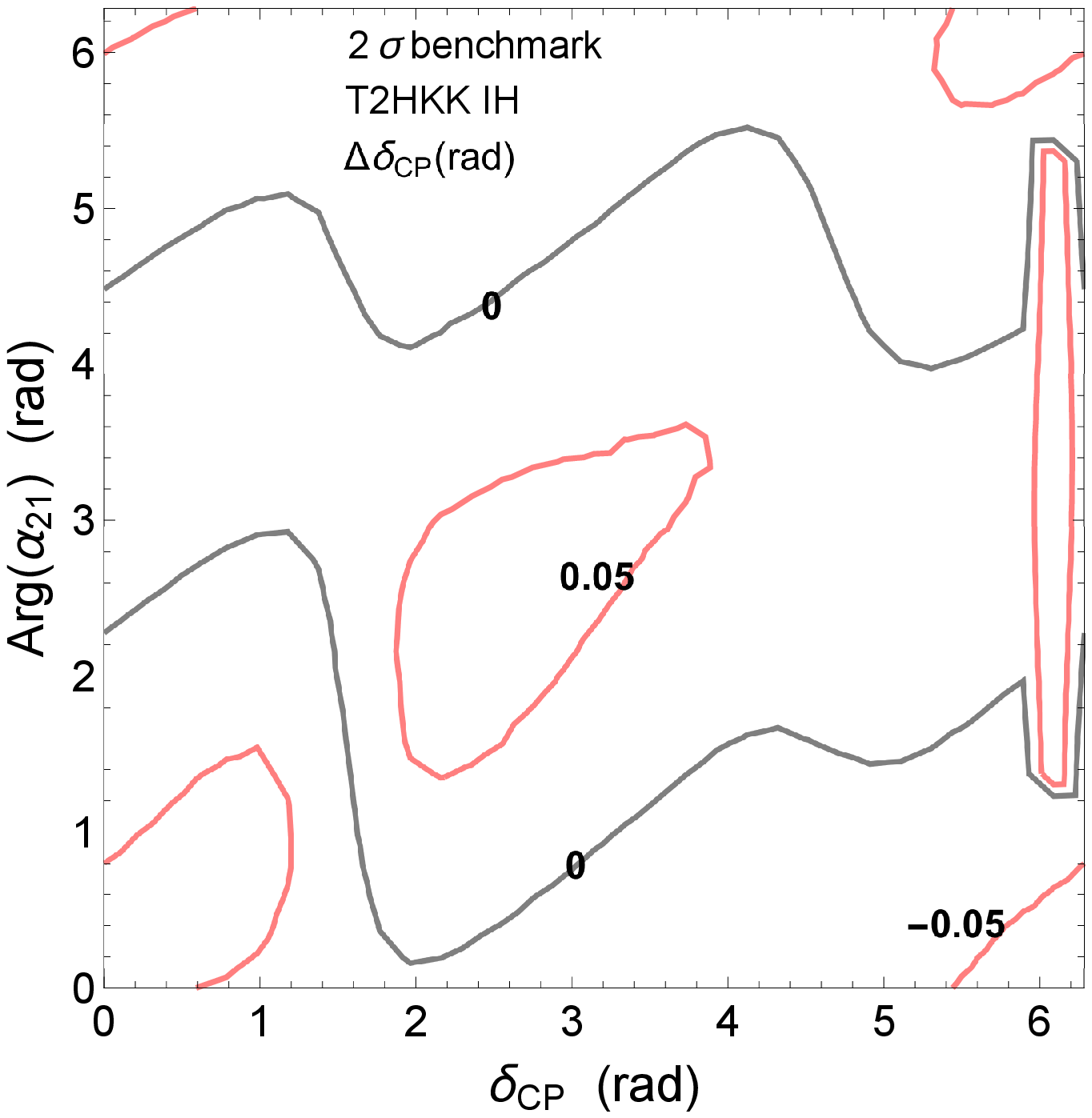}
    \\
    \includegraphics[width=80mm]{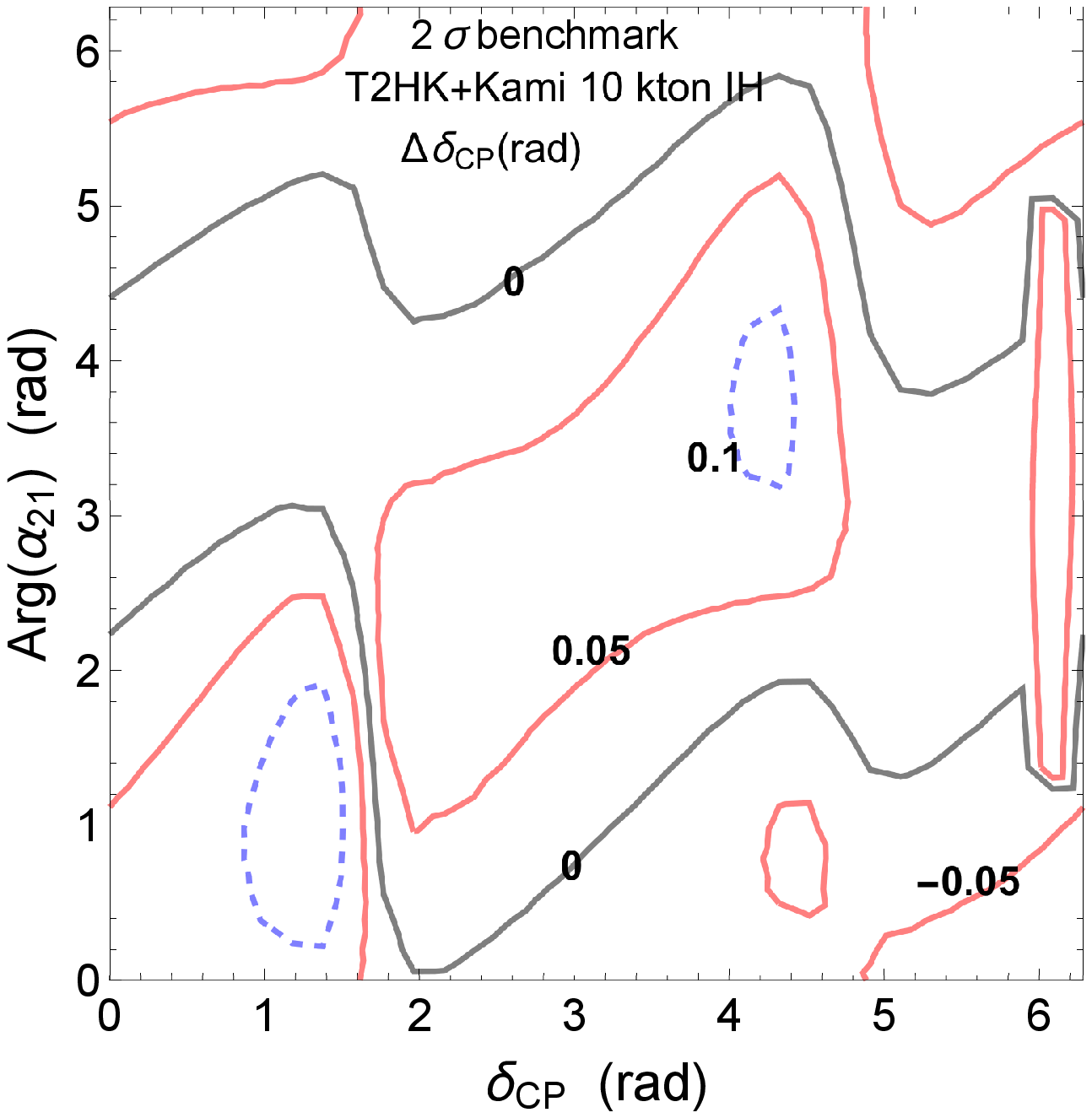}
    \includegraphics[width=80mm]{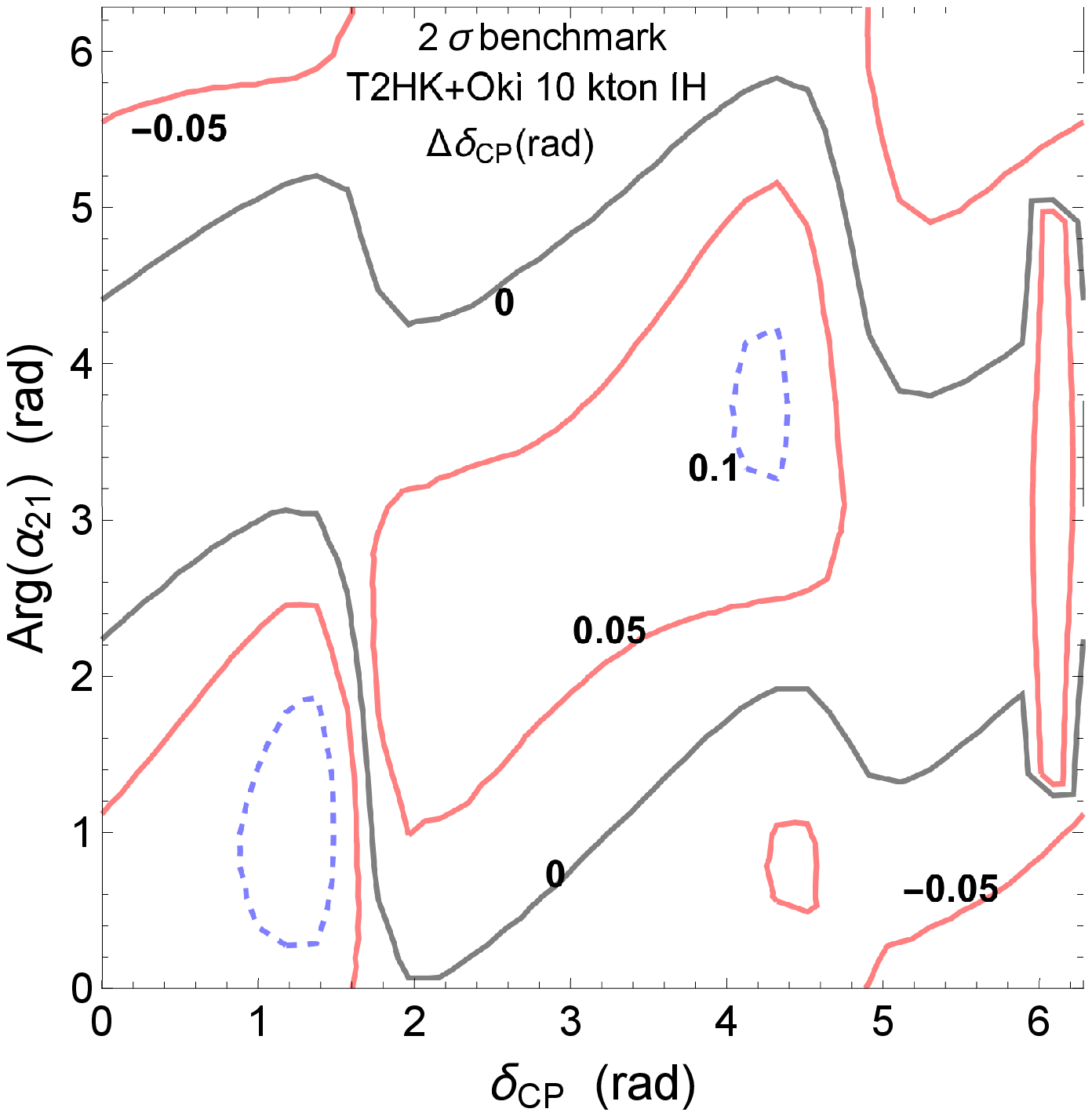}
    \caption{
       The difference between the true value of $\delta_{CP}$, and the value that minimizes $\chi^2(\Pi)$ of Eq.~(\ref{chi22})$\Delta\delta_{CP}$ with the true mass hierarchy.
    The benchmark with Eq.~(\ref{altnu}) is assumed, and the true mass hierarchy is inverted.
    The upper-left, upper-right and lower-right subplots correspond to the T2HK, the T2HKK, 
     and the plan of the T2HK plus a 10~kton water Cerenkov detector at Oki, respectively.
    For comparative study, we show in the lower-left a subplot for a plan of the T2HK plus a 10~kton water Cerenkov detector at Kamioka.
   $\vert\Delta\delta_{CP}\vert=0, \ 0.05{\rm rad}, \ 0.1{\rm rad}$, and $0.15{\rm rad}$ on the black solid, red sold, blue dashed, and purple dot-dashed contours, respectively.    
    }
    \label{ihdch2s}
  \end{center}
\end{figure}

We consider an alternative benchmark with a smaller non-unitary mixing that satisfies the experimental bounds Eqs.~(\ref{bound1}),~(\ref{bound2}) at 2$\sigma$ level
 as well as the mathematical inequality~Eq.~(\ref{bound3}).
The non-unitary mixing matrix in the new benchmark is given by
\begin{align}
N_{NU} &= \left(
\begin{array}{ccc}
\alpha_{11} & 0 & 0 \\
\alpha_{21} & \alpha_{22} & 0 \\
0 & 0 & 1
\end{array}
\right),
\nonumber \\
\alpha_{11} &=0.994486, \ \ \ \alpha_{22}=0.999897, \ \ \ \vert\alpha_{21}\vert=0.002343.
\label{altnu}
\end{align}
Based on the benchmark with Eq.~(\ref{altnu}), we repeat the same simulation and calculation as subsection~3.2 and 3.3,
 and obtain the results in Figures~\ref{nhch2s},~\ref{nhmass2s},~\ref{nhdch2s},~\ref{ihch2s},~\ref{ihmass2s},~\ref{ihdch2s}.

The plots for min$\chi^2$, Figures~\ref{nhch2s},~\ref{ihch2s}, show that the significance of non-unitary mixing is reduced below $3\sigma$ in most parameter regions in the T2HK and in the entire region in the T2HKK.
The dependence on $(\delta_{CP}, \, {\rm Arg}(\alpha_{21}))$ is similar to the plots for the original benchmark, Figures~\ref{nhch},~\ref{ihch}.

The plot for $\{\min\chi^2({\rm wrong \ H})-\min\chi^2({\rm true \ H})\}$ in the normal hierarchy case, Figure~\ref{nhmass2s},
 tells us that in the T2HK, the wrong mass hierarchy can still be favored if the true hierarchy is normal.
In contrast, Figure~\ref{ihmass2s} gives that this is no longer the case if the true hierarchy is inverted.

The plots for $\Delta_{CP}$, Figures~\ref{nhdch2s},~\ref{ihdch2s}, show that the value of $\delta_{CP}$ that minimizes $\chi^2(\Pi)$ of Eq.~(\ref{chi22}) 
 still possibly deviates from the true value by more than 0.1~radian in the T2HK, and more than 0.05~radian in the T2HKK, even with the alternative benchmark with the smaller non-unitary mixing.
This augments our argument that the non-unitary mixing terms must be incorporated in the $\delta_{CP}$ measurement when a hint of non-unitarity has been discovered.



\end{document}